\documentclass[twoside,11pt]{article}

\usepackage[accepted]{melba} %

\usepackage{amsmath,amsfonts}

\melbaheading{2021:019}{https://www.melba-journal.org/papers/2021:019.html}{2021}{1-60}{06/2021}{12/2021}{Marinescu, Oxtoby, Young, Bron, Golland, Klein and Alexander }{}{}

\ShortHeadings{The Alzheimer's Disease Prediction Of Longitudinal Evolution (TADPOLE) Challenge}{Marinescu et al.}
\firstpageno{1}

\usepackage{longtable} 

\usepackage{array}
\usepackage[a4paper, margin=1.0in]{geometry}
\usepackage{mathtools}
\usepackage{graphicx}
\usepackage{placeins}
\usepackage{booktabs}
\usepackage{multirow}
\usepackage{enumitem} 

\usepackage{epstopdf}

\usepackage[font=normalfont,labelfont=bf]{caption}

\usepackage{xcolor}

\usepackage{authblk}
\makeatletter
\renewcommand\AB@affilsepx{, \protect\Affilfont}
\makeatother

\usepackage{hyperref}
\definecolor{links}{HTML}{01368e}
\hypersetup{colorlinks,linkcolor=links,urlcolor=links,citecolor=links}

\renewcommand\Affilfont{\itshape\scriptsize} 

\title{The Alzheimer's Disease Prediction Of Longitudinal Evolution (TADPOLE) Challenge: Results after 1 Year Follow-up}
\author[1,2]{Razvan V. Marinescu}
\author[1]{Neil P. Oxtoby}
\author[1,39]{Alexandra L. Young}
\author[3]{Esther E. Bron}
\author[4]{Arthur W. Toga}
\author[5]{Michael W. Weiner}
\author[1,6,7]{Frederik Barkhof}
\author[7]{Nick C. Fox}
\author[45,1]{Arman Eshaghi}
\author[$\dagger$]{Tina Toni}
\author[$\dagger$]{Marcin Salaterski}
\author[$\dagger$]{Veronika Lunina}
\author[8]{Manon Ansart}
\author[8]{Stanley Durrleman}
\author[8]{Pascal Lu}
\author[9,10]{Samuel Iddi}
\author[9]{Dan Li}
\author[11]{Wesley K. Thompson}
\author[9]{Michael C. Donohue}
\author[12]{Aviv Nahon}
\author[12]{Yarden Levy}
\author[12]{Dan Halbersberg}
\author[12]{Mariya Cohen}
\author[13]{Huiling Liao}
\author[13]{Tengfei Li}
\author[13]{Kaixian Yu}
\author[13]{Hongtu Zhu}
\author[14]{Jos\'{e} G. Tamez-Pe\~{n}a}
\author[15]{Aya Ismail}
\author[15]{Timothy Wood}
\author[15]{Hector Corrada Bravo}
\author[16]{Minh Nguyen}
\author[16]{Nanbo Sun}
\author[16]{Jiashi Feng}
\author[16]{B.T. Thomas Yeo}
\author[17]{Gang Chen}
\author[18]{Ke Qi}
\author[18,19]{Shiyang Chen}
\author[18,19]{Deqiang Qiu}
\author[20]{Ionut Buciuman}
\author[20]{Alex Kelner}
\author[20]{Raluca Pop}
\author[20]{Denisa Rimocea}
\author[21,22,23,1]{Mostafa M. Ghazi}
\author[21,22,23]{Mads Nielsen}
\author[24,1]{Sebastien Ourselin}
\author[21,22,23]{Lauge S$\o$rensen}
\author[3]{Vikram Venkatraghavan}
\author[25]{Keli Liu}
\author[25]{Christina Rabe}
\author[25]{Paul Manser}
\author[26]{Steven M. Hill}
\author[26]{James Howlett}
\author[26]{Zhiyue Huang}
\author[26]{Steven Kiddle}
\author[42]{Sach Mukherjee}
\author[26]{Ana\"{i}s Rouanet}
\author[42]{Bernd Taschler}
\author[26]{Brian D. M. Tom}
\author[26]{Simon R. White}
\author[27]{Noel Faux}
\author[27]{Suman Sedai}
\author[14]{Javier de Velasco Oriol}
\author[14]{Edgar E. V. Clemente}
\author[28,44]{Karol Estrada}
\author[1]{Leon Aksman}
\author[1]{Andre Altmann}
\author[29]{Cynthia M. Stonnington}
\author[40]{Yalin Wang}
\author[40]{Jianfeng Wu}
\author[41]{Vivek Devadas}
\author[8]{Clementine Fourrier}
\author[30,31]{Lars Lau Raket}
\author[32]{Aristeidis Sotiras}
\author[32]{Guray Erus}
\author[32]{Jimit Doshi}
\author[32]{Christos Davatzikos}
\author[33]{Jacob Vogel}
\author[33]{Andrew Doyle}
\author[33]{Angela Tam}
\author[33]{Alex Diaz-Papkovich}
\author[34]{Emmanuel Jammeh}
\author[8]{Igor Koval}
\author[35]{Paul Moore}
\author[35]{Terry J. Lyons}
\author[43]{John Gallacher}
\author[36]{Jussi Tohka}
\author[36]{Robert Ciszek}
\author[37]{Bruno Jedynak}
\author[37]{Kruti Pandya}
\author[38]{Murat Bilgel}
\author[37]{William Engels}
\author[37]{Joseph Cole}
\author[2]{Polina Golland}
\author[3]{Stefan Klein}
\author[1]{Daniel C. Alexander}
\author[$\dagger$]{the EuroPOND Consortium}
\author[ ]{for the Alzheimer's Disease Neuroimaging Initiative\footnote{Data used in preparation of this article were obtained from the Alzheimer's Disease Neuroimaging Initiative (ADNI) database (adni.loni.usc.edu). As such, the investigators within the ADNI contributed to the design and implementation of ADNI and/or provided data but did not participate in analysis or writing of this report. A complete listing of ADNI investigators can be found at: \url{http://adni.loni.usc.edu/wp-content/uploads/how_to_apply/ADNI_Acknowledgement_List.pdf}}\\}
\affil[1]{Centre for Medical Image Computing, University College London, UK}
\affil[2]{Computer Science and Artificial Intelligence Laboratory, Massachusetts Institute of Technology, USA}
\affil[3]{Biomedical Imaging Group Rotterdam, Department of Radiology and Nuclear Medicine, Erasmus MC, Netherlands}
\affil[4]{Laboratory of NeuroImaging, University of Southern California, USA}
\affil[5]{Center for Imaging of Neurodegenerative Diseases, University of California San Francisco, USA}
\affil[6]{Department of Radiology and Nuclear Medicine, VU Medical Centre, Netherlands}
\affil[7]{Dementia Research Centre and the UK Dementia Research Institute, UCL Queen Square Institute of Neurology, UK}
\affil[8]{Institut du Cerveau et de la Moelle \'{e}pini\`{e}re, Paris, France}
\affil[9]{Alzheimer's Therapeutic Research Institute, University of Southern California, USA}
\affil[10]{Department of Statistics and Actuarial Science, University of Ghana, Ghana}
\affil[11]{Department of Family Medicine and Public Health, University of California San Diego, USA}
\affil[12]{Ben Gurion University of the Negev, Beersheba, Israel}
\affil[13]{The University of Texas Health Science Center at Houston, Houston, USA}
\affil[14]{Instituto Tecnol\'{o}gico y de Estudios Superiores de Monterrey, Monterrey, Mexico}
\affil[15]{University of Maryland, College Park, USA}
\affil[16]{National University of Singapore, Singapore, Singapore}
\affil[17]{Medical College of Wisconsin, Milwaukee, USA}
\affil[18]{Emory University, Atlanta, USA}
\affil[19]{Georgia Institute of Technology, Atlanta, USA}
\affil[20]{Vasile Lucaciu National College, Baia Mare, Romania}
\affil[21]{Biomediq A/S, Denmark}
\affil[22]{Cerebriu A/S, Denmark}
\affil[23]{University of Copenhagen, Denmark}
\affil[24]{School of Biomedical Engineering and Imaging Sciences, King's College London, UK}
\affil[25]{Genentech, USA}
\affil[26]{MRC Biostatistics Unit, University of Cambridge, UK}
\affil[27]{IBM Research Australia, Melbourne, Australia}
\affil[28]{Brandeis University, Waltham, USA}
\affil[29]{Mayo Clinic, Scottsdale, AZ, USA}
\affil[30]{H. Lundbeck A/S, Denmark}
\affil[31]{Clinical Memory Research Unit, Department of Clinical Sciences Malmo, Lund University, Lund, Sweden}
\affil[32]{Center for Biomedical Image Computing and Analytics, University of Pennsylvania}
\affil[33]{McGill University, Montreal, Canada}
\affil[34]{University of Plymouth, UK}
\affil[35]{Mathematical Institute, University of Oxford, UK}
\affil[36]{A.I. Virtanen Institute for Molecular Sciences, University of Eastern Finland, Finland}
\affil[37]{Portland State University, Portland, USA}
\affil[38]{Laboratory of Behavioral Neuroscience, National Institute on Aging,
National Institutes of Health, Baltimore, MD, USA}
\affil[39]{Department of Neuroimaging, Institute of Psychiatry, Psychology and Neuroscience, King's College London, London, UK}
\affil[40]{School of Computing, Informatics and Decision Systems Engineering, Arizona State University, Tempe, USA}
\affil[41]{Banner Alzheimer's Institute, Phoenix, USA}
\affil[42]{German Center for Neurodegenerative Diseases, Bonn, Germany}
\affil[43]{Department of Psychiatry, University of Oxford, UK}
\affil[44]{Department of Statistical Genetics, Biomarin, San Rafael, USA}
\affil[45]{Queen Square Multiple Sclerosis Centre, UCL Queen Square Institute of Neurology, UK}
\affil[$\dagger$]{Authors not affiliated with any research institution}
\affil[ ]{\normalfont{Correspondence email: \url{tadpole@cs.ucl.ac.uk}}}
\affil[ ]{\normalfont{Website: \url{https://tadpole.grand-challenge.org}}}

\renewcommand\Affilfont{\itshape\small}

\usepackage{filecontents}

\newcommand\crule[3][black]{\textcolor{#1}{\rule[0ex]{#2}{#3}}}

\newcommand{\co}[1]{\textcolor{black}{#1}}

\begin{document}

\maketitle

\sloppy 

\begin{abstract}
\end{abstract}
Accurate prediction of progression in subjects at risk of Alzheimer's disease is crucial for enrolling the right subjects in clinical trials. However, a prospective comparison of state-of-the-art algorithms for predicting disease onset and progression is currently lacking. We present the findings of \emph{The Alzheimer's Disease Prediction Of Longitudinal Evolution} (TADPOLE) Challenge, which compared the performance of 92 algorithms from 33 international teams at predicting the future trajectory of 219 individuals at risk of Alzheimer's disease. Challenge participants were required to make a prediction, for each month of a 5-year future time period, of three key outcomes: clinical diagnosis, Alzheimer's Disease Assessment Scale Cognitive Subdomain (ADAS-Cog13), and total volume of the ventricles. \co{The methods used by challenge participants included multivariate linear regression, machine learning methods such as support vector machines and deep neural networks, as well as disease progression models.} No single submission was best at predicting all three outcomes. For clinical diagnosis and ventricle volume prediction, the best algorithms strongly outperform simple baselines in predictive ability. However, for ADAS-Cog13 no single submitted prediction method was significantly better than random guesswork. Two ensemble methods based on taking the mean and median over all predictions, obtained top scores on almost all tasks. Better than average performance at diagnosis prediction was generally associated with the additional inclusion of features from cerebrospinal fluid (CSF) samples and diffusion tensor imaging (DTI). On the other hand, better performance at ventricle volume prediction was associated with inclusion of summary statistics, such as the slope or maxima/minima of patient-specific biomarkers. On a limited, cross-sectional subset of the data emulating clinical trials, performance of the best algorithms at predicting clinical diagnosis decreased only slightly (2 percentage points) compared to the full longitudinal dataset. The submission system remains open via the website \url{https://tadpole.grand-challenge.org}, while TADPOLE SHARE (\url{https://tadpole-share.github.io/}) collates code for submissions. TADPOLE's unique results suggest that current prediction algorithms provide sufficient accuracy to exploit biomarkers related to clinical diagnosis and ventricle volume, for cohort refinement in clinical trials for Alzheimer's disease. However, results call into question the usage of cognitive test scores for patient selection and as a primary endpoint in clinical trials. 


\FloatBarrier
\section{Introduction}
\label{intro}

Accurate prediction of the onset of Alzheimer's disease (AD) and its longitudinal progression is important for care planning and for patient selection in clinical trials. Current opinion holds that early detection will be critical for the successful administration of disease modifying treatments during presymptomatic phases of the disease prior to widespread brain damage, e.g. when pathological amyloid and tau start to accumulate (\cite{mehta2017trials}). Moreover, accurate prediction of the progression of at-risk subjects will help select homogenous patient groups for clinical trials, thus reducing variability in outcome measures that can obscure positive effects on patients at the right stage to benefit. 

A variety of mathematical and computational methods have been developed to predict the onset and progression of AD. Traditional approaches leverage statistical regression to model relationships between target variables (e.g. clinical diagnosis or cognitive/imaging markers) with other known markers (\cite{scahill2002mapping,sabuncu2011dynamics}) or measures derived from these markers such as the rate of cognitive decline (\cite{doody2010erratum}). \co{More recent approaches involve supervised machine learning techniques such as support vector machines (\cite{kloppel2008automatic,salas2010computer,morra2009comparison,alvarez2009alzheimer}), random forests (\cite{sarica2017random,ramirez2018ensemble,lebedev2014random,huang2016longitudinal,gray2013random}) and artificial neural networks (\cite{jo2019deep,lee2019predicting,lin2018convolutional,spasov2019parameter,li2019deep,duc20203d,cui2019rnn}). These approaches have been used to discriminate AD patients from cognitively normal individuals (\cite{kloppel2008automatic,zhang2011multimodal}), and for discriminating at-risk individuals who convert to AD in a certain time frame from those who do not (\cite{young2013accurate,mattila2011disease}). The emerging approach of disease progression modelling aims to reconstruct biomarker trajectories or other disease signatures across the disease progression timeline, without relying on clinical diagnoses or estimates of time to symptom onset. Examples include models built on a set of scalar biomarkers to produce discrete (\cite{fonteijn2012event,young2014data}) or continuous (\cite{jedynak2012computational,donohue2014estimating,lorenzi2017probabilistic,oxtoby2018data,schiratti2017bayesian,wang2014unsupervised,villemagne2013amyloid,lorenzi2019probabilistic}) biomarker trajectories; spatio-temporal models that focus on evolving image structure (\cite{bilgel2016multivariate,marinescu2019dive,abi2020monotonic,bone2018learning}), potentially conditioned by non-imaging variables (\cite{koval2018simulating}); and models that emulate putative disease mechanisms to estimate trajectories of change (\cite{raj2012network,iturria2016early,zhou2012predicting,garbarino2019differences})}. All these models show promise for predicting AD biomarker progression at group and individual levels. However, previous evaluations within individual publications provide limited information because: (1) they use different data sets or subsets of the same dataset, different processing pipelines, and different evaluation metrics and (2) over-training can occur due to heavy use of popular training datasets. Currently, the field lacks a comprehensive comparison of the capabilities of these methods on standardised tasks relevant to real-world applications.

Community challenges have consistently proved effective in moving forward the state of the art in technology to address specific data-analysis problems by providing platforms for unbiased comparative evaluation and incentives to maximise performance on key tasks (\cite{maier2018rankings}). In medical image analysis, for example, such challenges have provided important benchmarks in tasks such as registration (\cite{murphy2011evaluation}) and segmentation (\cite{menze2014multimodal}), and revealed fundamental insights about the problem studied, for example in structural brain-connectivity mapping (\cite{maier2017challenge}). Previous challenges in AD include the CADDementia challenge (\cite{bron2015standardized}), which aimed to identify clinical diagnosis from MRI scans. A similar challenge, the \emph{International challenge for automated prediction of MCI from MRI data} (\cite{castiglioni2018machine}), asked participants to predict diagnosis and conversion status from extracted MRI features of subjects from the Alzheimer's Disease Neuroimaging Initiative (ADNI) study (\cite{weiner2017recent}). Yet another challenge, \emph{The Alzheimer's Disease Big Data DREAM Challenge} (\cite{allen2016crowdsourced}), asked participants to predict cognitive decline from genetic and MRI data. These challenges have however several limitations: (i) they did not evaluate the ability of algorithms to predict biomarkers at future timepoints (with the exception of one sub-task of DREAM), which is important for patient stratification in clinical trials; (ii) the test data was available to organisers when the competitions were launched, leaving room for potential biases in the design of the challenges; (iii) the training data was drawn from a limited set of modalities.

\emph{The Alzheimer's Disease Prediction Of Longitudinal Evolution} (TADPOLE) Challenge (\url{https://tadpole.grand-challenge.org}) aims to identify the data, features and approaches that are the most predictive of future progression of subjects at risk of AD. In contrast to previous challenges, our challenge is designed to inform clinical trials through identification of patients most likely to benefit from an effective treatment, i.e., those at early stages of disease who are likely to progress over the short-to-medium term (1-5 years). The challenge focuses on forecasting the trajectories of three key features: clinical status, cognitive decline, and neurodegeneration (brain atrophy), over a five-year timescale. It uses ``rollover'' subjects from the ADNI study (\cite{weiner2017recent}) for whom a history of measurements (imaging, psychology, demographics, genetics) is available, and who are expected to continue in the study, providing future measurements for testing. TADPOLE participants were required to predict future measurements from these individuals and submit their predictions before a given submission deadline. Since the test data \emph{did not exist} at the time of forecast submissions, the challenge provides a performance comparison substantially less susceptible to many forms of potential bias than previous studies and challenges. The design choices were published (\cite{marinescu2018tadpole}) before the test set was acquired and analysed. TADPOLE also goes beyond previous challenges by drawing on a vast set of multimodal measurements from ADNI which might support prediction of AD progression.

This article presents the results of the TADPOLE Challenge and documents its key findings. We summarise the challenge design and present the results of the 92 prediction algorithms contributed by 33 participating teams worldwide, evaluated after an 18-month follow-up period. We discuss the results obtained by TADPOLE participants, which represent the current state-of-the-art in Alzheimer's disease prediction. We also report results on which input data features were most informative, and which feature selection strategies, data imputation methods and classes of algorithms were most effective. 

\section{Methods}
\label{methods}

\subsection{Predictions}
\label{pred}

TADPOLE Challenge asked participants to forecast three key biomarkers: (1) clinical diagnosis, which can be either cognitively normal (CN), mild cognitive impairment (MCI), or probable AD; (2) Alzheimer's Disease Assessment Scale Cognitive Subdomain (ADAS-Cog13) score; and (3) ventricle volume (divided by intra-cranial volume) from MRI. \co{Ventricle volume increase has been shown to be a good predictor of Alzheimer's disease diagnosis and progression \cite{nestor2008ventricular}, notably because portions of the lateral ventricles lie close to the medial  temporal lobe, which atrophy during early stages (\cite{ferrarini2006shape,giesel2006temporal}), and because it is correlated with increases in senile plaques and neurofibrillary tangles (\cite{silbert2003changes}).}

The exact time of future data acquisitions for any given individual was unknown at forecast time, so participants submitted month-by-month predictions for every individual. Predictions of clinical status comprise relative likelihoods of each option (CN, MCI, and AD) for each individual at each month. Predictions of ADAS-Cog13 and ventricle volume comprise a best-guess estimate as well as a 50\% confidence interval for each individual at each month. Full details on challenge design are given in the TADPOLE white paper (\cite{marinescu2018tadpole}).

\subsection{Data}
\label{data}

The challenge uses data from the Alzheimer's Disease Neuroimaging Initiative (ADNI) (\cite{weiner2017recent}). Specifically, the TADPOLE Challenge made four key data sets available to the challenge participants: 
\begin{itemize}
 \item \textbf{D1}: The TADPOLE standard training set draws on longitudinal data from the entire ADNI history. The data set contains measurements for every individual that has provided data to ADNI in at least two separate visits (different dates) across three phases of the study: ADNI1, ADNI GO, and ADNI2. 
 \item \textbf{D2}: The TADPOLE longitudinal prediction set contains as much available data as we could gather from the ADNI rollover individuals for whom challenge participants are asked to provide predictions. D2 includes data from all available time-points for these individuals. It defines the set of individuals for which participants are required to provide forecasts.
 \item \textbf{D3}: The TADPOLE cross-sectional prediction set contains a single (most recent) time point and a limited set of variables from each rollover individual in D2. Although we expect worse predictions from this data set than D2, D3 represents the information typically available when selecting a cohort for a clinical trial. 
 \item \textbf{D4}: The TADPOLE test set contains visits from ADNI rollover subjects that occurred after 1 Jan 2018 and contain at least one of the three outcome measures: diagnostic status, ADAS-Cog13 score, or ventricle volume. 
\end{itemize}

While participants were free to use any training datasets they wished, we provided the D1-D3 datasets in order to remove the need for participants to pre-process the data themselves, and also to be able to evaluate the performance of different algorithms on the same standardised datasets. Participants that used custom training data sets were asked also to submit results using the standard training data sets to enable direct performance comparison. We also included the D3 cross-sectional prediction set in order to simulate a clinical trial scenario. For information on how we created the D1-D4 datasets, see section \ref{creatingD14}. The software code used to generate the standard datasets is openly available on Github: \url{https://github.com/noxtoby/TADPOLE}. 

Table \ref{tab:demographics} shows the demographic breakdown of each TADPOLE data set as well as the proportion of biomarker data available in each dataset. Many entries are missing data, especially for certain biomarkers derived from exams performed on only subsets of subjects, such as tau imaging (AV1451). D1 and D2 also included demographic data typically available in ADNI (e.g. education, marital status) as well as standard genetic markers (e.g. Alipoprotein E -- APOE epsilon 4 status).


\begin{table}
 \centering
 
\setlength{\tabcolsep}{3pt} 
\begin{tabular}{c|c|c|c|c|c}
\multicolumn{6}{c}{\textbf{Demographics}}\\ 
\toprule
\multicolumn{2}{c|}{}  &          D1 &            D2 &           D3 &           D4 \\
\midrule
 \multicolumn{2}{c|}{Overall number of subjects} &          1667 &           896 &          896 &          219 \\
\midrule 
\multirow{6}{*}{Controls$^\dagger$} & Number (\% all subjects) &   508 (30.5\%) &   369 (41.2\%) &  299 (33.4\%) &   94 (42.9\%) \\
 & Visits per subject &     8.3 $\pm$ 4.5 &    8.5 $\pm$ 4.9 &    1.0 $\pm$ 0.0 &    1.0 $\pm$ 0.2 \\
 &                Age &    74.3 $\pm$ 5.8 &    73.6 $\pm$ 5.7 &   72.3 $\pm$ 6.2 &   78.4 $\pm$ 7.0 \\
 &    Gender (\% male) &         48.6\% &      47.2\% &        43.5\% &        47.9\% \\
 &               MMSE &    29.1 $\pm$ 1.1 &    29.0 $\pm$ 1.2 &   28.9 $\pm$ 1.4 &   29.1 $\pm$ 1.1 \\
  &        Converters* &   18 &      9 &       -       &       -       \\
\midrule
  \multirow{6}{*}{MCI$^\dagger$}  &        Number (\% all subjects) &   841 (50.4\%) &   458 (51.1\%) &  269 (30.0\%) &   90 (41.1\%) \\
 & Visits per subject &     8.2 $\pm$ 3.7 &    9.1 $\pm$ 3.6 &    1.0 $\pm$ 0.0 &    1.1 $\pm$ 0.3 \\
 &                Age &    73.0 $\pm$ 7.5 &    71.6 $\pm$ 7.2 &   71.9 $\pm$ 7.1 &   79.4 $\pm$ 7.0 \\
 &    Gender (\% male) &         59.3\% &      56.3\% &        58.0\% &        64.4\% \\
 &               MMSE &    27.6 $\pm$ 1.8 &    28.0 $\pm$ 1.7 &   27.6 $\pm$ 2.2 &   28.1 $\pm$ 2.1 \\
  &        Converters* &   117 &     37 &      -        &    9 \\
\midrule
  \multirow{6}{*}{AD$^\dagger$} &         Number (\% all subjects) &   318 (19.1\%) &     69 (7.7\%) &  136 (15.2\%) &   29 (13.2\%) \\
 & Visits per subject &     4.9 $\pm$ 1.6 &     5.2 $\pm$ 2.6 &    1.0 $\pm$ 0.0 &    1.1 $\pm$ 0.3 \\
 &                Age &    74.8 $\pm$ 7.7 &    75.1 $\pm$ 8.4 &   72.8 $\pm$ 7.1 &   82.2 $\pm$ 7.6 \\
 &    Gender (\% male) &         55.3\% &         68.1\% &        55.9\% &        51.7\% \\
 &               MMSE &    23.3 $\pm$ 2.0 &    23.1 $\pm$ 2.0 &   20.5 $\pm$ 5.9 &   19.4 $\pm$ 7.2 \\
  &        Converters* &       -        &        -       &      -        &    9 \\
 \multicolumn{6}{c}{}\\
 \multicolumn{6}{c}{\textbf{Number of clinical visits for all subjects with data available (\% of total visits)}}\\
\toprule
 \multicolumn{2}{c|}{}   &            D1 &            D2 &           D3 &           D4 \\
\midrule
\multicolumn{2}{c|}{ Cognitive } &  8862 (69.9\%) &  5218 (68.1\%) &  753 (84.0\%) &  223 (95.3\%) \\
\multicolumn{2}{c|}{MRI} &  7884 (62.2\%) &  4497 (58.7\%) &  224 (25.0\%) &  150 (64.1\%) \\
\multicolumn{2}{c|}{FDG-PET} &  2119 (16.7\%) &  1544 (20.2\%) &     - &     - \\
               \multicolumn{2}{c|}{AV45} &  2098 (16.6\%) &  1758 (23.0\%) &     - &     - \\
             \multicolumn{2}{c|}{AV1451} &     89 (0.7\%) &     89 (1.2\%) &     - &     - \\
                \multicolumn{2}{c|}{DTI} &    779 (6.1\%) &    636 (8.3\%) &     - &     - \\
                \multicolumn{2}{c|}{CSF} &  2347 (18.5\%) &  1458 (19.0\%) &     - &     - \\
\bottomrule
\end{tabular}
\caption{Summary of TADPOLE datasets D1-D4. ($^\dagger$) Diagnosis at first visit with available data. For D3 and D4, 192 and 6 subjects respectively did not have a diagnosis at any clinical visit, so numbers don't add up to 100\%. (*) For D4, converters are ADNI3 subjects who are MCI, but were previously CN, or who are AD, but were previously CN or MCI in their last visit in ADNI2. For D1, D2 and D3, converters are CN or MCI at their earliest available visit, who progress to a later classification of MCI/AD within 1.4 years (same duration as D4)}. 
\label{tab:demographics}
\end{table}

\subsection{Forecast Evaluation}
\label{eval}

For evaluation of clinical status predictions, we used similar metrics to those that proved effective in the CADDementia challenge (\cite{bron2015standardized}): (i) the multiclass area under the receiver operating characteristic curve (MAUC) and (ii) the overall balanced classification accuracy (BCA). For ADAS-Cog13 and ventricle volume, we used three metrics: (i) mean absolute error (MAE), weighted error score (WES) and coverage probability accuracy (CPA). BCA and MAE focus purely on prediction accuracy ignoring confidence, MAUC and WES account for accuracy and confidence, while CPA assesses the confidence interval only. The formulas for each performance metric are summarised in Table \ref{tab:perfMetrics}. See the TADPOLE white paper (\cite{marinescu2018tadpole}) for further rationale for choosing these performance metrics. In order to characterise the distribution of these metric scores, we compute scores based on 50 bootstraps with replacement on the test dataset.


\begin{table}
 \centering
 \begin{tabular}{>{\centering\arraybackslash}m{6.4cm} | >{\centering\arraybackslash}m{8.3cm}}
\textbf{Formula} & \textbf{Definitions}\\
\hline
$MAUC =\frac{1}{L(L-1)}\sum_{i=2}^L\sum_{j=1}^{i}\hat{A}(c_i|c_j)+\hat{A}(c_j|c_i)$ \vspace{5pt} 

where $\hat{A}(c_i|c_j)=\frac{S_i-n_i(n_i+1)/2}{n_i n_j}$
& $n_i$, $n_j$ -- number of points from class $i$ and $j$. $S_{ij}$ -- the sum of the ranks of the class $i$ test points, after ranking all the class $i$ and $j$ data points in increasing likelihood of belonging to class $i$. $L$ -- number of classes. $c_i$ -- class $i$.\\
\hline
$BCA = \frac{1}{2L}\sum_{i=1}^L \left[\frac{TP_i}{TP_i+FN_i}+\frac{TN_i}{TN_i+FP_i}\right]$ & $TP_i$, $FP_i$, $TN_i$, $FN_i$ – the number of true positives, false positives, true negatives and false negatives for class $i$.
$L$ – number of classes \\
\hline
$MAE = \frac{1}{N}\sum_{i=1}^{N}\left|{\tilde{M}_i-M_i}\right|$ & $M_i$ is the actual value in individual $i$ in future data. $\tilde{M}_i$ is the participant's best guess at $M_i$ and $N$ is the number of data points \\
\hline
$WES =\frac{\sum_{i=1}^{N}\tilde{C}_i\left|\tilde{M}_i-M_i\right|}{\sum_{i=1}^{N}\tilde{C}_i}$ & $M_i$, $\tilde{M}_i$ and $N$ defined as above. 
$\tilde{C}_i = (C_{+} - C_{-})^{-1}$, where $\left[C_{-}, C_{+}\right]$ is the 50\% confidence interval\\
\hline
$CPA = |ACP - 0.5|$ & actual coverage probability (ACP) - the proportion of measurements that fall within the 50\% confidence interval.\\
 \end{tabular}
 \caption{TADPOLE performance metric formulas and definitions for the terms.}
 \label{tab:perfMetrics}
\end{table}

%

\subsection{Statistical Analysis of Method Attributes with Performance}
\label{statAnalysis}

To identify which features and types of algorithms enable good predictions, we annotated each TADPOLE submission with a set of 21 attributes related to (i) feature selection (manual/automatic and large vs. small number of features), (ii) feature types (e.g. ``uses Amyloid PET''), (iii) strategy for data imputation (e.g. ``patient-wise forward-fill'') and (iv) prediction method (e.g. ``neural network'') for clinical diagnosis and ADAS/Ventricles separately. To understand which of these annotations were associated with increased performance, we applied a general linear model (\cite{kiebel2007general}), $Y = X\beta + \epsilon$, where $Y$ is the performance metric (e.g. diagnosis MAUC), $X$ is the nr\_submissions x 21 design matrix of binary annotations, and $\beta$ show the contributions of each of the 21 attributes towards achieving the performance measure $Y$.


\subsection{Algorithms}
\label{algos}
\newcolumntype{C}[1]{>{\centering\arraybackslash}m{#1}} 

A total of 33 participating teams submitted a total of 58 predictions from the longitudinal prediction set (D2), 34 predictions from the cross-sectional prediction set (D3), and 6 predictions from custom prediction sets (see section \ref{data} for description of D2/D3 datasets). A total of 8 D2/D3 submissions from 6 teams did not have predictions for all three target variables, so we computed the performance metrics for only the submitted target variables. Another 3 submissions lacked confidence intervals for either ADAS-Cog13 or ventricle volume, which we imputed using default low-width confidence ranges of 2 for ADAS-Cog13 and 0.002 for Ventricles normalised by intracranial volume (ICV).

\def\le{\\\hline}
\fontsize{7}{8}\selectfont
\setlength{\tabcolsep}{3pt} 

\newcommand{\hei}{0.16cm}
\newcommand{\wid}{0.16cm}


\begin{longtable}{>{\raggedleft}C{3.7cm}|C{1.4cm}|C{1.2cm}|C{1.5cm}|C{1.8cm}|C{1.8cm}|C{1.3cm}|C{1.4cm}}



\toprule
\textbf{Submission} & \textbf{Feature selection} & \textbf{Number of features} & \textbf{Missing data imputation} & \textbf{Diagnosis prediction model} & \textbf{ADAS/Vent. prediction model} & \textbf{Training time} & \textbf{Prediction time (one subject)}\\
\midrule
 AlgosForGood \crule[blue]{\hei}{\wid}\crule[blue]{\hei}{\wid}  & manual & 16+5* & forward-filling & Aalen model & linear regression & 1 min. & 1 sec.\le
 Apocalypse \crule[red]{\hei}{\wid}\crule[blue]{\hei}{\wid}  & manual & 16 & population average & SVM & linear regression & 40 min. & 3 min.\le
 ARAMIS-Pascal \crule[blue]{\hei}{\wid}\crule[gray]{\hei}{\wid}  & manual & 20 & population average & Aalen model & - & 16 sec. & 0.02 sec.\le
 ATRI-Biostat-JMM \crule[magenta]{\hei}{\wid}\crule[blue]{\hei}{\wid} & automatic & 15 & random forest & random forest & linear mixed effects model & 2 days & 1 sec.\le
 ATRI-Biostat-LTJMM \crule[magenta]{\hei}{\wid}\crule[green]{\hei}{\wid}  & automatic & 15 & random forest & random forest & DPM & 2 days & 1 sec.\le
 ATRI-Biostat-MA \crule[magenta]{\hei}{\wid}\crule[green]{\hei}{\wid}  & automatic & 15 & random forest & random forest & DPM + linear mixed effects model & 2 days & 1 sec.\le
BGU-LSTM \crule[cyan]{\hei}{\wid}\crule[cyan]{\hei}{\wid}  & automatic & 67 & none & feed-forward NN & LSTM & 1 day & millisec.\le
 BGU-RF/ BGU-RFFIX \crule[magenta]{\hei}{\wid}\crule[magenta]{\hei}{\wid} & automatic & $\approx$ 67+1340* & none & semi-temporal RF & semi-temporal RF & a few min. & millisec.\le
 BIGS2 \crule[magenta]{\hei}{\wid}\crule[blue]{\hei}{\wid} & automatic & all & Iterative Thresholded SVD & RF & linear regression & 2.2 sec. & 0.001 sec.\le
 Billabong (all) \crule[green]{\hei}{\wid}\crule[green]{\hei}{\wid} & manual & 15-16 & linear regression & linear scale & non-parametric SM & 7 hours & 0.13 sec.\le
 BORREGOSTECMTY \crule[blue]{\hei}{\wid}\crule[blue]{\hei}{\wid} & automatic & $\approx$100 + 400* & nearest-neighbour & regression ensemble & ensemble of regression + hazard models  & 18 hours & 0.001 sec.\le
 BravoLab \crule[cyan]{\hei}{\wid}\crule[cyan]{\hei}{\wid} & automatic & 25 & hot deck & LSTM & LSTM & 1 hour & a few sec.\le
 CBIL \crule[cyan]{\hei}{\wid}\crule[cyan]{\hei}{\wid} & manual & 21 & linear interpolation & LSTM & LSTM & 1 hour & one min.\le
 Chen-MCW \crule[blue]{\hei}{\wid}\crule[green]{\hei}{\wid} & manual & 9 & none & linear regression & DPM & 4 hours & $<$ 1 hour\le
 CN2L-NeuralNetwork \crule[cyan]{\hei}{\wid}\crule[cyan]{\hei}{\wid} & automatic & all & forward-filling & RNN & RNN & 24 hours & a few sec.\le
 CN2L-RandomForest \crule[magenta]{\hei}{\wid}\crule[magenta]{\hei}{\wid} & manual & $>$200 & forward-filling & RF & RF & 15 min. & $<$ 1 min.\le
 CN2L-Average \crule[cyan]{\hei}{\wid}\crule[cyan]{\hei}{\wid} & automatic & all & forward-filling & RNN/RF & RNN/RF & 24 hours & $<$ 1 min.\le
 CyberBrains \crule[blue]{\hei}{\wid}\crule[blue]{\hei}{\wid} & manual & 5 & population average & linear regression & linear regression & 20 sec. & 20 sec.\le
 DIKU (all) \crule[blue]{\hei}{\wid}\crule[green]{\hei}{\wid} & semi-automatic & 18 & none & Bayesian classifier/LDA + DPM & DPM & 290 sec. & 0.025 sec.\le
 DIVE \crule[green]{\hei}{\wid}\crule[green]{\hei}{\wid} & manual & 13 & none & KDE+DPM & DPM & 20 min. & 0.06 sec.\le
 EMC1 \crule[red]{\hei}{\wid}\crule[green]{\hei}{\wid} & automatic & 250 & nearest neighbour & DPM + 2D spline + SVM & DPM + 2D spline & 80 min. & a few sec.\le
 EMC-EB \crule[red]{\hei}{\wid}\crule[red]{\hei}{\wid} & automatic & 200-338 & nearest-neighbour & SVM classifier & SVM regressor & 20 sec. & a few sec.\le
 FortuneTellerFish-Control \crule[red]{\hei}{\wid}\crule[blue]{\hei}{\wid} & manual & 19 & nearest neighbour & multiclass ECOC SVM & linear mixed effects model & 1 min. & $<$ 1 sec.\le
 FortuneTellerFish-SuStaIn \crule[red]{\hei}{\wid}\crule[green]{\hei}{\wid} & manual & 19 & nearest neighbour & multiclass ECOC SVM + DPM & linear mixed effects model + DPM & 5 hours & $<$ 1 sec.\le
 Frog \crule[red]{\hei}{\wid}\crule[red]{\hei}{\wid} & automatic & $\approx$ 70+420* & none & gradient boosting & gradient boosting & 1 hour & -\le
 GlassFrog-LCMEM-HDR \crule[blue]{\hei}{\wid}\crule[green]{\hei}{\wid} & semi-automatic & all & forward-fill/nearest-neigh. & multi-state model & DPM + regression & 15 min. & 2 min.\le
 GlassFrog-SM \crule[blue]{\hei}{\wid}\crule[green]{\hei}{\wid} & manual & 7 & linear model & multi-state model & parametric SM & 93 sec. & 0.1 sec.\le
 GlassFrog-Average \crule[blue]{\hei}{\wid}\crule[green]{\hei}{\wid} & semi-automatic & all & forward-fill/linear & multi-state model & DPM + SM + regression & 15 min. & 2 min.\le
 IBM-OZ-Res \crule[blue]{\hei}{\wid}\crule[blue]{\hei}{\wid} & manual & Oct-15 & filled with zero & stochastic gradient boosting & stochastic gradient boosting & 20 min. & 0.1 sec.\le
 ITESMCEM \crule[magenta]{\hei}{\wid}\crule[blue]{\hei}{\wid} & manual & 48 & mean of previous values & RF & LASSO + Bayesian ridge regression & 20 min. & 0.3 sec.\le
 lmaUCL (all) \crule[blue]{\hei}{\wid}\crule[blue]{\hei}{\wid}  & manual & 5 & regression & multi-task learning & multi-task learning & 2 hours & millisec.\le
 Mayo-BAI-ASU \crule[blue]{\hei}{\wid}\crule[blue]{\hei}{\wid} & manual & 15 & population average & linear mixed effects model & linear mixed effects model & 20 min. & 1.3 sec.\le
 Orange \crule[gray]{\hei}{\wid}\crule[gray]{\hei}{\wid}  & manual & 17 & none & clinician's decision tree & clinician's decision tree & none & 0.2 sec.\le
 Rocket \crule[blue]{\hei}{\wid}\crule[green]{\hei}{\wid} & manual & 6 & median of diagnostic group & linear mixed effects model & DPM & 5 min. & 0.3 sec.\le
 SBIA \crule[red]{\hei}{\wid}\crule[blue]{\hei}{\wid}  & manual & 30-70 & dropped visits with missing data & SVM + density estimator & linear mixed effects model & 1 min. & a few sec.\le
 SPMC-Plymouth (all) \crule[gray]{\hei}{\wid}\crule[gray]{\hei}{\wid} & automatic & 20 & none & unknown & - & unknown & 1 min.\le
 SmallHeads-NeuralNetwork \crule[cyan]{\hei}{\wid}\crule[cyan]{\hei}{\wid} & automatic & 376 & nearest neighbour & deep fully -connected NN & deep fully -connected NN & 40 min. & 0.06 sec.\le
 SmallHeads-LinMixedEffects \crule[gray]{\hei}{\wid}\crule[blue]{\hei}{\wid} & automatic & unknown & nearest neighbour & - & linear mixed effects model & 25 min. & 0.13 sec.\le
 Sunshine (all) \crule[red]{\hei}{\wid}\crule[blue]{\hei}{\wid} & semi-automatic & 6 & population average & SVM & linear model & 30 min. & $<$ 1 min.\le
 Threedays \crule[magenta]{\hei}{\wid}\crule[gray]{\hei}{\wid} & manual & 16 & none & RF & - & 1 min. & 3 sec.\le
 Tohka-Ciszek-SMNSR \crule[gray]{\hei}{\wid}\crule[blue]{\hei}{\wid} & manual & $\approx$ 32 & nearest neighbour & - & SMNSR & several hours & a few sec.\le
 Tohka-Ciszek-RandomForestLin \crule[magenta]{\hei}{\wid}\crule[magenta]{\hei}{\wid} & manual & ~32 & mean patient value & RF & linear model & a few min. & a few sec.\le
 VikingAI (all) \crule[green]{\hei}{\wid}\crule[green]{\hei}{\wid} & manual & 10 & none & DPM + ordered logit model & DPM & 10 hours & 8 sec.\le
 BenchmaskLastVisit \crule[orange]{\hei}{\wid}\crule[orange]{\hei}{\wid} & None & 3 & none & constant model & constant model & 7 sec. & millisec.\le
 BenchmarkMixedEffects \crule[orange]{\hei}{\wid}\crule[orange]{\hei}{\wid} & None & 3 & none & Gaussian model & linear mixed effects model & 30 sec. & 0.003 sec.\le
 BenchmarkMixedEffects-APOE \crule[orange]{\hei}{\wid}\crule[orange]{\hei}{\wid} & None & 4 & none & Gaussian model & linear mixed effects model & 30 sec. & 0.003 sec.\le
 BenchmarkSVM \crule[orange]{\hei}{\wid}\crule[orange]{\hei}{\wid} & manual & 6 & mean of previous values & SVM & support vector regressor (SVR) & 20 sec. & 0.001 sec.\le
  \bottomrule

 \caption{Summary of prediction methods used in the TADPOLE submissions. Keywords: SVM -- Support Vector Machine, RF -- random forest, LSTM -- long short-term memory network, NN -- neural network, RNN -- recurrent neural network, SMNSR -- Sparse Multimodal Neighbourhood Search Regression, DPM -- disease progression model, KDE -- kernel density estimation, LDA -- linear discriminant analysis, SM -- slope model, ECOC -- error-correcting output codes, SVD -- singular value decomposition (*) Augmented features, or summary statistics, such as trends, slope, min/max, moments, generally derived patient-wise using longitudinal data. Color tags denote prediction method category: \crule[blue]{0.3cm}{0.3cm} regression/proportional hazards model, \crule[magenta]{0.3cm}{0.3cm} random forest, \crule[cyan]{0.3cm}{0.3cm} neural networks, \crule[green]{0.3cm}{0.3cm} disease progression model, \crule[red]{0.3cm}{0.3cm} machine learning (other), \crule[orange]{0.3cm}{0.3cm} benchmark, \crule[gray]{0.3cm}{0.3cm} other. The left-side box denotes the category for diagnosis prediction method, while the right-side box denotes the category for ADAS/Ventricle prediction method.}
 \label{tabSummary}
\end{longtable}
\normalsize
\setlength{\tabcolsep}{6pt} 

Table \ref{tabSummary} summarises the methods used in the submissions in terms of feature selection, handling of missing data, predictive models for clinical diagnosis and ADAS/Ventricles biomarkers, as well as training and prediction times. A detailed description of each method is in section \ref{predMethods}. In particular, some entries constructed augmented features (i.e. summary statistics), which are extra features such as slope, min/max or moments that are derived from existing features.

In addition to the forecasts submitted by participants, we also evaluated four benchmark methods, which were made available to participants during the submission phase of the challenge: (i) \emph{BenchmaskLastVisit} uses the measurement of each target from the last available clinical visit as the forecast, (ii) \emph{BenchmarkMixedEffects} uses a mixed effects model with age as predictor variable for ADAS and Ventricle predictions, and Gaussian likelihood model for diagnosis prediction, (iii) \emph{BenchmarkMixedEffectsAPOE} is as (ii) but adds APOE status as a covariate and (iv) \emph{BenchmarkSVM} uses an out-of-the-box support vector machine (SVM) classifier and regressor (SVR) to provide forecasts. More details on these methods can be found in section \ref{predMethods}. We also evaluated two ensemble methods based on taking the mean (\emph{ConsensusMean}) and median (\emph{ConsensusMedian}) of the forecasted variables over all submissions. We further evaluated 100 random predictions by adding Gaussian noise to the forecasts of the simplest benchmark model (\emph{BenchmarkLastVisit}), to control for potentially spurious strong performance arising from multiple comparisons. In the subsequent results tables we will show, for each performance metric, only the best score obtained by any of these 100 random predictions (\emph{RandomisedBest}) -- See end of section \ref{predMethods} for more information on \emph{RandomisedBest}.

\section{Results}
\label{results}

\subsection{Forecasts from the longitudinal prediction set (D2)}
\label{resultsD2}

\begin{table}
\resizebox{0.95\textwidth}{!}{%
\begin{tabular}{rc|ccc|cccc|cccc}
\toprule
  & Overall & \multicolumn{3}{c|}{Diagnosis} & \multicolumn{4}{c|}{ADAS-Cog13}  & \multicolumn{4}{c}{Ventricles (\% ICV)}\\
                       Submission & Rank &  Rank & MAUC & BCA  & Rank & MAE & WES & CPA & Rank & MAE & WES & CPA\\
\midrule
ConsensusMedian \crule[gray]{0.3cm}{0.3cm}\crule[gray]{0.3cm}{0.3cm}  & - & - & 0.925 & \textbf{0.857} & - & 5.12 & 5.01 & 0.28 & \textbf{-} & \textbf{0.38} & 0.33 & 0.09\\ 
Frog \crule[red]{0.3cm}{0.3cm}\crule[red]{0.3cm}{0.3cm}  & \textbf{1} & \textbf{1} & \textbf{0.931} & 0.849 & 4 & 4.85 & 4.74 & 0.44 & 10 & 0.45 & 0.33 & 0.47\\ 
ConsensusMean \crule[gray]{0.3cm}{0.3cm}\crule[gray]{0.3cm}{0.3cm}  & - & - & 0.920 & 0.835 & \textbf{-} & \textbf{3.75} & \textbf{3.54} & \textbf{0.00} & - & 0.48 & 0.45 & 0.13\\ 
EMC1-Std \crule[red]{0.3cm}{0.3cm}\crule[green]{0.3cm}{0.3cm}  & 2 & 8 & 0.898 & 0.811 & 23-24 & 6.05 & 5.40 & 0.45 & 1-2 & 0.41 & \textbf{0.29} & 0.43\\ 
VikingAI-Sigmoid \crule[green]{0.3cm}{0.3cm}\crule[green]{0.3cm}{0.3cm}  & 3 & 16 & 0.875 & 0.760 & 7 & 5.20 & 5.11 & 0.02 & 11-12 & 0.45 & 0.35 & 0.20\\ 
EMC1-Custom \crule[red]{0.3cm}{0.3cm}\crule[green]{0.3cm}{0.3cm}  & 4 & 11 & 0.892 & 0.798 & 23-24 & 6.05 & 5.40 & 0.45 & 1-2 & 0.41 & \textbf{0.29} & 0.43\\ 
CBIL \crule[cyan]{0.3cm}{0.3cm}\crule[cyan]{0.3cm}{0.3cm}  & 5 & 9 & 0.897 & 0.803 & 15 & 5.66 & 5.65 & 0.37 & 13 & 0.46 & 0.46 & 0.09\\ 
Apocalypse \crule[red]{0.3cm}{0.3cm}\crule[blue]{0.3cm}{0.3cm}  & 6 & 7 & 0.902 & 0.827 & 14 & 5.57 & 5.57 & 0.50 & 20 & 0.52 & 0.52 & 0.50\\ 
GlassFrog-Average \crule[blue]{0.3cm}{0.3cm}\crule[green]{0.3cm}{0.3cm}  & 7 & 4-6 & 0.902 & 0.825 & 8 & 5.26 & 5.27 & 0.26 & 29 & 0.68 & 0.60 & 0.33\\ 
GlassFrog-SM \crule[blue]{0.3cm}{0.3cm}\crule[green]{0.3cm}{0.3cm}  & 8 & 4-6 & 0.902 & 0.825 & 17 & 5.77 & 5.92 & 0.20 & 21 & 0.52 & 0.33 & 0.20\\ 
BORREGOTECMTY \crule[blue]{0.3cm}{0.3cm}\crule[blue]{0.3cm}{0.3cm}  & 9 & 19 & 0.866 & 0.808 & 20 & 5.90 & 5.82 & 0.39 & 5 & 0.43 & 0.37 & 0.40\\ 
BenchmarkMixedEffects \crule[orange]{0.3cm}{0.3cm}\crule[orange]{0.3cm}{0.3cm}  & - & - & 0.846 & 0.706 & - & 4.19 & 4.19 & 0.31 & - & 0.56 & 0.56 & 0.50\\ 
EMC-EB \crule[red]{0.3cm}{0.3cm}\crule[red]{0.3cm}{0.3cm}  & 10 & 3 & 0.907 & 0.805 & 39 & 6.75 & 6.66 & 0.50 & 9 & 0.45 & 0.40 & 0.48\\ 
lmaUCL-Covariates \crule[blue]{0.3cm}{0.3cm}\crule[blue]{0.3cm}{0.3cm}  & 11-12 & 22 & 0.852 & 0.760 & 27 & 6.28 & 6.29 & 0.28 & 3 & 0.42 & 0.41 & 0.11\\ 
CN2L-Average \crule[cyan]{0.3cm}{0.3cm}\crule[cyan]{0.3cm}{0.3cm}  & 11-12 & 27 & 0.843 & 0.792 & 9 & 5.31 & 5.31 & 0.35 & 16 & 0.49 & 0.49 & 0.33\\ 
VikingAI-Logistic \crule[green]{0.3cm}{0.3cm}\crule[green]{0.3cm}{0.3cm}  & 13 & 20 & 0.865 & 0.754 & 21 & 6.02 & 5.91 & 0.26 & 11-12 & 0.45 & 0.35 & 0.20\\ 
lmaUCL-Std \crule[blue]{0.3cm}{0.3cm}\crule[blue]{0.3cm}{0.3cm}  & 14 & 21 & 0.859 & 0.781 & 28 & 6.30 & 6.33 & 0.26 & 4 & 0.42 & 0.41 & 0.09\\ 
RandomisedBest \crule[gray]{0.3cm}{0.3cm}\crule[gray]{0.3cm}{0.3cm}  & - & - & 0.800 & 0.803 & - & 4.52 & 4.52 & 0.27 & - & 0.46 & 0.45 & 0.33\\ 
CN2L-RandomForest \crule[magenta]{0.3cm}{0.3cm}\crule[magenta]{0.3cm}{0.3cm}  & 15-16 & 10 & 0.896 & 0.792 & 16 & 5.73 & 5.73 & 0.42 & 31 & 0.71 & 0.71 & 0.41\\ 
FortuneTellerFish-SuStaIn \crule[red]{0.3cm}{0.3cm}\crule[green]{0.3cm}{0.3cm}  & 15-16 & 40 & 0.806 & 0.685 & 3 & 4.81 & 4.81 & 0.21 & 14 & 0.49 & 0.49 & 0.18\\ 
CN2L-NeuralNetwork \crule[cyan]{0.3cm}{0.3cm}\crule[cyan]{0.3cm}{0.3cm}  & 17 & 41 & 0.783 & 0.717 & 10 & 5.36 & 5.36 & 0.34 & 7 & 0.44 & 0.44 & 0.27\\ 
BenchmarkMixedEffectsAPOE \crule[orange]{0.3cm}{0.3cm}\crule[orange]{0.3cm}{0.3cm}  & 18 & 35 & 0.822 & 0.749 & 2 & 4.75 & 4.75 & 0.36 & 23 & 0.57 & 0.57 & 0.40\\ 
Tohka-Ciszek-RandomForestLin \crule[magenta]{0.3cm}{0.3cm}\crule[magenta]{0.3cm}{0.3cm}  & 19 & 17 & 0.875 & 0.796 & 22 & 6.03 & 6.03 & 0.15 & 22 & 0.56 & 0.56 & 0.37\\ 
BGU-LSTM \crule[cyan]{0.3cm}{0.3cm}\crule[cyan]{0.3cm}{0.3cm}  & 20 & 12 & 0.883 & 0.779 & 25 & 6.09 & 6.12 & 0.39 & 25 & 0.60 & 0.60 & 0.23\\ 
DIKU-GeneralisedLog-Custom \crule[blue]{0.3cm}{0.3cm}\crule[green]{0.3cm}{0.3cm}  & 21 & 13 & 0.878 & 0.790 & 11-12 & 5.40 & 5.40 & 0.26 & 38-39 & 1.05 & 1.05 & 0.05\\ 
DIKU-GeneralisedLog-Std \crule[blue]{0.3cm}{0.3cm}\crule[green]{0.3cm}{0.3cm}  & 22 & 14 & 0.877 & 0.790 & 11-12 & 5.40 & 5.40 & 0.26 & 38-39 & 1.05 & 1.05 & 0.05\\ 
CyberBrains \crule[blue]{0.3cm}{0.3cm}\crule[blue]{0.3cm}{0.3cm}  & 23 & 34 & 0.823 & 0.747 & 6 & 5.16 & 5.16 & 0.24 & 26 & 0.62 & 0.62 & 0.12\\ 
AlgosForGood \crule[blue]{0.3cm}{0.3cm}\crule[blue]{0.3cm}{0.3cm}  & 24 & 24 & 0.847 & 0.810 & 13 & 5.46 & 5.11 & 0.13 & 30 & 0.69 & 3.31 & 0.19\\ 
lmaUCL-halfD1 \crule[blue]{0.3cm}{0.3cm}\crule[blue]{0.3cm}{0.3cm}  & 25 & 26 & 0.845 & 0.753 & 38 & 6.53 & 6.51 & 0.31 & 6 & 0.44 & 0.42 & 0.13\\ 
BGU-RF \crule[magenta]{0.3cm}{0.3cm}\crule[magenta]{0.3cm}{0.3cm}  & 26 & 28 & 0.838 & 0.673 & 29-30 & 6.33 & 6.10 & 0.35 & 17-18 & 0.50 & 0.38 & 0.26\\ 
Mayo-BAI-ASU \crule[blue]{0.3cm}{0.3cm}\crule[blue]{0.3cm}{0.3cm}  & 27 & 52 & 0.691 & 0.624 & 5 & 4.98 & 4.98 & 0.32 & 19 & 0.52 & 0.52 & 0.40\\ 
BGU-RFFIX \crule[magenta]{0.3cm}{0.3cm}\crule[magenta]{0.3cm}{0.3cm}  & 28 & 32 & 0.831 & 0.673 & 29-30 & 6.33 & 6.10 & 0.35 & 17-18 & 0.50 & 0.38 & 0.26\\ 
FortuneTellerFish-Control \crule[red]{0.3cm}{0.3cm}\crule[blue]{0.3cm}{0.3cm}  & 29 & 31 & 0.834 & 0.692 & 1 & 4.70 & 4.70 & 0.22 & 50 & 1.38 & 1.38 & 0.50\\ 
GlassFrog-LCMEM-HDR \crule[blue]{0.3cm}{0.3cm}\crule[green]{0.3cm}{0.3cm}  & 30 & 4-6 & 0.902 & 0.825 & 31 & 6.34 & 6.21 & 0.47 & 51 & 1.66 & 1.59 & 0.41\\ 
SBIA \crule[red]{0.3cm}{0.3cm}\crule[blue]{0.3cm}{0.3cm}  & 31 & 43 & 0.776 & 0.721 & 43 & 7.10 & 7.38 & 0.40 & 8 & 0.44 & 0.31 & 0.13\\ 
Chen-MCW-Stratify \crule[blue]{0.3cm}{0.3cm}\crule[green]{0.3cm}{0.3cm}  & 32 & 23 & 0.848 & 0.783 & 36-37 & 6.48 & 6.24 & 0.23 & 36-37 & 1.01 & 1.00 & 0.11\\ 
Rocket \crule[blue]{0.3cm}{0.3cm}\crule[green]{0.3cm}{0.3cm}  & 33 & 54 & 0.680 & 0.519 & 18 & 5.81 & 5.71 & 0.34 & 28 & 0.64 & 0.64 & 0.29\\ 
BenchmarkSVM \crule[orange]{0.3cm}{0.3cm}\crule[orange]{0.3cm}{0.3cm}  & 34-35 & 30 & 0.836 & 0.764 & 40 & 6.82 & 6.82 & 0.42 & 32 & 0.86 & 0.84 & 0.50\\ 
Chen-MCW-Std \crule[blue]{0.3cm}{0.3cm}\crule[green]{0.3cm}{0.3cm}  & 34-35 & 29 & 0.836 & 0.778 & 36-37 & 6.48 & 6.24 & 0.23 & 36-37 & 1.01 & 1.00 & 0.11\\ 
DIKU-ModifiedMri-Custom \crule[blue]{0.3cm}{0.3cm}\crule[green]{0.3cm}{0.3cm}  & 36 & 36-37 & 0.807 & 0.670 & 32-35 & 6.44 & 6.44 & 0.27 & 34-35 & 0.92 & 0.92 & \textbf{0.01}\\ 
DIKU-ModifiedMri-Std \crule[blue]{0.3cm}{0.3cm}\crule[green]{0.3cm}{0.3cm}  & 37 & 38-39 & 0.806 & 0.670 & 32-35 & 6.44 & 6.44 & 0.27 & 34-35 & 0.92 & 0.92 & \textbf{0.01}\\ 
DIVE \crule[green]{0.3cm}{0.3cm}\crule[green]{0.3cm}{0.3cm}  & 38 & 51 & 0.708 & 0.568 & 42 & 7.10 & 7.10 & 0.34 & 15 & 0.49 & 0.49 & 0.13\\ 
ITESMCEM \crule[magenta]{0.3cm}{0.3cm}\crule[blue]{0.3cm}{0.3cm}  & 39 & 53 & 0.680 & 0.657 & 26 & 6.26 & 6.26 & 0.35 & 33 & 0.92 & 0.92 & 0.43\\ 
BenchmarkLastVisit \crule[orange]{0.3cm}{0.3cm}\crule[orange]{0.3cm}{0.3cm}  & 40 & 44-45 & 0.774 & 0.792 & 41 & 7.05 & 7.05 & 0.45 & 27 & 0.63 & 0.61 & 0.47\\ 
Sunshine-Conservative \crule[red]{0.3cm}{0.3cm}\crule[blue]{0.3cm}{0.3cm}  & 41 & 25 & 0.845 & 0.816 & 44-45 & 7.90 & 7.90 & 0.50 & 43-44 & 1.12 & 1.12 & 0.50\\ 
BravoLab \crule[cyan]{0.3cm}{0.3cm}\crule[cyan]{0.3cm}{0.3cm}  & 42 & 46 & 0.771 & 0.682 & 47 & 8.22 & 8.22 & 0.49 & 24 & 0.58 & 0.58 & 0.41\\ 
DIKU-ModifiedLog-Custom \crule[blue]{0.3cm}{0.3cm}\crule[green]{0.3cm}{0.3cm}  & 43 & 36-37 & 0.807 & 0.670 & 32-35 & 6.44 & 6.44 & 0.27 & 47-48 & 1.17 & 1.17 & 0.06\\ 
DIKU-ModifiedLog-Std \crule[blue]{0.3cm}{0.3cm}\crule[green]{0.3cm}{0.3cm}  & 44 & 38-39 & 0.806 & 0.670 & 32-35 & 6.44 & 6.44 & 0.27 & 47-48 & 1.17 & 1.17 & 0.06\\ 
Sunshine-Std \crule[red]{0.3cm}{0.3cm}\crule[blue]{0.3cm}{0.3cm}  & 45 & 33 & 0.825 & 0.771 & 44-45 & 7.90 & 7.90 & 0.50 & 43-44 & 1.12 & 1.12 & 0.50\\ 
Billabong-UniAV45 \crule[green]{0.3cm}{0.3cm}\crule[green]{0.3cm}{0.3cm}  & 46 & 49 & 0.720 & 0.616 & 48-49 & 9.22 & 8.82 & 0.29 & 41-42 & 1.09 & 0.99 & 0.45\\ 
Billabong-Uni \crule[green]{0.3cm}{0.3cm}\crule[green]{0.3cm}{0.3cm}  & 47 & 50 & 0.718 & 0.622 & 48-49 & 9.22 & 8.82 & 0.29 & 41-42 & 1.09 & 0.99 & 0.45\\ 
ATRI-Biostat-JMM \crule[magenta]{0.3cm}{0.3cm}\crule[blue]{0.3cm}{0.3cm}  & 48 & 42 & 0.779 & 0.710 & 51 & 12.88 & 69.62 & 0.35 & 54 & 1.95 & 5.12 & 0.33\\ 
Billabong-Multi \crule[green]{0.3cm}{0.3cm}\crule[green]{0.3cm}{0.3cm}  & 49 & 56 & 0.541 & 0.556 & 55 & 27.01 & 19.90 & 0.46 & 40 & 1.07 & 1.07 & 0.45\\ 
ATRI-Biostat-MA \crule[magenta]{0.3cm}{0.3cm}\crule[green]{0.3cm}{0.3cm}  & 50 & 47 & 0.741 & 0.671 & 52 & 12.88 & 11.32 & 0.19 & 53 & 1.84 & 5.27 & 0.23\\ 
BIGS2 \crule[magenta]{0.3cm}{0.3cm}\crule[blue]{0.3cm}{0.3cm}  & 51 & 58 & 0.455 & 0.488 & 50 & 11.62 & 14.65 & 0.50 & 49 & 1.20 & 1.12 & 0.07\\ 
Billabong-MultiAV45 \crule[green]{0.3cm}{0.3cm}\crule[green]{0.3cm}{0.3cm}  & 52 & 57 & 0.527 & 0.530 & 56 & 28.45 & 21.22 & 0.47 & 45 & 1.13 & 1.07 & 0.47\\ 
ATRI-Biostat-LTJMM \crule[magenta]{0.3cm}{0.3cm}\crule[green]{0.3cm}{0.3cm}  & 53 & 55 & 0.636 & 0.563 & 54 & 16.07 & 74.65 & 0.33 & 52 & 1.80 & 5.01 & 0.26\\ 
Threedays \crule[magenta]{0.3cm}{0.3cm}\crule[gray]{0.3cm}{0.3cm}  & - & 2 & 0.921 & 0.823 & - & - & - & - & - & - & - & -\\ 
ARAMIS-Pascal \crule[blue]{0.3cm}{0.3cm}\crule[gray]{0.3cm}{0.3cm}  & - & 15 & 0.876 & 0.850 & - & - & - & - & - & - & - & -\\ 
IBM-OZ-Res \crule[blue]{0.3cm}{0.3cm}\crule[blue]{0.3cm}{0.3cm}  & - & 18 & 0.868 & 0.766 & - & - & - & - & 46 & 1.15 & 1.15 & 0.50\\ 
Orange \crule[gray]{0.3cm}{0.3cm}\crule[gray]{0.3cm}{0.3cm}  & - & 44-45 & 0.774 & 0.792 & - & - & - & - & - & - & - & -\\ 
SMALLHEADS-NeuralNet \crule[cyan]{0.3cm}{0.3cm}\crule[cyan]{0.3cm}{0.3cm}  & - & 48 & 0.737 & 0.605 & 53 & 13.87 & 13.87 & 0.41 & - & - & - & -\\ 
SMALLHEADS-LinMixedEffects \crule[gray]{0.3cm}{0.3cm}\crule[blue]{0.3cm}{0.3cm}  & - & - & - & - & 46 & 8.09 & 7.94 & 0.04 & - & - & - & -\\ 
Tohka-Ciszek-SMNSR \crule[gray]{0.3cm}{0.3cm}\crule[blue]{0.3cm}{0.3cm}  & - & - & - & - & 19 & 5.87 & 5.87 & 0.14 & - & - & - & -\\ 

  \bottomrule
\end{tabular}
}
\caption{Ranked scores for all TADPOLE submissions and benchmarks using the longitudinal prediction data set (D2). Best scores in each category are bolded. Missing numerical entries indicate that submissions did not include forecasts for the corresponding target variable. The ``Diagnosis'' ranking uses multiclass area under the receiver operating characteristic curve (MAUC), those of ADAS-Cog13 and Ventricles use mean absolute error (MAE). The overall ranking on the left uses the sum of the ranks from the three target variables. The table also lists the secondary metrics: BCA -- balanced classification accuracy, WES -- weighted error score, CPA -- coverage probability accuracy. }
\label{tab:resultsD2}
\end{table}

Table \ref{tab:resultsD2} compiles all metrics for all TADPOLE submitted forecasts, as well as benchmarks and ensemble forecasts, from the longitudinal D2 prediction set. For details on datasets D2 and D3, see section \ref{data}, while for details on performance metrics see section \ref{eval}. Box-plots showing the distribution of scores, computed on 50 bootstraps of the test set, are shown in Supplementary Fig. \ref{confIntD2}, while the distribution of ranks is shown in Supplementary Figs. \ref{visRanksMAUC} -- \ref{visRanksVENTS}. Among the benchmark methods, \emph{BenchmarkMixedEffectsAPOE} had the best overall rank of 18, obtaining rank 35 on clinical diagnosis prediction, rank 2 on ADAS-Cog13 and rank 23 on Ventricle volume prediction. Removing the APOE status as covariate proved to significantly increase the predictive performance (\emph{BenchmarkMixedEffects}), although we do not show ranks for this entry as it was found during the evaluation phase. Among participant methods, the submission with the best overall rank was \emph{Frog}, obtaining rank 1 for prediction of clinical diagnosis, rank 4 for ADAS-Cog13 and rank 10 for Ventricle volume prediction. 

\textcolor{black}{For clinical diagnosis, the best submitted forecasts (team \emph{Frog}) scored better than all benchmark methods, reducing the error of the best benchmark methods by 0.085 (8.5 percentage points) for the multiclass area under the receiver operating characteristic curve (MAUC) and by 0.058 (5.8 p.p.) for balanced classification accuracy (BCA)}. Here, the best benchmarks obtained a MAUC of 0.846 (\emph{BenchmarkMixedEffects}) and a BCA of 0.792 (\emph{BenchmarkLastVisit}). Among participant methods, \emph{Frog} had the best MAUC score of 0.931, significantly better than all entries other than \emph{Threedays} according to the bootstrap test (p-value = 0.24, see section \ref{statTestingMAUC} for details on significance testing). Supplementary Figure \ref{visRanksMAUC} further shows the variability in performance ranking over bootstrap samples and highlights that the top two entries consistently remain at the top of the ranking. In terms of BCA, \emph{ARAMIS-Pascal} had the best score of 0.850. Moreover, ensemble methods (\emph{ConsensusMedian}) achieved the second best MAUC score of 0.925 and the best BCA score of 0.857. In contrast, the best randomised prediction (\emph{RandomisedBest}) achieved a much lower MAUC of 0.800 and a BCA of 0.803, suggesting entries below these scores did not perform significantly better than random guessing according to the bootstrap test (p-value = 0.01). MAUC and BCA performance metrics had a relatively high correlation across all submissions ($r$ = 0.88, Supplementary Fig. \ref{corrMetricsD2}). 

\textcolor{black}{For Ventricle volume, the best submitted forecasts among participants (team \emph{EMC1}) obtained substantially lower error scores than all benchmark methods, scoring 73\% of the lowest benchmark MAE (BenchmarkMixedEffects MAE=0.56) and 52\% of the lowest benchmark WES (BenchmarkMixedEffects WES=0.56)}. Among participant submissions, \emph{EMC1-Std/-Custom} had the best MAE of 0.41 (\% ICV), significantly lower than all entries other than \emph{lmaUCL-Covariates/-Std/-half-D1}, \emph{BORREGOTECMTY} and \emph{SBIA} according to the Wilcoxon signed-rank test (see section \ref{statTestingMAE}) -- this is also confirmed in Supplementary Fig. \ref{visRanksVENTS} by the variability in performance ranking over bootstrap samples. Team \emph{EMC1} also had the best Ventricle WES of 0.29, while \emph{DIKU-ModifiedMri-Custom/-Std} had the best Ventricle coverage probability accuracy (CPA) of 0.01. Ensemble methods (\emph{ConsensusMean}) achieved the best Ventricle MAE of 0.38. In contrast, the best randomised prediction (\emph{RandomisedBest}) achieved a higher MAE of 0.46, WES of 0.45 and CPA of 0.33. MAE and WES scores showed high correlation ($r$ = 0.99, Supplementary Fig. \ref{corrMetricsD2}) and were often of equal value for many submissions ($n=24$), as teams set equal weights for all subjects analysed. CPA did not correlate ($r \approx - 0.01$, Supplementary Fig. \ref{corrMetricsD2}) with either MAE or WES. 

For ADAS-Cog13, the best submitted forecasts did not score significantly better than the simple benchmarks. Here, the simple \emph{BenchmarkMixedEffects} model obtained the second-best MAE of 4.19, which was significantly lower than all other submitted entries according to the Wilcoxon signed-rank test. \emph{BenchmarkMixedEffects} also had the best ADAS-Cog13 WES of 4.19, while \emph{VikingAI-Sigmoid} had the best ADAS-Cog13 CPA of 0.02. Among participants' submissions, \emph{FortuneTellerFish-Control} ranked first in ADAS-Cog13 prediction with a MAE of 4.70 \textcolor{black}{(112\% of the lowest benchmark score)}. Moreover, all participants' forecasts scored worse than the best randomised prediction (\emph{RandomisedBest}), which here achieved a MAE of 4.52 and WES of 4.52. Nevertheless, the ensemble method \emph{ConsensusMean} obtained the best ADAS scores for MAE (3.75), WES (3.54) and CPA (0.0), which along with \emph{BenchmarkMixedEffects} were the only entries that performed significantly better than random guesswork (p-value = 0.01). The MAE and WES scores for ADAS-Cog13 had relatively high correlation ($r = 0.97$, Supplementary Fig. \ref{corrMetricsD2}) and were often of equal value for many submissions ($n = 25$). CPA had a weak but significant correlation with MAE ($r = 0.37$, p-value $<$ 0.02) and WES ($r = 0.35$, p-value $<$ 0.02).

\subsection{Forecasts from the cross-sectional prediction set (D3) and custom prediction sets}
\label{resultsD3}

\begin{table}
\resizebox{1\textwidth}{!}{%
\begin{tabular}{rc|ccc|cccc|cccc}
\toprule
  & Overall & \multicolumn{3}{c|}{Diagnosis} & \multicolumn{4}{c|}{ADAS-Cog13}  & \multicolumn{4}{c}{Ventricles (\% ICV)}\\
                       Submission & Rank &  Rank & MAUC & BCA  & Rank & MAE & WES & CPA & Rank & MAE & WES & CPA\\
\midrule
ConsensusMean \crule[gray]{0.3cm}{0.3cm}\crule[gray]{0.3cm}{0.3cm}  & - & \textbf{-} & \textbf{0.917} & 0.821 & - & 4.58 & 4.34 & 0.12 & - & 0.73 & 0.72 & 0.09\\ 
ConsensusMedian \crule[gray]{0.3cm}{0.3cm}\crule[gray]{0.3cm}{0.3cm}  & - & - & 0.905 & 0.817 & - & 5.44 & 5.37 & 0.19 & - & 0.71 & 0.65 & 0.10\\ 
GlassFrog-Average \crule[blue]{0.3cm}{0.3cm}\crule[green]{0.3cm}{0.3cm}  & \textbf{1} & 2-4 & 0.897 & 0.826 & 5 & 5.86 & 5.57 & 0.25 & 3 & 0.68 & 0.55 & 0.24\\ 
GlassFrog-LCMEM-HDR \crule[blue]{0.3cm}{0.3cm}\crule[green]{0.3cm}{0.3cm}  & 2 & 2-4 & 0.897 & 0.826 & 9 & 6.57 & 6.56 & 0.34 & \textbf{1} & \textbf{0.48} & \textbf{0.38} & 0.24\\ 
GlassFrog-SM \crule[blue]{0.3cm}{0.3cm}\crule[green]{0.3cm}{0.3cm}  & 3 & 2-4 & 0.897 & 0.826 & 4 & 5.77 & 5.77 & 0.19 & 9 & 0.82 & 0.55 & 0.07\\ 
Tohka-Ciszek-RandomForestLin \crule[magenta]{0.3cm}{0.3cm}\crule[magenta]{0.3cm}{0.3cm}  & 4 & 11 & 0.865 & 0.786 & 2 & 4.92 & 4.92 & 0.10 & 10 & 0.83 & 0.83 & 0.35\\ 
RandomisedBest \crule[gray]{0.3cm}{0.3cm}\crule[gray]{0.3cm}{0.3cm}  & - & - & 0.811 & 0.783 & - & 4.54 & 4.50 & 0.26 & - & 0.92 & 0.50 & \textbf{0.00}\\ 
lmaUCL-Std \crule[blue]{0.3cm}{0.3cm}\crule[blue]{0.3cm}{0.3cm}  & 5-9 & 12-14 & 0.854 & 0.698 & 16-18 & 6.95 & 6.93 & 0.05 & 5-7 & 0.81 & 0.81 & 0.22\\ 
lmaUCL-Covariates \crule[blue]{0.3cm}{0.3cm}\crule[blue]{0.3cm}{0.3cm}  & 5-9 & 12-14 & 0.854 & 0.698 & 16-18 & 6.95 & 6.93 & 0.05 & 5-7 & 0.81 & 0.81 & 0.22\\ 
lmaUCL-halfD1 \crule[blue]{0.3cm}{0.3cm}\crule[blue]{0.3cm}{0.3cm}  & 5-9 & 12-14 & 0.854 & 0.698 & 16-18 & 6.95 & 6.93 & 0.05 & 5-7 & 0.81 & 0.81 & 0.22\\ 
Rocket \crule[blue]{0.3cm}{0.3cm}\crule[green]{0.3cm}{0.3cm}  & 5-9 & 10 & 0.865 & 0.771 & 3 & 5.27 & 5.14 & 0.39 & 23 & 1.06 & 1.06 & 0.27\\ 
VikingAI-Logistic \crule[green]{0.3cm}{0.3cm}\crule[green]{0.3cm}{0.3cm}  & 5-9 & 8 & 0.876 & 0.768 & 6 & 5.94 & 5.91 & 0.22 & 22 & 1.04 & 1.01 & 0.18\\ 
EMC1-Std \crule[red]{0.3cm}{0.3cm}\crule[green]{0.3cm}{0.3cm}  & 10 & 30 & 0.705 & 0.567 & 7 & 6.29 & 6.19 & 0.47 & 4 & 0.80 & 0.62 & 0.48\\ 
BenchmarkMixedEffects \crule[orange]{0.3cm}{0.3cm}\crule[orange]{0.3cm}{0.3cm}  & - & - & 0.839 & 0.728 & \textbf{-} & \textbf{4.23} & \textbf{4.23} & 0.34 & - & 1.13 & 1.13 & 0.50\\ 
SBIA \crule[red]{0.3cm}{0.3cm}\crule[blue]{0.3cm}{0.3cm}  & 11 & 28 & 0.779 & 0.782 & 10 & 6.63 & 6.43 & 0.40 & 8 & 0.82 & 0.75 & 0.18\\ 
BGU-LSTM \crule[cyan]{0.3cm}{0.3cm}\crule[cyan]{0.3cm}{0.3cm}  & 12-14 & 5-7 & 0.877 & 0.776 & 13-15 & 6.75 & 6.17 & 0.39 & 26-28 & 1.11 & 0.79 & 0.17\\ 
BGU-RFFIX \crule[magenta]{0.3cm}{0.3cm}\crule[magenta]{0.3cm}{0.3cm}  & 12-14 & 5-7 & 0.877 & 0.776 & 13-15 & 6.75 & 6.17 & 0.39 & 26-28 & 1.11 & 0.79 & 0.17\\ 
BGU-RF \crule[magenta]{0.3cm}{0.3cm}\crule[magenta]{0.3cm}{0.3cm}  & 12-14 & 5-7 & 0.877 & 0.776 & 13-15 & 6.75 & 6.17 & 0.39 & 26-28 & 1.11 & 0.79 & 0.17\\ 
BravoLab \crule[cyan]{0.3cm}{0.3cm}\crule[cyan]{0.3cm}{0.3cm}  & 15 & 18 & 0.813 & 0.730 & 28 & 8.02 & 8.02 & 0.47 & 2 & 0.64 & 0.64 & 0.42\\ 
BORREGOTECMTY \crule[blue]{0.3cm}{0.3cm}\crule[blue]{0.3cm}{0.3cm}  & 16-17 & 15 & 0.852 & 0.748 & 8 & 6.44 & 5.86 & 0.46 & 30 & 1.14 & 1.02 & 0.49\\ 
CyberBrains \crule[blue]{0.3cm}{0.3cm}\crule[blue]{0.3cm}{0.3cm}  & 16-17 & 17 & 0.830 & 0.755 & 1 & 4.72 & 4.72 & 0.21 & 35 & 1.54 & 1.54 & 0.50\\ 
ATRI-Biostat-MA \crule[magenta]{0.3cm}{0.3cm}\crule[green]{0.3cm}{0.3cm}  & 18 & 19 & 0.799 & 0.772 & 26 & 7.39 & 6.63 & \textbf{0.04} & 11 & 0.93 & 0.97 & 0.10\\ 
DIKU-GeneralisedLog-Std \crule[blue]{0.3cm}{0.3cm}\crule[green]{0.3cm}{0.3cm}  & 19-20 & 20 & 0.798 & 0.684 & 20-21 & 6.99 & 6.99 & 0.17 & 16-17 & 0.95 & 0.95 & 0.05\\ 
EMC-EB \crule[red]{0.3cm}{0.3cm}\crule[red]{0.3cm}{0.3cm}  & 19-20 & 9 & 0.869 & 0.765 & 27 & 7.71 & 7.91 & 0.50 & 21 & 1.03 & 1.07 & 0.49\\ 
DIKU-GeneralisedLog-Custom \crule[blue]{0.3cm}{0.3cm}\crule[green]{0.3cm}{0.3cm}  & 21 & 21 & 0.798 & 0.681 & 20-21 & 6.99 & 6.99 & 0.17 & 16-17 & 0.95 & 0.95 & 0.05\\ 
DIKU-ModifiedLog-Std \crule[blue]{0.3cm}{0.3cm}\crule[green]{0.3cm}{0.3cm}  & 22-23 & 22-23 & 0.798 & 0.688 & 22-25 & 7.10 & 7.10 & 0.17 & 12-15 & 0.95 & 0.95 & 0.05\\ 
DIKU-ModifiedMri-Std \crule[blue]{0.3cm}{0.3cm}\crule[green]{0.3cm}{0.3cm}  & 22-23 & 22-23 & 0.798 & 0.688 & 22-25 & 7.10 & 7.10 & 0.17 & 12-15 & 0.95 & 0.95 & 0.05\\ 
DIKU-ModifiedMri-Custom \crule[blue]{0.3cm}{0.3cm}\crule[green]{0.3cm}{0.3cm}  & 24-25 & 24-25 & 0.798 & 0.691 & 22-25 & 7.10 & 7.10 & 0.17 & 12-15 & 0.95 & 0.95 & 0.05\\ 
DIKU-ModifiedLog-Custom \crule[blue]{0.3cm}{0.3cm}\crule[green]{0.3cm}{0.3cm}  & 24-25 & 24-25 & 0.798 & 0.691 & 22-25 & 7.10 & 7.10 & 0.17 & 12-15 & 0.95 & 0.95 & 0.05\\ 
Billabong-Uni \crule[green]{0.3cm}{0.3cm}\crule[green]{0.3cm}{0.3cm}  & 26 & 31 & 0.704 & 0.626 & 11-12 & 6.69 & 6.69 & 0.38 & 19-20 & 0.98 & 0.98 & 0.48\\ 
Billabong-UniAV45 \crule[green]{0.3cm}{0.3cm}\crule[green]{0.3cm}{0.3cm}  & 27 & 32 & 0.703 & 0.620 & 11-12 & 6.69 & 6.69 & 0.38 & 19-20 & 0.98 & 0.98 & 0.48\\ 
ATRI-Biostat-JMM \crule[magenta]{0.3cm}{0.3cm}\crule[blue]{0.3cm}{0.3cm}  & 28 & 26 & 0.794 & 0.781 & 29 & 8.45 & 8.12 & 0.34 & 18 & 0.97 & 1.45 & 0.37\\ 
CBIL \crule[cyan]{0.3cm}{0.3cm}\crule[cyan]{0.3cm}{0.3cm}  & 29 & 16 & 0.847 & 0.780 & 33 & 10.99 & 11.65 & 0.49 & 29 & 1.12 & 1.12 & 0.39\\ 
BenchmarkLastVisit \crule[orange]{0.3cm}{0.3cm}\crule[orange]{0.3cm}{0.3cm}  & 30 & 27 & 0.785 & 0.771 & 19 & 6.97 & 7.07 & 0.42 & 33 & 1.17 & 0.64 & 0.11\\ 
Billabong-MultiAV45 \crule[green]{0.3cm}{0.3cm}\crule[green]{0.3cm}{0.3cm}  & 31 & 33 & 0.682 & 0.603 & 30-31 & 9.30 & 9.30 & 0.43 & 24-25 & 1.09 & 1.09 & 0.49\\ 
Billabong-Multi \crule[green]{0.3cm}{0.3cm}\crule[green]{0.3cm}{0.3cm}  & 32 & 34 & 0.681 & 0.605 & 30-31 & 9.30 & 9.30 & 0.43 & 24-25 & 1.09 & 1.09 & 0.49\\ 
ATRI-Biostat-LTJMM \crule[magenta]{0.3cm}{0.3cm}\crule[green]{0.3cm}{0.3cm}  & 33 & 29 & 0.732 & 0.675 & 34 & 12.74 & 63.98 & 0.37 & 32 & 1.17 & 1.07 & 0.40\\ 
BenchmarkSVM \crule[orange]{0.3cm}{0.3cm}\crule[orange]{0.3cm}{0.3cm}  & 34 & 36 & 0.494 & 0.490 & 32 & 10.01 & 10.01 & 0.42 & 31 & 1.15 & 1.18 & 0.50\\ 
DIVE \crule[green]{0.3cm}{0.3cm}\crule[green]{0.3cm}{0.3cm}  & 35 & 35 & 0.512 & 0.498 & 35 & 16.66 & 16.74 & 0.41 & 34 & 1.42 & 1.42 & 0.34\\ 
IBM-OZ-Res \crule[blue]{0.3cm}{0.3cm}\crule[blue]{0.3cm}{0.3cm}  & - & 1 & 0.905 & \textbf{0.830} & - & - & - & - & 36 & 1.77 & 1.77 & 0.50\\ 

  \bottomrule
\end{tabular}
}
\caption{Ranked prediction scores for all TADPOLE submissions that used the cross-sectional prediction data set (D3). Best scores in each category are bolded. Missing numerical entries indicate that submissions did not include predictions for the corresponding target variable. The ``Diagnosis'' ranking uses multiclass area under the receiver operating characteristic curve (MAUC), those of ADAS-Cog13 and Ventricles use mean absolute error (MAE). The overall ranking on the left uses the sum of the ranks from the three target variables. The table also lists the secondary metrics: BCA -- balanced classification accuracy, WES -- weighted error score, CPA -- coverage probability accuracy. See section \ref{eval} for details on performance metrics.}
\label{tab:resultsD3}
\end{table}

Table \ref{tab:resultsD3} shows the ranking of the forecasts from the cross-sectional D3 prediction set. Box-plots showing the distribution of scores, computed on 50 bootstraps of the test set, are shown in Supplementary Fig \ref{confIntD3}, while the distribution of ranks is shown in Supplementary Figs. \ref{visRanksMAUCD3} -- \ref{visRanksVENTSD3}. Due to the lack of longitudinal data, most submissions had lower performance compared to their equivalents from the D2 longitudinal prediction set. Among submitted forecasts, \emph{GlassFrog-Average} had the best overall rank, as well as rank 2-4 on diagnosis prediction, rank 5 on ADAS-Cog13 prediction and rank 3 on ventricle prediction. 


\textcolor{black}{For clinical diagnosis prediction on D3, the best prediction among TADPOLE participants (team \emph{IBM-OZ-Res}) scored better than all benchmarks, improving over the best benchmark MAUC (BenchmarkMixedEffects, MAUC=0.839) by 6.6 percentage points and the best benchmark BCA (BenchmarkLastVisit, BCA=0.771) by 5.9 percentage points.} Among participant methods, \emph{IBM-OZ-Res} had the best MAUC score of 0.905, significantly better than all entries other than \emph{GlassFrog-SM/-Average/-LCMEM-HDR}, \emph{BGU-RF/-RFFIX/-LSTM}, \emph{VikingAI-Logistic}, \emph{EMC-EB}, \emph{Rocket} and \emph{Tohka-Ciszek-RandomForestLin} according to the bootstrap hypothesis test (same methodology as in D2). This is further confirmed in Supplementary Fig. \ref{visRanksMAUCD3} by the variability of ranks under boostrap samples of the dataset, as these teams often remain at the top of the ranking. \emph{IBM-OZ-Res} also had the best BCA score of 0.830 among participants. Among ensemble methods, \emph{ConsensusMean} obtained the best Diagnosis MAUC of 0.917. In contrast, the best randomised prediction (\emph{RandomisedBest}) obtained an MAUC of 0.811 and a BCA of 0.783. MAUC and BCA performance metrics had a relatively high correlation across all submissions ($r = 0.9$, Supplementary Fig. \ref{corrMetricsD3}).


\textcolor{black}{For Ventricle volume prediction on D3, the best prediction (\emph{GlassFrog-LCMEM-HDR}) obtained substantially lower error scores than all benchmark methods, scoring 42\% of the lowest benchmark MAE (BenchmarkMixedEffects MAE=1.13) and 59\% of the lowest benchmark WES (BenchmarkLastVisit WES=0.64)}, and achieving error rates comparable to the best predictions of D2. Among participant submissions, \emph{GlassFrog-LCMEM-HDR} had the best MAE of 0.48, significantly lower than all other submitted entries according to the Wilcoxon signed-rank test -- this is also confirmed in Supplementary Fig. \ref{visRanksVENTSD3} by the rank distribution under dataset boostraps.  \emph{GlassFrog-LCMEM-HDR} also had the best Ventricle WES of 0.38, while submissions by team \emph{DIKU} had the best Ventricle CPA of 0.05. Among ensemble methods, \emph{ConsensusMedian} obtained a Ventricle MAE of 0.71 (4th best) and WES of 0.65 (7th best). In contrast, the best randomised prediction (\emph{RandomisedBest}) obtained a Ventricle MAE of 0.92, WES of 0.50 and CPA of 0. As in D2, MAE and WES scores in D3 for Ventricles had very high correlation ($r = 0.99$, Supplementary Fig. \ref{corrMetricsD3}), while CPA showed weak correlation with MAE ($r = 0.24$, p-value = 0.17) and WES ($r = 0.37$, p-value $< 0.032$).

For ADAS-Cog13 on D3, the predictions submitted by participants again did not perform better than the best benchmark methods. \emph{BenchmarkMixedEffects} had the best MAE of 4.23, which was significantly lower than all entries by other challenge participants. Moreover, the MAE of 4.23 was only marginally worse than the equivalent MAE (4.19) by the same model on D2. \emph{BenchmarkMixedEffects} also had the best ADAS-Cog13 WES of 4.23, while \emph{ATRI-Biostat-MA} had the best ADAS-Cog13 CPA of 0.04. Among participants' submissions, \emph{CyberBrains} ranked first in ADAS-Cog13 prediction with MAE/WES scores of 4.72 \textcolor{black}{(111\% of the lowest benchmark score)}. Among ensemble methods, \emph{ConsensusMean} obtained an ADAS-Cog13 MAE of 4.58, WES of 4.34, better than all participants' entries. As in D2, the best randomised predictions (\emph{RandomisedBest}) obtained an ADAS-Cog13 MAE of 4.54 (2nd best) and WES of 4.50 (3rd best). As in D2, MAE and WES scores for ADAS-Cog13 had high correlation ($r$ = 0.97, Supplementary Fig. \ref{corrMetricsD3}), while CPA showed weak, non-significant correlation with MAE ($r$ = 0.34, p-value $\approx$ 0.052) or WES ($r$ = 0.33, p-value $\approx$ 0.057). 

Results on the custom prediction sets are presented in Supplementary Table \ref{resDcustom}.

\subsection{Algorithm characteristics associated with increased performance}

\begin{figure}
	  \includegraphics[width=\textwidth]{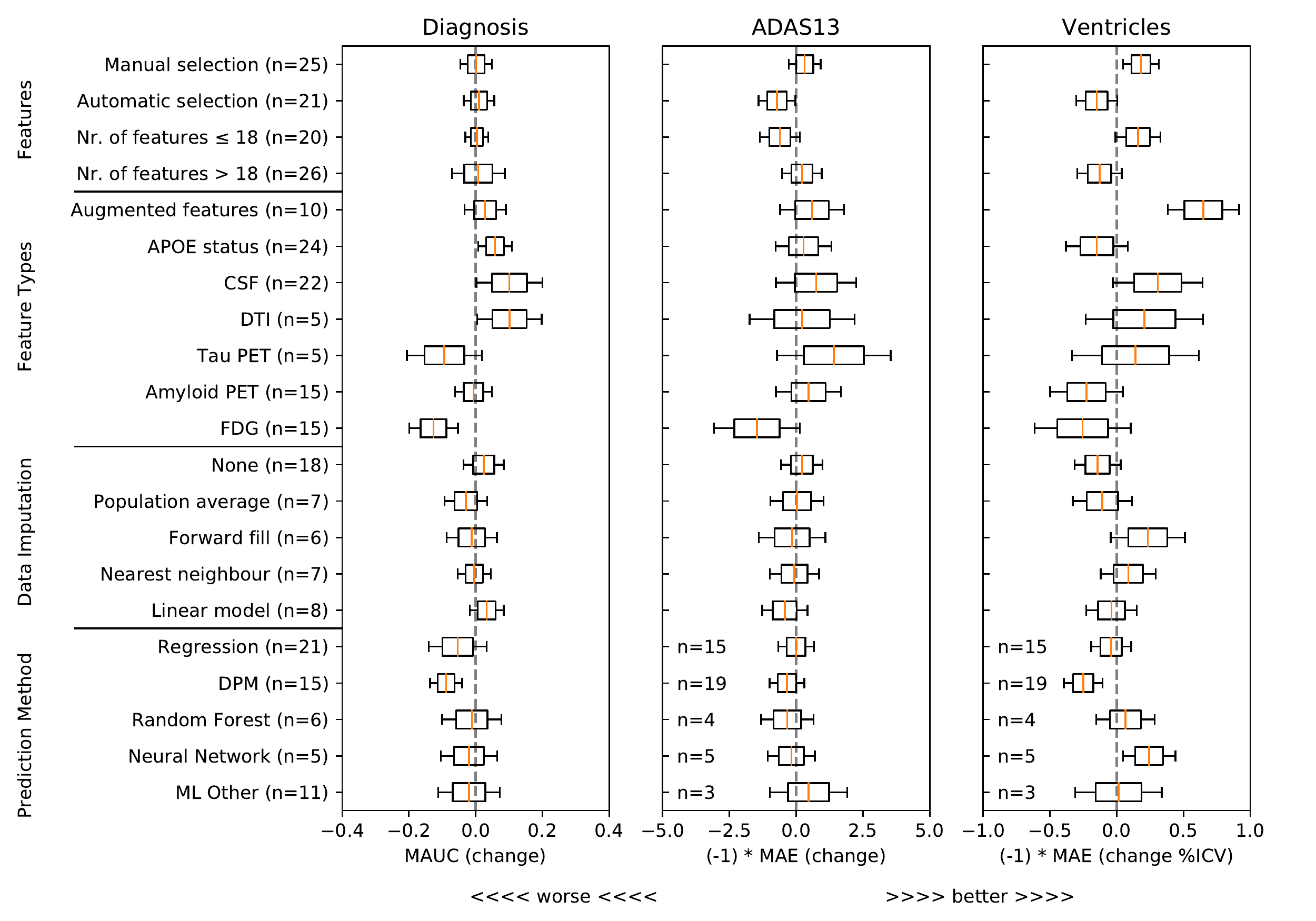}
 \caption{Associations between the prediction of clinical diagnosis, ADAS-Cog13 and Ventricle volume and different strategies of (top) feature selection, (upper-middle) types of features, (lower-middle) data imputation strategies and (bottom) prediction methods for the target variables. For each type of feature/method (rows) and each target variable (columns), we show the distribution of estimated coefficients from a general linear model. Positive coefficients, where distributions lie to the right of the dashed vertical line, indicate better performance than baseline (vertical dashed line). For ADAS-Cog13 and Ventricle prediction, we flipped the sign of the coefficients, to consistently show better performance to the right of the vertical line.}
 \label{metaAnalysis}
\end{figure}

To understand what characteristics of algorithms could have yielded higher performance, we show in Figure \ref{metaAnalysis} associations from a general linear model between predictive performance and feature selection methods, different types of features, methods for data imputation, and methods for forecasting of target variables. For each type of feature/method and each target variable (clinical diagnosis, ADAS-Cog13 and Ventricles), we show the distribution of estimated coefficients from a general linear model, derived from the approximated inverse Hessian matrix at the maximum likelihood estimator (see section \ref{statAnalysis}). From this analysis we removed outliers, defined as submissions with ADAS MAE higher than 10 and Ventricle MAE higher than 1.15 (\%ICV). For all plots, distributions to the right of the gray dashed vertical line denote increased performance compared to baseline (i.e. when those characteristics are not used). 

For feature selection, Figure \ref{metaAnalysis} shows that methods with manual selection of features tend to be associated with better predictive performance in ADAS-Cog13 and Ventricles. In terms of feature types, CSF and DTI features were generally associated with an increase in predictive performance for clinical diagnosis, while augmented features were associated with performance improvements for ventricle prediction. In terms of data imputation methods, while some differences can be observed, no clear conclusions can be drawn. In terms of prediction models, the only positive association that indicates increased performance is in the neural networks for ventricle prediction. \co{However, given the small number of methods tested (just under 50) and the large number of degrees of freedom ($=21$), these results should be interpreted with care. }

{\color{black} 
\subsection{External Validation}

To verify the performance trends such as ensemble models outperforming individual entries and benchmarks, we performed, in section \ref{secExternal}, external validation experiments on data from two separate studies: the Australian Imaging, Biomarkers, and Lifestyle Flagship Study of Ageing (AIBL) (\cite{ellisAIBL2009}), and the DHA clinical trial (\cite{quinn2010DHA}).
}

\section{Discussion}

The results of the TADPOLE Challenge provide unique and important insights into how well state-of-the-art algorithms can predict progression of AD diagnoses and markers of disease progression both from rich longitudinal data sets and, comparatively, from sparser cross-sectional data sets typical of a clinical trial scenario. The challenge further highlights the algorithms, features and data-handling strategies that tend to lead to improved forecasts. In the following sections we discuss the key conclusions that we draw from our study and highlight important limitations.

\subsection{TADPOLE pushed forward the performance on AD clinical diagnosis prediction}

In comparison to previous state-of-the-art results in the literature, the best TADPOLE methods show similar or higher performance in AD diagnostic classification while also tackling a harder problem than most previous studies of predicting future, rather than estimating current, classification. A comparison of 15 studies presented by (\cite{moradi2015machine}) reported lower performance (maximum AUC of 0.902 vs 0.931 obtained by the best TADPOLE method) for the simpler two-class classification problem of separating MCI-stable from MCI-converters in ADNI. A more recent method by (\cite{long2017prediction}) reported a maximum AUC of 0.932 and accuracy of 0.88 at the same MCI-stable vs -converter classification task. However, a) TADPOLE's discrimination of CN-converters from CN-stable subjects is harder as disease signal is weaker at such early stages, and b) the predictive performance drops in three-class problems like TADPOLE compared to two-class. Furthermore, the best out of 19 algorithms in the CADDementia Challenge (\cite{bron2015standardized}) obtained an MAUC of 0.78. 

We are unaware of previous studies forecasting future ventricle volume or ADAS-Cog13, so TADPOLE sets a new benchmark state-of-the-art performance on these important prediction tasks.

\subsection{No one-size-fits-all prediction algorithm}

The results on the longitudinal D2 prediction set suggest no clear winner on predicting all target variables -- no single method performed best on all tasks. While \emph{Frog} had the best overall submission with the lowest sum of ranks, for each performance metric individually different winners emerge: \emph{Frog} (clinical diagnosis MAUC of 0.931), \emph{ARAMIS-Pascal} (clinical diagnosis BCA of 0.850), \emph{BenchmarkMixedEffects} (ADAS-Cog13 MAE and WES of 4.19), \emph{VikingAI-Sigmoid} (ADAS-Cog13 CPA of 0.02), \emph{EMC1-Std/EMC1-Custom} (ventricle MAE of 0.41 and WES or 0.29), and \emph{DIKU-ModifiedMri-Std/-Custom} (ventricle CPA of 0.01). Moreover, on the cross-sectional D3 prediction set, the methods by \emph{Glass-Frog} had the best performance. Associations of method-type with increased performance in Fig. \ref{metaAnalysis} confirm no clear increase in performance for any types of prediction methods (with the exception of neural networks for ventricle volume prediction). \textcolor{black}{This suggests performance depends more on data quality and feature choice. Substantially larger data sets may reveal differences arising from algorithmic choices, but the results we present here are representative of realistic clinical-trial scenarios.} 

\subsection{Characteristics of top-5 algorithms}



\co{The top-5 algorithms had several common characteristics. In the clinical diagnosis prediction category (top 5: \emph{Frog}, \emph{Threedays}, \emph{EMC-EB}, \emph{GlassFrog} and \emph{Apocalypse}), 4/5 used Machine Learning methods, 4/5 used APOE status as an input feature and 4/5 predicted the clinical diagnosis directly from the input at each future timestep, whereas 1/5 first predicted the ADAS-Cog/Ventricle measures at each future timestep, then predicted the clinical diagnosis from the ADAS-Cog/Ventricle measures. In the ADAS-Cog13 prediction category (top 5: \emph{BenchmarkMixedEffects}, \emph{FortuneTellerFish}, \emph{Frog}, \emph{Mayo-BAI-ASU} and \emph{CyberBrains}), we find that 4/5 used manual feature selection and 4/5 used linear regression methods. One potential reason why linear regression models performed the best out of all models here is due to their implicit regularization, although we note that none of them performed significantly better than random guessing (with the exception of \emph{BenchmarkMixedEffects} -- see section \ref{predADAS}). In the Ventricle prediction category (top 5: \emph{EMC1}, \emph{lmaUCL}, \emph{BORREGOTECMTY}, \emph{CN2L}, \emph{SBIA}), 4/5 used APOE status as an input feature, 3/5 used an automatic feature selection mechanism (which selected $>$250 features), 3/5 generated augmented features, and 3/5 used (parametric) regression (while 1/5 used a neural network and 1/5 a disease progression model).}

%
%

\subsection{Ensemble methods perform strongly}

Consistently strong results from ensemble methods (\emph{ConsensusMean/ConsensusMedian} outperformed all others on most tasks) might suggest that the varying assumptions in different methods cause different biases: some consistently over-estimate, some consistently under-estimate, and thus averaging aligns more closely with the truth. This is confirmed by plots of the difference between true and estimated measures (Supplementary Figures \ref{visGridMAUC}--\ref{visGridVENTS}), where most methods systematically under- or over-estimate in \emph{all subjects}. \co{The consistent over-estimation and under-estimation of individual methods is likely due to the effect of their inductive biases on the predictions, in the presence of domain shifts from the D1-D2 training sets to the D4 test set (e.g. older individuals in D4 compared to D1-D3)}. However, even if methods were completely unbiased, averaging over all methods could also help predictions by reducing the variance in the estimated target variables.

\subsection{Predictability of ADAS-Cog13 scores}
\label{predADAS}

ADAS-Cog13 scores were more difficult to forecast than clinical diagnosis or ventricle volume. The only single method able to forecast ADAS-Cog13 better than informed random guessing (\emph{RandomisedBest}) was the \emph{BenchmarkMixedEffects}, a simple mixed effects model with no covariates and age as a regressor. \textcolor{black}{One possible explanation is the complex multi-effect relationship between the acquired data (imaging, protein markers, etc.) and the composite cognitive test score. However, that relationship is no less complex for clinical diagnosis where prediction appears much more feasible. Alternatively, the difficulty may arise from variability in administering the cognitive tests, or practice effects}. Treatment trials often use a change of 4 or more in ADAS-Cog13 as a threshold to identify responders/non-responders (\cite{grochowalski2016examining}), so error scores of 4 or less provide a sensible target performance level. With the exception of the ensemble method, all submitted forecasts failed to produce mean error below 4, highlighting the substantial challenge of estimating change in ADAS-Cog13 over the 1.4 year interval. \textcolor{black}{The difficulty in forecasting ADAS-Cog13 calls into question the usage of cognitive test scores in patient selection and as primary endpoint.} 
 
 
\subsection{Prediction errors from limited cross-sectional dataset mimicking clinical trials are similar to those from longitudinal dataset}

For clinical diagnosis, the best performance on the limited, cross-sectional D3 prediction set was similar to the best performance on the D2 longitudinal prediction set: 0.917 vs 0.931 for MAUC (p-value = 0.14), representing 2 p.p. decrease for D3 compared to D2. Slightly larger and significant differences were observed for ADAS MAE (3.75 vs 4.23, p-value $<$ 0.01) and Ventricle MAE (0.38 vs 0.48, p-value $<$ 0.01). It should be noted that Ventricle predictions for D3 were extremely difficult, given that only 25\% of subjects to be forecasted had MRI data in D3. This suggests that, for clinical diagnosis, current forecast algorithms are reasonably robust to lack of longitudinal data and missing inputs, while for ADAS and Ventricle volume prediction, some degree of performance is lost. Future work is also required to determine the optimal balance of input data quality and quantity versus cost of acquisition.

\subsection{DTI and CSF features appear informative for clinical diagnosis prediction, augmented features appear informative for ventricle prediction} 

DTI and CSF features are most associated with increases in clinical diagnosis forecast performance. CSF, in particular, is well established as an early marker of AD (\cite{jack2010hypothetical}) and likely to help predictions for early-stage subjects, while DTI, measuring microstructure damage, may be informative for middle-stage subjects. On the other hand, for prediction of ventricle volume, augmented features had the highest association with increases in prediction performance. Future work is required to confirm the added value of these features and others in a more systematic way. 





\subsection{Challenge design and limitations}

TADPOLE Challenge has several limitations that future editions of the challenge may consider addressing. One limitation is the reliability of the three target variables: clinical diagnosis, ADAS-Cog13 and Ventricle volume. First of all, clinical diagnosis has only moderate agreement with gold-standard neuropathological post-mortem diagnosis. In particular, one study (\cite{beach2012accuracy}) has shown that a clinical diagnosis of probable AD has sensitivity between 70.9\% and 87.3\% and specificity between 44.3\% and 70.8\%. With the advent of post-mortem confirmation in ADNI, future challenges might address this by evaluating the algorithms on subjects with pathological confirmation. Similarly, ADAS-Cog13 is known to suffer from low reliability across consecutive visits (\cite{grochowalski2016examining}), and TADPOLE algorithms fail to forecast it reliably. However, this might be related to the short time-window (1.4 years), and more accurate predictions might be possible over longer time-windows, when there is more significant cognitive decline. Ventricle volume measurements depend on MRI scanner factors such as field strength, manufacturer and pulse sequences (\cite{han2006reliability}), although these effects have been removed to some extent by ADNI through data preprocessing and protocol harmonization. TADPOLE Challenge also assumes all subjects either remain stable or convert to Alzheimer's disease, whereas in practice some of them might develop other types of neurodegenerative diseases.

For performance evaluation, we elected to use very simple yet reliable metrics as the primary performance scores: the multiclass area under the curve (mAUC) for the clinical categorical variable and the mean absolute error (MAE) for the two numerical variables. While the mAUC accounts for decision confidence, the MAE does not, which means that the confidence intervals submitted by participants do not contribute to the rankings computed in Tables \ref{tab:resultsD2} and \ref{tab:resultsD3}. While the weighted error score (WES) takes confidence intervals into account, we consider it susceptible to ``hacking'', e.g. participants might assign high confidence to only one or two data points and thereby skew the score to ignore most of the predictions -- in practice, we did not observe this behaviour in any submission. For clinical relevance, we believe that confidence intervals are an extremely important part of such predictions and urge future studies to consider performance metrics that require and take account of participant-calculated confidence measures.

TADPOLE has some limitations related to the algorithms' comparability and generalisability. We could only compare full methods submissions and not different types of features, and strategies for data imputation and prediction used within the full method. \co{While we tried to evaluate the effect of these characteristics in Figure \ref{metaAnalysis}, in practice the numbers were small and hence most effects did not reach statistical significance. The analysis in Figure \ref{metaAnalysis} also assumes a linear correspondence between method characteristics and performance, which was necessitated due to the small number of methods tested (just under 50) and the large number of degrees of freedom ($=21$).} Moreover, the challenge format does not provide an exhaustive comparison of all combinations of data processing, predictive model, features, etc., so does not lead to firm conclusions on the best combinations but rather provides hypotheses for future testing. In future work, we plan to test inclusion of features and strategies for data imputation and prediction independently, by changing one such characteristic at a time. Future challenges might also consider how to provide stronger external validation of findings, e.g. by evaluating all submissions directly on prescribed independent data sets. However, this presents substantial difficulties, as comprehensive external data consistent with the internal, especially for which follow-up occurs on the same timescale, is difficult to find, and extra demands on participants present barriers to entry so the trade-off with engagement must be considered carefully.

Another limitation is that the number of controls and MCI converters in the D4 test set is low (9 MCI converters and 9 control converters). However, these numbers will increase over time as ADNI acquires more data, and we plan to re-run the evaluation at a later stage with the additional data acquired after April 2019. A subsequent evaluation will also enable us to evaluate the TADPOLE methods on longer time-horizons, over which the effects of putative drugs would be higher.

\section{Conclusion}

In this work we presented the results of the TADPOLE Challenge. The results of the challenge provide important insights into the current state of the art in AD forecasting, such as performance levels achievable with current data and technology as well as specific algorithms, features and data-handling strategies that support the best forecasts. The developments and outcomes of TADPOLE Challenge can aid refinement of cohorts and endpoint assessment for clinical trials, and can support accurate prognostic information in clinical settings. The challenge website (\url{https://tadpole.grand-challenge.org}) will stay open for submissions, which can be added to our current ranking. The open test set remains available on the ADNI LONI website and also allows individual participants to evaluate future submissions. Through TADPOLE-SHARE \url{https://tadpole-share.github.io/}, we further plan to implement many TADPOLE methods in a common framework, to be made publicly available. TADPOLE provides a standard benchmark for evaluation of future AD prediction algorithms.


\section{Prediction Algorithms}
\label{predMethods}

\setlist[itemize]{noitemsep, topsep=0pt}

 \textbf{Team:} AlgosForGood \crule[blue]{0.3cm}{0.3cm}\crule[blue]{0.3cm}{0.3cm} (Members: Tina Toni, Marcin Salaterski, Veronika Lunina, Institution: N/A)\\
 \textbf{Overall Ranking:} 24\\
 \textbf{Feature selection:} Manual + Automatic: Manual selection of uncorrelated variables from correlation matrix and automatic selection of variables that have highest cumulative hazard rates in survival regression from MCI to AD.\\
 \textbf{Selected features:} Demographics (age, education, gender, race, marital status), cognitive tests (ADAS-Cog13, RAVLT immediate, RAVLT forgetting, CDRSOB, ADAS11, FDG), Ventricles, AV45, ICV, APOE4.\\
 \textbf{Missing data:} Fill-in using last available value from corresponding patient\\
 \textbf{Confounder correction:} none\\
 \textbf{Method category:} Statistical Regression / Proportional hazards model\\
 \textbf{Prediction method:}
 \begin{itemize}
  \item Diagnosis: Aalen additive regression constructing a cumulative hazard rate for progressing to AD.
  \item ADAS-Cog13: regression using change in ventricles/ICV as predictive variable, stratified by last known diagnosis.
  \item Ventricles: regression over month, with several pre-processing steps: 1. Enforced monotonicity by accumulating maximum value, 2. For APOE positive patients used only last three visits due to non-linearity 3. Stratified by diagnosis\\\\
 \end{itemize} 
 \textbf{Team:} Apocalypse \crule[red]{0.3cm}{0.3cm}\crule[blue]{0.3cm}{0.3cm} (Members: Manon Ansart, Stanley Durrleman, Institution: Institut du Cerveau et de la Moelle \'{e}pini\`{e}re, ICM, Paris, France)\\
 \textbf{Overall Ranking:} 6\\
 \textbf{Feature selection:} Manual -- important features were identified by looking at the correlations with the diagnosis. Personal knowledge of the disease was also used to complement those results and select relevant features. Different feature sets were compared using cross-validation.\\
 \textbf{Selected features:}  Cognitive features (ADAS-Cog13, MMSE, RAVLT immediate, FAQ, CDRSOB), MRI features (WholeBrain, Entorhinal, Fusiform, MidTemp, Ventricles, Entorhinal, Hippocampus), APOE4, education, age, clinical diagnosis\\
 \textbf{Missing data:} Filled in using the mean feature value\\
 \textbf{Confounder correction:} none\\
 \textbf{Method category:} Machine learning / Regression\\
 \textbf{Prediction method:} Linear regression is used to first predict the future of a set of features (MMSE, ADAS-Cog13, CDRSOB, RAVLT, Ventricles) at the prediction dates. Afterwards, an SVM is used to predict the current diagnosis for each prediction date, based on the forecasted features as well as other features which are constant for the subject (APOE4, education, age at last known visit).\\
 \\
  \\
 \textbf{Team:} ARAMIS-Pascal \crule[blue]{0.3cm}{0.3cm}\crule[gray]{0.3cm}{0.3cm} (Members: Pascal Lu, Institution: Institut du Cerveau et de la Moelle \'{e}pini\`{e}re, ICM, Paris, France)\\
 \textbf{Clinical Diagnosis Ranking:} 15\\
 \textbf{Feature selection:} Manual, based on known biomarkers from the literature.\\
 \textbf{Selected features:} APOE4, cognitive tests (CDRSB, ADAS11, ADAS-Cog13, MMSE, RAVLT immediate, FAQ), volumetric MRI (hippocampus, ventricles, whole brain, entorhinal, fusiform, middle temporal, ICV), whole brain FDG, CSF biomarkers (amyloid-beta, tau, phosphorylated tau), education and age.\\
 \textbf{Missing data:} Imputed using the average biomarker values across the population.\\
 \textbf{Confounder correction:} none\\
 \textbf{Method category:}  Statistical Regression / Proportional hazards model\\
 \textbf{Prediction method:} For diagnosis prediction, the Aalen model for survival analysis was used to predict the conversion from MCI to AD, which returns the probability of a subject remaining MCI as a function of time. The method assumes cognitively normal and dementia subjects will not convert and thus will remain constant. The method did not predict ADAS-Cog13 or Ventricles.\\
 \textbf{Publication link:} \url{https://hal.inria.fr/tel-02433613/document}
 \\
  \\
  \\
 \textbf{Team:} ATRI\_Biostat (Members: Samuel Iddi$^{1,2}$, Dan Li$^1$, Wesley K. Thompson$^3$ and Michael C. Donohue$^1$. Institutions: $^1$Alzheimer's Therapeutic Research Institute, USC, USA; $^2$Department of Statistics and Actuarial Science, University of Ghana, Ghana; $^3$Department of Family Medicine and Public Health, University of California, USA)\\
 \textbf{Overall Ranking:} 48-53\\
 \textbf{Feature selection:} Automatic - features were ranked by their importance in classifying diagnostic status using a random forest algorithm. All cognitive tests, imaging biomarkers, demographic information and APOE status were considered as potential features\\
 \textbf{Selected features:} ADAS-Cog13, EcogTotal, CDRSOB, FAQ, MOCA, MMSE, RAVLT immediate, Ventricles/ICV, Entorhinal, Hippocampus/ICV and FDG Pet. Age, gender and APOE status were included as covariates. The interaction between diagnosis at first available visit and years since first visit was also considered.\\
 \textbf{Missing data:} Imputed using the MissForest Algorithm, based on a non-parametric random forest methodology (\cite{stekhoven2011missforest}). The algorithm was chosen based on its ability to handle mixed-type outcomes, complex interactions and non-linear relationships between variables.\\
 \textbf{Confounder correction:} APOE status, last known clinical status, age and gender.\\
 \textbf{Method category:} Machine learning and data-driven disease progression models\\
 \textbf{Prediction method:} The method applied different types of mixed-effects models to forecast ADAS-Cog13 and Ventricles, and then used a Random Forest classifier to predict the clinical diagnosis from the forecasted continuous scores. 
 \begin{itemize}
 \item JMM \crule[magenta]{0.3cm}{0.3cm}\crule[blue]{0.3cm}{0.3cm}  -- Joint Mixed-effect Modelling with subject-specific intercept and slope.
 \item LTJMM \crule[magenta]{0.3cm}{0.3cm}\crule[green]{0.3cm}{0.3cm} -- Latent Time Joint Mixed-effect Modelling with subject-specific intercept, slope and time-shift
 \item MA \crule[magenta]{0.3cm}{0.3cm}\crule[green]{0.3cm}{0.3cm} -- Model average of the two models above, as well as a third model where random intercepts are shared across outcomes.
 \end{itemize}
\textbf{Confidence Intervals:} The 50\% prediction intervals for ADAS-Cog13 and Ventricles were obtained by taking the 25th and 75th percentile of the posterior predicted samples.    \\
 \textbf{Publication link:} \url{https://braininformatics.springeropen.com/articles/10.1186/s40708-019-0099-0}\\
  \\
  \\
 \textbf{Team:}  BGU (Members: Aviv Nahon, Yarden Levy, Dan Halbersberg, Mariya Cohen, Institution: Ben Gurion University of the Negev, Beersheba, Israel)\\
 \textbf{Overall Ranking:} 20-28\\
 \textbf{Feature selection:} Automatic -- used the following algorithm: 1. Find the two variables with highest correlation (Spearman for continuous variables and Mutual information for discrete variables). 2. Compute the correlation of each variable with the target variables separately and remove the variable with the lower correlation. 3. If there are still pairs of variables with a correlation of more than 80\%, repeat from step 1.\\
 \textbf{Selected features:} Cognitive tests (CDRSOB, MMSE, RAVLT, MOCA, all Ecog), MRI biomarkers (Freesurfer cross-sectional and longitudinal), FDG- PET (hypometabolic convergence index),  AV45 PET (Ventricles, Corpus Callosum, Hippocampus), White-matter hypointensities' volume, CSF biomarkers (amyloid-beta, tau, phosphorylated tau). For each continuous variable, an additional set of 20 augmented features was used, representing changes and trends in variables (e.g. mean, standard deviation, trend mean, trend standard deviation, minimum, mean minus global mean, baseline value, last observed value). This resulted in 233 features, which were used for prediction.\\
 \textbf{Missing data:} Random forest can deal automatically with missing data. LSTM network used indicator that was set to zero for missing data.\\
 \textbf{Confounder correction:} None\\
 \textbf{Method category:} Machine learning\\
 \textbf{Prediction method:}
 \begin{itemize}
 \item BGU-LSTM \crule[cyan]{0.3cm}{0.3cm}\crule[cyan]{0.3cm}{0.3cm}: This model consisted of two integrated neural networks: an LSTM network for modelling continuous variables and a feed-forward neural network for the static variables.
 \item BGU-RF \crule[magenta]{0.3cm}{0.3cm}\crule[magenta]{0.3cm}{0.3cm}: A semi-temporal Random Forest was used which contained the augmented features.
 \item BGU-RFFIX \crule[magenta]{0.3cm}{0.3cm}\crule[magenta]{0.3cm}{0.3cm}: Same as BGU-RF, but with small correction for the prediction of diagnosis: whenever the model predicted AD with probability higher than 80\%, the probability of CN was changed to zero and vice versa.\\\\
 \end{itemize}
 \textbf{Team:} BIGS2 \crule[magenta]{0.3cm}{0.3cm}\crule[blue]{0.3cm}{0.3cm} (Members: Huiling Liao, Tengfei Li, Kaixian Yu, Hongtu Zhu, Yue Wang, Binxin Zhao, Institution: University of Texas, Houston, USA)\\
 \textbf{Overall Ranking:} 51\\
 \textbf{Feature selection:} Automatic -- used auto-encoder to extract aggregated features.\\
 \textbf{Selected features:} All continuous features in D1/D2, which represented the input for the autoencoder. Apart from the autoencoder-extracted features, other features used for the classifier were demographic information, APOE status, whole brain biomarkers from MRI (volume) and PET (FDG, PIB and AV45), and MMSE.\\
 \textbf{Missing data:} SoftImpute method (\cite{mazumder2010spectral}) was used for imputing missing data. The complete dataset was then used as input to the autoencoder.\\
 \textbf{Confounder correction:} none\\
 \textbf{Method category:} Regression and Machine Learning\\
 \textbf{Prediction method:} Linear models were used to predict ADAS-Cog13 and Ventricle scores independently. For prediction of clinical diagnosis, a random forest was used based on the autoencoder-extracted features and the other selected features.\\
 \\
  \\
 \textbf{Team:} Billabong \crule[green]{0.3cm}{0.3cm}\crule[green]{0.3cm}{0.3cm} (Members: Neil Oxtoby, Institution: University College London, UK)\\
 \textbf{Overall Ranking:} 46-52\\
 \textbf{Feature selection:} Manual, using knowledge from literature\\
 \textbf{Selected features:} MRI biomarkers normalised by ICV (ventricles, hippocampus, whole brain, entorhinal, fusiform, middle temporal), FDG, AV45, CSF biomarkers (amyloid beta, tau, phosphorylated tau) and cognitive tests (ADAS-Cog13, MMSE, MOCA, RAVLT immediate). Separate submissions (Billabong-UniAV45, Billabong-MultiAV45) were made which also included AV45, that was initially excluded due to noise.\\
 \textbf{Missing data:} Only imputed during staging via linear regression against age. The method can deal with missing data during training.\\
 \textbf{Confounder correction:} None\\
 \textbf{Method category:} Data-driven disease progression model\\
 \textbf{Prediction method:} For each selected feature independently, a data-driven longitudinal trajectory was estimated using a differential equation model based on Gaussian Process Regression (\cite{oxtoby2018data}). Subjects were staged using either a multivariate or univariate approach:
 \begin{itemize}
 \item Billabong-Uni: Univariate staging which estimates disease stage for each target variable independently.
 \item Billabong-Multi: Multivariate staging that combines all selected features, producing an average disease stage. 
 \end{itemize}
 For the prediction of clinical diagnosis, the historical ADNI diagnoses were mapped to a linear scale using partially-overlapping squared-exponential distribution functions. The linear scale and the three distributions were used to forecast the future diagnoses.\\
 Custom prediction set: Predictions were made also for a custom dataset, which was similar to D3 but missing data was filled in using the last available biomarker data.\\
 \textbf{Confidence Intervals:} The 25th and 75th percentiles of the GPR posterior were each integrated into a trajectory to obtain 50\% confidence (credible) intervals for the forecasts of ADAS-Cog13 and Ventricles/ICV.\\
 \textbf{Publication link:} \url{https://doi.org/10.1093/brain/awy050}\\
  \\
  \\
 \textbf{Team:} BORREGOSTECMTY \crule[blue]{0.3cm}{0.3cm}\crule[blue]{0.3cm}{0.3cm} (Members: José Gerardo Tamez-Peña, Institution: Tecnologico de Monterrey, Monterrey, Mexico)\\
 \textbf{Overall Ranking:} 9\\
 \textbf{Feature selection:} Automatic, using bootstrapped stage-wise selection.\\
 \textbf{Selected features:} Main cognitive tests (excluding subtypes), MRI biomarkers, APOE status, demographic information (age, gender, education) and diagnosis status.  Augmented features were further constructed from the MRI set: the cubic root of all volumes, the square root of all surface areas, the compactness, the coefficient of variation, as well as the mean value and absolute difference between the left and right measurements.\\
 \textbf{Missing data:} Imputed using nearest-neighbourhood strategy based on L1 norm.\\
 \textbf{Confounder correction:} Gender and intracranial volume (ICV) adjustments relative to controls.\\
 \textbf{Method category:} Regression (ensemble of statistical models)\\
 \textbf{Prediction method:} ADAS-Cog13 and Ventricles were predicted using an ensemble of 50 linear regression models, one set for each diagnostic category. The best models were selected using Bootstrap Stage-Wise Model Selection, using statistical fitness (\cite{pencina2008evaluating}) tests to evaluate models and features to use within the models. All selected models were then averaged in a final prediction using bagging. For prediction, the last known diagnosis of the subject was used to select the category of models for forecasting.\\
 For the prediction of clinical diagnosis, a two-stage approach was used based on prognosis and time-to-event estimation. The prognosis approach used an ensemble of 50 regression models to estimate the future diagnosis, while the time-to-event method used an ensemble of 25 models to estimate the square root of the time it took for a subject to convert to MCI or AD. These approaches were performed independently for CN-to-MCI, MCI-to-AD and CN-to-AD conversion.\\
 \textbf{Confidence Intervals:} The 50\% confidence intervals for ADAS-Cog13 and Ventricle volume were estimated by extracting the interquartile range of the 50 regression estimates. \\
 \textbf{Repository link:} \url{https://github.com/joseTamezPena/TADPOLE}\\
  \\
  \\
 \textbf{Team:} BravoLab \crule[cyan]{0.3cm}{0.3cm}\crule[cyan]{0.3cm}{0.3cm} (Members: Aya Ismail, Timothy Wood, Hector Corrada Bravo, Institution: University of Maryland, USA)\\
 \textbf{Overall Ranking:} 42\\
 \textbf{Feature selection:} Automatic, using a random forest to select features with highest cross-entropy or GINI impurity reduction.\\
 \textbf{Selected features:}
 \begin{itemize}
 \item Ventricle prediction: MRI volumes of ventricular sub-regions (Freesurfer cross-sectional and longitudinal)
 \item ADAS-Cog13 prediction: RAVLT, Diagnosis, MMSE, CDRSOB
 \item Diagnosis prediction: ADAS-Cog13, ADAS11, MMSE, CSRSOB
 \end{itemize}
 \textbf{Missing data:} Imputation using Hot Deck (\cite{andridge2010review}) was done only for data missing at random. \\
 \textbf{Confounder correction:} None\\
 \textbf{Method category:} Machine learning\\
 \textbf{Prediction method:} A long-short term memory network (LSTM) with target replication was trained independently for each category: Diagnosis, ADA13 and Ventricles. All existing data was used for the first forecast, after which the output of the last prediction was used as input for the next prediction, along with other features that remain constant over time. Since subjects had a different number of visits and available biomarker data, the network was adapted to accept inputs of variable length. For predictions, the network used a weighted mean absolute error as a loss function. In addition, for the prediction of the clinical diagnosis, a soft-max function was used to get the final prediction.\\
 \\
  \\
  \\
 \textbf{Team:} CBIL \crule[cyan]{0.3cm}{0.3cm}\crule[cyan]{0.3cm}{0.3cm} (Members: Minh Nguyen, Nanbo Sun, Jiashi Feng, Thomas Yeo, Institution: National University of Singapore, Singapore)\\
 \textbf{Overall Ranking:} 5\\
 \textbf{Feature selection:} Manual, based on model performance on D1 subset.\\
 \textbf{Selected features:} Cognitive tests (CDRSOB, ADAS11, ADAS-Cog13, MMSE, RAVLT immediate, learning, forgetting and percent forgetting, MOCA, FAQ), MRI biomarkers (entorhinal, fusiform, hippocampus, ICV, middle temporal, ventricles, whole brain), whole brain AV45 and FDG, CSF biomarkers (amyloid-beta, tau, phosphorylated tau).\\
 \textbf{Missing data:} Imputation using interpolation.\\
 \textbf{Confounder correction:} None\\
 \textbf{Method category:} Machine learning and data-driven disease progression model\\
 \textbf{Prediction method:} Recurrent neural network adapted for variable duration between time-points. A special loss function was designed, which ensured forecasts at timepoints close together are more correlated than those at timepoints further apart.\\
 \textbf{Confidence Intervals:} hardcoded values\\
 \textbf{Publication link:} \url{https://www.biorxiv.org/content/10.1101/755058v1}\\
 \textbf{Repository link:} \url{https://github.com/ThomasYeoLab/CBIG/tree/master/stable_projects/predict_phenotypes/Nguyen2020_RNNAD}\\
  \\
  \\
 \textbf{Team:} Chen-MCW \crule[blue]{0.3cm}{0.3cm}\crule[green]{0.3cm}{0.3cm} (Members: Gang Chen, Institution: Medical College of Wisconsin, Milwaukee, USA)\\
 \textbf{Overall Ranking:} 32-35\\
 \textbf{Feature selection:} Manual\\
 \textbf{Selected features:} ADAS-Cog13, MMSE, MRI volumes (hippocampus, whole brain, entorhinal, fusiform and middle temporal), APOE status, gender and education.\\
 \textbf{Missing data:} No imputation performed.\\
 \textbf{Confounder correction:} None\\
 \textbf{Method category:} Regression and data-driven disease progression model\\
 \textbf{Prediction method:} Prediction of ADAS-Cog13 and Ventricles was made using linear regression using age, APOE status, gender and education as covariates. Different models were estimated for CN, MCI and AD subjects. For diagnosis prediction, an AD risk stage was calculated based on the Event-based probability (EBP) model (\cite{chen2016staging}). Prediction of clinical diagnosis was then made based on two approaches:
 \begin{itemize}
\item Chen-MCW-Std: Predict diagnosis based on AD stage as well as APOE4, gender and education using a Cox proportional hazards model.
\item Chen-MCW-Stratify: As above, but the model was further stratified based on AD risk stages, into low risk and high-risk.\\\\
 \end{itemize}
 \textbf{Team:} CN2L (Members: Ke Qi$^1$, Shiyang Chen$^{1,2}$, Deqiang Qiu$^{1,2}$, Institutions: $^1$Emory University, $^2$Georgia Institute of Technology)\\
 \textbf{Overall Ranking:} 11-17\\
 \textbf{Feature selection:} Automatic\\
 \textbf{Selected features:} For the neural network, all features in D1/D2 are used; For the random forest, the main cognitive tests, MRI biomarkers (cross-sectional only), FDG, AV45, AV1451, DTI, CSF, APOE, demographics and clinical diagnosis are used.\\
 \textbf{Missing data:} Imputation in a forward filled manner (i.e. using last available value).\\
 \textbf{Confounder correction:} None\\
 \textbf{Method category:} Machine learning\\
 \textbf{Prediction method:}
 \begin{itemize}
  \item CN2L-NeuralNetwork \crule[cyan]{0.3cm}{0.3cm}\crule[cyan]{0.3cm}{0.3cm} : 3-layer recurrent neural network, 1024 units/layer. Dropout layers (dropout rate:0.1) were added to output and state connections to prevent overfitting. Adam method was used for training. Validation was performed using a leave-last-time-point-out approach.
  \item CN2L-RandomForest \crule[magenta]{0.3cm}{0.3cm}\crule[magenta]{0.3cm}{0.3cm}: Random forest method was used, where features of small importance for diagnosis prediction were filtered out. For the prediction of clinical diagnosis, an ensemble of 200 trees was trained on a class-balanced bootstrap sample of the training set. For the prediction of ADAS-Cog13 and Ventricles, an ensemble of 100 trees was used.  Different predictions are made for different previous visits of a patient, and the final prediction is taken as the average of all predictions.
  \item CN2L-Average \crule[cyan]{0.3cm}{0.3cm}\crule[cyan]{0.3cm}{0.3cm}: The average of the above two methods.
 \end{itemize}
 \textbf{Confidence Intervals:} Confidence intervals are estimated based on probabilities output of the model.\\
 \textbf{Publication link:} \url{https://cds.ismrm.org/protected/18MPresentations/abstracts/3668.html} Chen et al., ISMRM, 2018 (\cite{chen2018is})\\
 \\
 \textbf{Team:} CyberBrains \crule[blue]{0.3cm}{0.3cm}\crule[blue]{0.3cm}{0.3cm} (Members: Ionut Buciuman, Alex Kelner, Raluca Pop, Denisa Rimocea, Kruk Zsolt, Institution: Vasile Lucaciu College, Baia Mare, Romania)\\
 \textbf{Overall Ranking:} 23\\
 \textbf{Feature selection:} Manual\\
 \textbf{Selected features:} MRI volumes (Ventricles, middle temporal), ADAS-Cog13, APOE status\\
 \textbf{Missing data:} For subjects with no ventricle measurements, authors computed an average value based on ADAS-Cog13 tests. This was used especially for D3 predictions.\\
 \textbf{Confounder correction:} None\\
 \textbf{Method category:} Regression\\
 \textbf{Prediction method:} Fit a linear model of monthly difference in ventricle volume, as a function of ventricle volume, stratified by clinical diagnosis and ventricle volumes smaller and larger than 140,000 mm3. A similar model is applied for ADAS-Cog13 prediction, but stratified by APOE status and middle temporal volume smaller or greater than 16,000 mm3. Prediction of clinical diagnosis also used a linear model that was stratified based on the ADAS-Cog13, for ADAS-Cog13 ranges between 10 and 45. For ADAS-Cog13 greater than 45 and smaller than 10, pre-defined values were used for the probabilities of each diagnosis.\\
 \\
  \\
  \\
 \textbf{Team:} DIKU \crule[blue]{0.3cm}{0.3cm}\crule[green]{0.3cm}{0.3cm} (Members: Mostafa Mehdipour Ghazi$^{1,2,3,5}$, Mads Nielsen$^{1,2,3}$, Akshay Pai$^{1,2,3}$, Marc Modat$^{4,5}$, M. Jorge Cardoso$^{4,5}$, Sebastien Ourselin$^{4,5}$, Lauge S$\o$rensen$^{1,2,3}$; Institutions: $^1$Biomediq A/S, $^2$Cerebriu A/S, $^3$University of Copenhagen, Denmark, $^4$King's College London, UK, $^5$University College London, UK)\\
 \textbf{Overall Ranking:} 21-44\\
 \textbf{Feature selection:} Semi-automatic; linear discriminant analysis (LDA) was applied to select the top most-informative biomarkers, and ventricular volume and a few other MRI measures were subsequently manually added.\\
 \textbf{Selected features:} Cognitive tests (CDR-SB, ADAS-11, ADAS-13, MMSE, FAQ, MOCA, RAVLT-Immediate, RAVLT-Learning, RAVLT-Percent-Forgetting), CSF measures (amyloid-beta, phosphorylated tau), MRI volumetric measures divided by ICV  (ventricles, hippocampus, whole brain, entorhinal, fusiform, middle temporal).\\
 \textbf{Missing data:} Method automatically deals with missing data.\\
 \textbf{Confounder correction:} Linear transformation of age as part of the algorithm.\\
 \textbf{Method category:} Data-driven disease progression model.\\
 \textbf{Prediction method:} For predicting ADAS-13 and Ventricles, a data-driven disease progression model was used, which estimated a parametric trajectory for each selected feature over a common disease progression axis reflecting an estimated latent disease progression score (DPS). The chosen parametric function was generalised logistic function (Richard's curve), and the DPS was a linear transformation of the age of subjects representing the subject-specific time shift and progression speed.  The constrained fitting was performed alternating between estimation of subject-specific DPS transformations and global biomarker trajectories using L2-norm loss functions. The authors made three submissions:
 \begin{itemize}
  \item DIKU-GeneralisedLog-Std: constrained, generalised logistic function for the trajectory model;
 \item DIKU-ModifiedLog-Std: constrained sigmoid function for the trajectory model;
 \item DIKU-ModifiedMri-Std: as above, but separately fitting MRI biomarkers for Ventricles prediction.
 \end{itemize}
 The above models were trained on D1 data only. Authors also made predictions from a custom training set (D1+D2 together), named DIKU-***-Custom. Clinical diagnosis was predicted based on the DPS scores using both a Bayesian classifier with likelihoods modeling using Gaussian mixture models, as well as an ensemble of LDAs. The final prediction was obtained through bagging of the two classifiers' predictions. The whole method and a robust extension developed post-TADPOLE is described in (\cite{ghazi2019robust}).\\
 \textbf{Confidence Intervals:} They were obtained by using bootstrapping via Monte Carlo resampling and evaluating the model performance assuming a Gaussian distribution.\\
 \textbf{Publication link:} \url{https://arxiv.org/abs/1908.05338}\\
 \\
  \\
 \\
 \textbf{Team:} DIVE \crule[green]{0.3cm}{0.3cm}\crule[green]{0.3cm}{0.3cm} (Members: Razvan Marinescu, Institution: University College London, UK, Massachusetts Institute of Technology, USA)\\
 \textbf{Overall Ranking:} 38\\
 \textbf{Feature selection:} Manual\\
 \textbf{Selected features:} FDG, AV45, CDRSOB, ADAS-Cog13, MRI volumes (ventricles, hippocampus, whole brain, entorhinal, middle temporal), CSF (amyloid-beta, tau, phosphorylated tau)\\
 \textbf{Missing data:} Method automatically deals with missing data\\
 \textbf{Confounder correction:} None\\
 \textbf{Method category:} Data-driven disease progression model\\
 \textbf{Prediction method:}\\
 For predicting the ADAS-Cog13 and Ventricle volume, the “Data-Driven Inference of Vertexwise Evolution” (DIVE) algorithm was used (\cite{marinescu2019dive}), which clusters the input biomarkers based on how similar their progression is over the disease time-course. While the original DIVE method was a spatio-temporal model, for TADPOLE it was applied on extracted features directly. The model estimates a parametric, sigmoidal trajectory of the biomarkers, which are a function of subjects' disease progression scores (DPS), representing a linear transformation of their age. Subject-specific parameters included the latent time-shift and progression speed, as well as an intercept. For the prediction of clinical diagnosis, the posterior probability of each class was computed given the future DPS scores using non-parametric Kernel Density Estimators (KDE), fitted on the DPS scores for each diagnostic class independently. Code for the model is available online:\\
 \textbf{Confidence Intervals:} The model estimates a variance parameter under a gaussian noise model, which was scaled accordingly to obtain the 50\% confidence intervals for ADAS-Cog13 and Ventricle volume.\\
 \textbf{Publication link:} \url{https://www.sciencedirect.com/science/article/pii/S1053811919301491}\\
 \textbf{Repository link:} \url{https://github.com/mrazvan22/dive}\\
 \\
  \\
 \textbf{Team:} EMC1 \crule[red]{0.3cm}{0.3cm}\crule[green]{0.3cm}{0.3cm} (Members: Vikram Venkatraghavan, Esther Bron, Stefan Klein, Institution: Erasmus MC, The Netherlands)\\
 \textbf{Overall Ranking:} 2-4\\
 \textbf{Feature selection:} Automatic -- Only the subjects who had converted to AD were used for feature selection. Features with the largest changes over time after correcting for age, gender, education and ICV were selected \\
 \textbf{Selected features:}  250 features from the set of FDG, AV45, DTI, MRI (cross-sectional Freesurfer volumes), Arterial Spin Labelling (ASL) MRI, CSF and cognitive tests.  \\
 \textbf{Missing data:}  Imputed using nearest-neighbour interpolation. For D2, visits with missing diagnosis were excluded. For the D3 subjects with no known diagnosis, this was estimated using a nearest-neighbour search based on disease severity \\
 \textbf{Confounder correction:} Corrected for age, gender, education and ICV using linear regression based on data from controls.\\
 \textbf{Method category:} Data-driven disease progression model and machine learning\\
 \textbf{Prediction method:}  Authors hypothesize that aging and progression of AD are the primary causes for the change in biomarker values with time and that these changes eventually lead to a change in clinical status. To predict biomarker values at future timepoints, the rate of AD progression is estimated in each subject. This is followed by estimating the interactions of aging and AD progression in the progression of different biomarkers. Lastly, authors use the biomarkers estimated at the future timepoint to predict the change in clinical status. These steps are elaborated below: \\
 Rate of Progression of AD: To assess the severity of AD, we estimated the sequence in which the selected features became abnormal in AD using a Discriminative Event-Based Model (\cite{venkatraghavan2019disease}) and used it to estimate the disease severity at all the timepoints for each subject. A linear mixed effect model was fit to estimate the rate of change of disease severity for different subjects. This model was used for predicting the disease severity at all the future timepoints.  \\
 Interactions of aging and AD progression: For predicting the biomarker values at the future timepoint, we fit linear mixed effect models for each biomarker considering interactions between the estimated disease severity and age, with gender and ICV as additional covariates. This model was used to forecast the future values of all 250 selected features, including ADAS-Cog13 scores and Ventricle volumes. \\
 Predicting the change in clinical status: For the diagnosis prediction, the forecasted values of the biomarkers and the last known clinical diagnosis of the subject were used as inputs for a soft-margin SVM classifier with a radial basis function kernel. Two separate submissions were made:
 \begin{itemize}
  \item EMC1-Std (ID 1): ASL based features were excluded in this model
 \item EMC1-Custom (ID 2): ASL based features were included in this model
 \end{itemize}
 \textbf{Confidence Intervals:}  Standard errors of the predicted values of Ventricles and ADAS-Cog-13 were estimated by repeating the prediction procedure, including the estimation of disease severity, for 10 repetitions of bootstrap sampling. These standard errors were used to define the confidence intervals. \\
 \textbf{Publication link:} \url{https://doi.org/10.1016/j.neuroimage.2018.11.024}\\
 \textbf{Repository link:} \url{https://github.com/88vikram/TADPOLE_submission_with_debm}\\
  \\
  \\
 \textbf{Team:} EMC-EB \crule[red]{0.3cm}{0.3cm}\crule[red]{0.3cm}{0.3cm} (Members: Esther E. Bron, Vikram Venkatraghavan, Stefan Klein, Institution: Erasmus MC, The Netherlands)\\
 \textbf{Overall Ranking:} 10\\
 \textbf{Feature selection:} Automatic -- For the D2 prediction, features were selected that had the largest change over time in subjects who converted to AD using corrections for age, gender, education and ICV, i.e. the same approach as EMC1.  For the D3 prediction, features with less than 50\% missing data were selected.\\
 \textbf{Selected features:} 200 (D2 prediction) and 338 (D3 prediction) from the set of clinical diagnosis, cognitive tests, MRI volumes (Freesurfer cross-sectional), FDG PET, DTI measures (FA, MD, RD, AD) and CSF features.\\
 \textbf{Missing data:} Imputation using nearest-neighbour interpolation based on the subject's earlier timepoints. If not possible, imputation by the mean of training set was used. Visits with no clinical diagnosis were excluded for classifier training.\\
 \textbf{Confounder correction:} None\\
 \textbf{Method category:} Machine learning\\
 \textbf{Prediction method:} All predictions were based on SVMs. For diagnosis and ADAS-Cog13 prediction, respectively a classifier with balanced class weights and a regressor were trained to predict the target measures at the next visit. These predictions do not explicitly take account of time, but assume that the times between two visits are roughly equal. For Ventricle prediction, a regressor was trained to predict the change of ventricle volume per year. Ventricle volumes were normalized using ICV at baseline (ICV at current time point for D3). Using the predicted change, the normalized ventricle volume at each future visit was computed. For all predictions, authors used a radial basis function (RBF) kernel SVM, of which the C-parameter was set to $C=0.5$ and gamma to the reciprocal of the number of features. All features were normalized to zero mean and unit standard deviation. \\
 \textbf{Confidence Intervals:} Bootstrap resampling ($n=100$) of the training set.\\
 \textbf{Repository link:} \url{https://github.com/tadpole-share/tadpole-algorithms/tree/master/tadpole_algorithms/models/ecmeb}\\
  \\
  \\
 \textbf{Team:} FortuneTellerFish (Members: Alexandra Young, Institutions: University College London, UK, King's College London, UK)\\
 \textbf{Overall Ranking:} 15-29\\
 \textbf{Feature selection:} Manual\\
 \textbf{Selected features:} Age at assessment, age at scan, APOE4 status, education, gender, MRI volumes (ventricles, hippocampus, whole brain, entorhinal, fusiform, middle temporal, all major lobes, insula, and basal ganglia). The probability of being amyloid positive, obtained from joint mixture modelling of CSF amyloid-beta and global AV45, was also included as a feature. Two key features, disease subtype and stage, were derived from the Subtype and Stage Inference (SuStaIn) model based on the MRI features (\cite{young2018uncovering}.\\
 \textbf{Missing data:} Imputed by averaging over the k-nearest neighbours with $k=5$\\
 \textbf{Confounder correction:} Brain volumes were corrected for age, intracranial volume and field strength using linear regression. Parameters for the linear regression were estimated based on amyloid-negative controls.\\
 \textbf{Method category:} Data-driven disease progression models + statistical regression\\
 \textbf{Prediction method:} For ADAS-Cog13 and Ventricle prediction, a linear mixed effects model was used which used a different set of fixed/random effects, depending on the submission:
 \begin{itemize}
  \item FortuneTellerFish-Control \crule[red]{0.3cm}{0.3cm}\crule[blue]{0.3cm}{0.3cm}: For Ventricles, fixed effects were age at scan and gender. For ADAS-Cog13 and MMSE the fixed effects were age at scan, education, APOE status and amyloid positivity. For all target measures, there was one random effect per subject.
  \item FortuneTellerFish-SuStaIn \crule[red]{0.3cm}{0.3cm}\crule[green]{0.3cm}{0.3cm}: two additional fixed effects from the SuStaIn model: subtype and stage.
 \end{itemize}
 For the prediction of clinical diagnosis, a multiclass error-correcting output codes (ECOC) classifier based on SVMs was trained with the following inputs: age at assessment, age at scan, APOE status, amyloid positivity, gender, education, SuStaIn subtype and stage, ADAS-Cog13, MMSE and ventricle volume. For diagnosis prediction at future timepoints, the forecasted values for ADAS-Cog13, MMSE and Ventricle volume were used as input to the classifier.\\
 \\
  \\
 \textbf{Team:} Frog \crule[red]{0.3cm}{0.3cm}\crule[red]{0.3cm}{0.3cm} (Members: Keli Liu, Christina Rabe, Paul Manser Institution: Genentech, USA)\\
 \textbf{Overall Ranking:} 1\\
 \textbf{Feature selection:} Automatic using the Xgboost package (\cite{chen2016xgboost})\\
 \textbf{Selected features:} Cognitive tests (ADAS-Cog13, CDRSB, MMSE, RAVLT), clinical diagnosis, MRI measurements, FDG PET measurements, APOE status and CSF measurements. For each longitudinal measurement (e.g. test scores and MRI), the following transformations were computed and used to augment the original feature set: most recent measurement, time since most recent measurement, the historical highest (lowest) measurement, time since the historical highest and lowest measurement, and the most recent change in measurement.\\
 \textbf{Missing data:} Xgboost package automatically deals with missing data through inference based on reduction of training loss.\\
 \textbf{Confounder correction:} None\\
 \textbf{Method category:} Statistical prediction using regression\\
 \textbf{Prediction method:} Flexible models and features were chosen automatically using gradient boosting (Xgboost package). Different models were trained for the following forecast windows: 0-8 months, 9-15, 16-27, 28-39, 40-60, $>$60 (given windows are for clinical status prediction, slightly different windows used for ADAS-Cog13 and ventricular volume prediction) . Variable importance scores from Xgboost suggest that MRI features play a bigger role in models trained for longer forecast windows.\\
 \textbf{Confidence Intervals:} Standard deviation for prediction error was estimated based on cross validation (on training set). Normality of prediction errors was then assumed to construct prediction intervals based on estimated standard deviation.\\
 \\
  \\
  \\
 \textbf{Team:} GlassFrog-LCMEM-HDR \crule[blue]{0.3cm}{0.3cm}\crule[green]{0.3cm}{0.3cm} (Members: Steven Hill$^1$, James Howlett$^1$, Robin Huang$^1$, Steven Kiddle$^1$, Sach Mukherjee$^2$, Anaïs Rouanet$^1$, Bernd Taschler$^2$, Brian Tom$^1$, Simon White$^1$, Institutions: $^1$MRC Biostatistics Unit, University of Cambridge, UK; $^2$German Center for Neurodegenerative Diseases (DZNE), Bonn, Germany)\\
 \textbf{Overall Ranking:} 30\\
 \textbf{Feature selection:}
 \begin{itemize}
 \item MSM: Automatic, by selecting features which passed a likelihood ratio test when compared against a model with no covariates.
 \item LCMEM: Manual
 \item HDR: Automatic, selected via sparse Lasso regression.
 \end{itemize}
 \textbf{Selected features:}
  \begin{itemize}
 \item MSM: D2 - gender, age, education, ADAS-Cog13, diagnosis, MMSE, CDRSOB, APOE status, first 5 principal components from imaging, amyloid positivity, tau level. D3 - gender, age, education, ADAS-Cog13, diagnosis, MMSE, Ventricles/ICV
 \item LCMEM: ADAS-Cog13,  gender, education, age at baseline
 \item HDR: all features were provided to the method, excluding some features with many missing values
\end{itemize}
 \textbf{Missing data:}
 \begin{itemize}
 \item MSM: Filling with last known value, or nearest neighbour if feature was never observed.
 \item LCMEM: Complete case analysis (assumption that missing data are missing at random)
 \item HDR: Imputation using within-subject interpolation and nearest neighbour matching
\end{itemize}
 \textbf{Confounder correction:} None\\
 \textbf{Method category:} Combination of statistical regression and data-driven disease progression models\\
 \textbf{Prediction method:} The prediction of clinical diagnosis was done using a Multi-State Model (MSM). Multi-state models (MSMs) (\cite{kalbfleisch1985analysis}) are continuous-time Markov chain models, here with states corresponding to CN, MCI, AD, and transition rates estimated from the data using covariates selected as described above. The model accounts for noise in the historical diagnostic labels. Predictions for a given forecast month were made using the last observed disease state and associated covariates.
 
 Prediction of ADAS-Cog13 was done using a Latent class mixed effects model (LCMEM).  The model used four latent classes, where class membership probability was modelled via a multinomial logistic. For each latent class a specific linear mixed effects model defined a Gaussian latent process, with class-specific fixed effects, random effects for intercept, slope and square slope, and Gaussian noise. Age at baseline, gender and education were also included as covariates. Finally, a Beta cumulative distribution link function was used (and estimated simultaneously) between ADAS-Cog 13 and the latent process, to account for the departure from the Gaussian assumption on the outcome. The number of latent classes was optimised with the Bayesian Information Criterion (BIC).
 
 Prediction of Ventricle volume was done using high-dimensional regression (HDR) and disease state-specific slope models: Subject-specific slopes were obtained by a combination of Lasso regression and shrinkage towards disease-state-specific shrinkage targets. Conversion times, from one disease state to another, were forecasted using the MSM model.\\
 \textbf{Confidence Intervals:} LCMEM: the 50\% confidence intervals for ADAS-Cog13 were obtained using  a bootstrap approach. HDR: Confidence intervals were set as percentages of the predicted values. \\
 \\
  \\
  \\
 \textbf{Team:} GlassFrog-SM \crule[blue]{0.3cm}{0.3cm}\crule[green]{0.3cm}{0.3cm} (members and affiliations as above)\\
 \textbf{Overall Ranking:} 8\\
 \textbf{Feature selection:} Manual\\
 \textbf{Selected features:} ADAS-Cog13, Ventricles/ICV, age at visit, APOE status, education, diagnosis\\
 \textbf{Missing data:} For training, complete case analysis was performed. For prediction, imputation was performed for missing outcomes using a linear model with age, education, diagnosis and APOE status as covariates.\\
 \textbf{Confounder correction:} None\\
 \textbf{Method category:} Combination of statistical regression and data-driven disease progression models\\
 \textbf{Prediction method:} The prediction of clinical diagnosis was using MSM models as in GlassFrog-LCMEM-HDR. The prediction for ADAS-Cog13 and Ventricles used a Slope Model (SM), which used a quadratic function to model the slope of the outcome variable as a function of the  current outcome value and covariates. Covariates used were age at visit, education and APOE status.\\
 \textbf{Confidence Intervals:} SM: Confidence intervals were set as percentages of the predicted values. The percentages used were manually selected and depended on the missingness of covariates for each individual.\\
 \\
  \\
 \textbf{Team:} GlassFrog-Average \crule[blue]{0.3cm}{0.3cm}\crule[green]{0.3cm}{0.3cm} (members and affiliations as above)\\
 \textbf{Overall Ranking:} 7\\
 \textbf{Prediction method:} The prediction of clinical diagnosis was using MSM models as in GlassFrog-LCMEM-HDR. For ADAS-Cog13 and Ventricles an ensemble approach was used that averaged the predictions from three methods: LCMEM, HDR and SM, as described above. Confidence interval bounds were also averaged. \\
  \\
 \textbf{Team:} IBM-OZ-Res (Members: Noel Faux, Suman Sedai, Institution: IBM Research Australia, Melbourne, Australia)\\
 \textbf{Clinical Diagnosis Ranking:} 18\\
 \textbf{Feature selection:} using boosting regression\\
 \textbf{Selected features:} Ventricle volume, AV45, FDG PET, cognitive tests, clinical diagnosis, age\\
 \textbf{Missing data:} Imputed with zero; observations with missing ventricle volume are dropped.\\
 \textbf{Confounder correction:} None\\
 \textbf{Method category:} Machine Learning\\
 \textbf{Prediction method:} A stochastic gradient boosting regression machine (GBM) was used to predict Ventricle Volume, with a Huber loss function. To reduce overfitting, a shrinkage mechanism was adopted, where the response of each tree is reduced by a factor of 0.01. Independent predictions were made for each individual visit, and averaged when a subject had more than one visit. For the prediction of clinical status, a similar GBM model was adopted, but with a multinomial deviance loss function.\\
 \\
  \\
 \textbf{Team:} ITESMCEM \crule[magenta]{0.3cm}{0.3cm}\crule[blue]{0.3cm}{0.3cm} (Members: Javier de Velasco Oriol$^1$, Edgar Emmanuel Vallejo Clemente$^1$, Karol Estrada$^2$. Institution: $^1$Instituto Tecnol\'{o}gico y de Estudios Superiores de Monterrey, $^2$Brandeis University)\\
 \textbf{Overall Ranking:} 39\\
 \textbf{Feature selection:} Manual\\
 \textbf{Selected features:} Demographics, MRI volumes, FDG PET and all cognitive tests\\
 \textbf{Missing data:} Imputation using the mean of previous values of that patient, otherwise mean across all patients.\\
 \textbf{Confounder correction:} None\\
 \textbf{Method category:} Machine learning\\
 \textbf{Prediction method:} ADAS-Cog13 is predicted with a Lasso model with $\alpha = 0.1$. Ventricles are predicted using a Bayesian ridge regression. Clinical diagnosis is predicted using two different Random Forest models, one which selected between CN and either MCI or AD and the second which in turn predicted between MCI and AD for those selected as non-CN. The predictions for all target variables made use of a “transition model”, which predicted the next timepoint given the current one, until all 60 monthly predictions were made. The transition model was implemented using a total of 29 Lasso models.\\
 \textbf{Confidence Intervals:} They were calculated by sampling the test samples multiple times, evaluating the performance of the model with those samples and analyzing the CIs supossing a Gaussian distribution and the corresponding t-distribution.\\
 \\
  \\
 \textbf{Team:} lmaUCL \crule[blue]{0.3cm}{0.3cm}\crule[blue]{0.3cm}{0.3cm} (Members: Leon Aksman, Institution: University College London, UK)\\
 \textbf{Overall Ranking:} 11-25\\
 \textbf{Feature selection:} Manual\\
 \textbf{Selected features:} Diagnosis, gender, education, APOE4 status and MMSE.\\
 \textbf{Missing data:} Imputation using regression over ventricles and demographics (age, gender, education)\\
 \textbf{Confounder correction:} None\\
 \textbf{Method category:} Statistical regression and machine learning\\
 \textbf{Prediction method:} For ADAS-Cog13 and Ventricles, a multi-task learning model was used with similar trajectories across subjects. The regression model was a linear model over age, but dependencies between different subjects were modelled through a special prior structure over the coefficients of the linear model. The prior structure has hyperparameters that control for the amount of coupling across subjects, and are optimised through empirical Bayes. Clinical diagnosis was predicted using the ADAS-Cog13 trajectory estimates plus a simple estimate of the mean and standard deviation of ADAS-Cog13 in each diagnostic group (AD/MCI/CN). Each ADAS-Cog13 prediction was then assigned a probability of belonging to each group. Three different submissions were made:
 \begin{itemize}
  \item lmaUCL-Std: used only last available diagnosis as covariate
 \item lmaUCL-Covariates: as above, but also used gender, education, APOE status and MMSE as covariates
 \item lmaUCL-halfD1: trained only on half of the D1 dataset, but allowed for longer training time of 10 hours.
  \end{itemize}
 \textbf{Confidence Intervals:} Both the multi-task learning and clinical diagnosis models are probabilistic, providing estimates of predictive mean and standard deviation assuming a normal distribution. These were converted to confidence intervals using the inverse normal CDF. \\
 \textbf{Publication link:}  \url{https://doi.org/10.1002/hbm.24682}\\
 \textbf{Repository link:} \url{https://github.com/LeonAksman/bayes-mtl-traj}\\
  \\
 \textbf{Team:} Mayo-BAI-ASU \crule[blue]{0.3cm}{0.3cm}\crule[blue]{0.3cm}{0.3cm} (Members: Cynthia M. Stonnington$^1$, Yalin Wang$^2$, Jianfeng Wu$^2$, Vivek Devadas$^3$, Institution: $^1$Mayo Clinic, Scottsdale, AZ, USA, $^2$School of Computing, Informatics and Decision Systems Engineering, Arizona State University, USA, $^3$Banner Alzheimer's Institute, Phoenix, AZ, USA)\\
 \textbf{Overall Ranking:} 27\\
 \textbf{Feature selection:} Manual, from clinical experience\\
 \textbf{Selected features:} Age, PET(AV45, AV1451, FDG), hippocampal volume/ICV, ventricle volume/ICV, diagnosis, cognitive tests (ADAS11, ADAS-Cog13, MMSE, RAVLT, MOCA, Ecog), amyloid-beta, tau, phosphorylated tau,  APOE status. All features except age were z-score normalised.\\
 \textbf{Missing data:} Imputation with zero.\\
 \textbf{Confounder correction:} None\\
 \textbf{Method category:} Statistical regression\\
 \textbf{Prediction method:} ADAS-Cog13 and Ventricles were forecasted using a linear mixed effects model, using all features as fixed effects and one random effect per subject (intercept). Training used all visits, but the forecasts only used the last visit. Clinical diagnosis was predicted with a similar model, by converting to CN/MCI/AD to a categorical variable (1/2/3).\\
 \\
  \\
 \textbf{Team:} Orange \crule[gray]{0.3cm}{0.3cm}\crule[gray]{0.3cm}{0.3cm} (Members: Clementine Fourrier, Institution: Institut du Cerveau et de la Moelle \'{e}pini\`{e}re, ICM, Paris, France)\\
 \textbf{Clinical Diagnosis Ranking:} 44-45\\
 \textbf{Feature selection:} Manual, through knowledge from literature\\
 \textbf{Selected features:} demographics (age, education, gender), cognitive tests (ADAS11, CDSRB, MMSE), imaging(AV45 PET, FDG PET, hippocampus size, cortical thickness) and molecular markers (phosphorylated tau to amyloid-beta ratio, total tau, CMRgl, HCI)\\
 \textbf{Method category:} Decision tree of a clinician\\
 \textbf{Prediction method:} This method is based on the decision tree of a clinician. It looks at the latest available visit for a patient, and based on the value of the selected features, it predicts a duration to conversion. The duration to conversion depends on the initial clinical diagnosis and the other available data. Depending on the initial diagnosis, the algorithm assumes the patient will convert within a certain time period, and this time period is modulated by the available data about the patient. In that regard, the algorithm does not need to account for missing values. The ADAS-Cog13 and Ventricle measures at each month are computed assuming a linear evolution between the current time point and the conversion date.\\
 \\
  \\
 \textbf{Team:} Rocket \crule[blue]{0.3cm}{0.3cm}\crule[green]{0.3cm}{0.3cm} (Members: Lars Lau Raket, Institutions: H. Lundbeck A/S, Denmark; Clinical Memory Research Unit, Department of Clinical Sciences Malmö, Lund University, Lund, Sweden )\\
 \textbf{Overall Ranking:} 33\\
 \textbf{Feature selection:} Manual\\
 \textbf{Selected features:} ADAS-Cog13, baseline diagnosis, and APOE4 carrier status.\\
 \textbf{Missing data:} APOE4: Imputed using median number of alleles per baseline diagnostic group. Missing ADAS-Cog13 scores were imputed using multivariate imputation by chained equations based on age, sex, diagnosis and cognitive tests.\\
 \textbf{Confounder correction:} None\\
 \textbf{Method category:} Statistical regression and data-driven disease progression modelling\\
 \textbf{Prediction method:} Prediction of ADAS-Cog13 is done through a latent-time non-linear mixed-effects model, where the trajectory is parameterised using an exponential function. The time shift is built from a fixed effect shift relative to time since baseline for different diagnostic groups (e.g. AD patients will be shifted to be later in the course of cognitive decline) and a random effect shift for each subject. The model also includes APOE status as a fixed effect that modifies rate of decline. This disease progression modeling methodology along with several extensions is presented in (\cite{raket2019disease}). Prediction of Ventricles/ICV uses a linear mixed-effects model using an integrated B-spline basis (5 knots + intercept) in predicted ADAS-Cog13 disease time. A random intercept is included per subject. Prediction of clinical diagnosis is based on kernel density estimation of the states (CN/MCI/AD) across the disease time from the ADAS-Cog13 model. \\
 \textbf{Confidence Intervals:} Prediction intervals conditioned on the predicted disease time of the subject were derived based on the estimated variance-covariance matrix of the model. Because of the monotone nature of cognitive decline, and to heuristically compensate for the conditioning on disease time, the upper limit of the prediction interval was multiplied by 1.5.    \\
  \textbf{Publication link:} \url{https://doi.org/10.1101/2019.12.13.19014860}
 \\
  \\
 \textbf{Team:} SBIA \crule[red]{0.3cm}{0.3cm}\crule[blue]{0.3cm}{0.3cm} (Members: Aristeidis Sotiras, Guray Erus, Jimit Doshi, Christos Davatzikos, Institution: Center for Biomedical Image Computing and Analytics, University of Pennsylvania)\\
 \textbf{Overall Ranking:} 31\\
 \textbf{Feature selection:} Manual\\
 \textbf{Selected features:} demographics, cognitive tests, diagnosis, MRI features (Freesurfer cross-sectional). Imaging indices (SPARE-AD and SPARE-MCI) that summarise brain atrophy patterns were estimated through support vector machines with linear kernels. Another index representing brain age (SPARE-BA) was estimated using a regressor model applied to imaging features. \\
 \textbf{Missing data:} Features and time-points with missing data were not included\\
 \textbf{Confounder correction:} Age, gender, APOE4, education, and SPARE scores were used as covariates in the linear mixed effects models. Also, the regression model was applied separately on different diagnosis groups.\\
 \textbf{Method category:} Statistical regression and machine learning\\
 \textbf{Prediction method:} A linear mixed effects model was used to forecast the SPARE indices for future timepoints. For diagnosis predictions, authors used class probability distribution estimations based on the forecasted SPARE-AD score.  For the prediction of ADAS-Cog13 and Ventricles, linear mixed effects models were used, with age, gender and SPARE scores as covariates.\\
 \\
  \\
 \textbf{Team:} SmallHeads -- BigBrains (Members: Jacob Vogel, Andrew Doyle, Angela Tam, Alex Diaz-Papkovich, Institution: McGill University, Montreal, Canada)\\
 \textbf{ADAS-Cog13 Ranking:} 46-53\\
 \textbf{Feature selection:} Automatic\\
 \textbf{Selected features:}
 \begin{itemize}
  \item SmallHeads-NeuralNetwork \crule[cyan]{0.3cm}{0.3cm}\crule[cyan]{0.3cm}{0.3cm}: All features in the TADPOLE spreadsheet were considered, as long as they had less than 50\% missing data. This resulted in a final set of 376 features. Features were normalised to zero mean and unit variance.
 \item SmallHeads-LinMixedEffects \crule[gray]{0.3cm}{0.3cm}\crule[blue]{0.3cm}{0.3cm}: All features were normalised to zero mean and unit variance. A LASSO feature selection algorithm with ADAS-Cog13 and “Y” variable was used to select the best features, which were required to have a weight greater than 0.001. 10-fold cross-validation was used to estimate the best LASSO parameters.
\end{itemize}
 \textbf{Missing data:} Imputed using 5-nearest neighbour method, using Euclidean distance (FancyImpute 0.0.4)\\
 \textbf{Confounder correction:} None\\
 \textbf{Method category:} Machine learning\\
 \textbf{Prediction method:}
 \begin{itemize}
 \item SmallHeads-NeuralNetwork: A Deep fully connected neural network was trained to predict the future diagnosis using the selected features and time until future timepoint as input. Network has 5 fully-connected layers with Leaky ReLU activations. Each layer has 512, 512, 1024, 1024 and 256 neurons, with softmax layer at output. Training used P(0.5) dropout using the Adam optimiser, based on a class-unweighted categorical cross-entropy loss function.
 \item SmallHeads-LinMixedEffects: Only ADAS-Cog13 was predicted with a linear mixed effects model, using months since baseline as an interaction term. A random (subject | time) effect was also added, allowing variable subject-specific slopes over time.
 \end{itemize}
 \textbf{Repository link:} \url{https://github.com/SmallHeads/tadpole}\\
  \\
  \\
 \textbf{Team:} SPMC-Plymouth \crule[gray]{0.3cm}{0.3cm}\crule[gray]{0.3cm}{0.3cm} (Members: Emmanuel Jammeh, Institution: University of Plymouth, UK)\\
 \textbf{Overall Ranking:} N/A\\
 \textbf{Feature selection:} Automatic -- Authors used the WEKA (\url{https://www.cs.waikato.ac.nz/ml/weka/}) machine learning tool.\\
 \textbf{Selected features:} Age, gender, education, ApoE4, CDRSB, ADAS11, MMSE, RAVLT, Moca, Ecog,  Hippocampus, WholeBrain, Entorhinal, MidTemp, FDG, AV45, PIB, ABETA, TAU, PTAU \\
 \textbf{Missing data:}\\
 \textbf{Confounder correction:} None\\
 \textbf{Method category:} Machine learning\\
 \textbf{Prediction method:} A machine learning classifier based on k-nearest neighbours, Naive Bayes, Random Forest and SVM was used to predict the clinical diagnosis. ADAS-Cog13 and Ventricle volume were not predicted. Two predictions were made, SPMC-Plymouth1 and  SPMC-Plymouth2, but the authors could not be contacted to provide details on the differences between the two. \\
  \\
 \textbf{Team:} Sunshine \crule[red]{0.3cm}{0.3cm}\crule[blue]{0.3cm}{0.3cm} (Members: Igor Koval, Stanley Durrleman, Institution: Institut du Cerveau et de la Moelle \'{e}pini\`{e}re, ICM, Paris, France)\\
 \textbf{Overall Ranking:} 41-45\\
 \textbf{Feature selection:} Semi-automatic -- An initial set of 60 features was selected by a clinical expert. Out of this set, the features that had more than 30\% missing data were removed. A final subset of features was chosen based on cross-validation results using trial and error.\\
 \textbf{Selected features:} Age, APOE status, MMSE, ADAS-Cog13, RAVLT immediate and CDRSOB\\
 \textbf{Missing data:} Imputed using mean value\\
 \textbf{Confounder correction:} None\\
 \textbf{Method category:} Statistical regression and machine learning\\
 \textbf{Prediction method:} A linear model was used to predict future values of ADAS-Cog13 and Ventricles, as well as other cognitive tests: MMSE, RAVLT and CDRSOB. For the prediction of clinical diagnosis, forecasted values of the previous five measures were used as input to an SVM. APOE4 status and education were also used as inputs to the SVM. Adding extra features did not seem to increase prediction scores based on cross-validation. Two submissions were made:
 \begin{itemize}
 \item Sunshine-Conservative: CN and AD subjects were forecasted to have the same diagnosis (i.e. no conversion) for all future timepoints, after observing that a small proportion of them convert after 1 year.
 \item Sunshine-Std: Without the above modification.\\\\
\end{itemize}
 \textbf{Team:} Threedays \crule[magenta]{0.3cm}{0.3cm}\crule[gray]{0.3cm}{0.3cm} (Members: Paul Moore$^1$, Terry J. Lyons$^1$, John Gallacher$^2$, Institution: $^1$Mathematical Institute, University of Oxford, $^2$Department of Psychiatry, University of Oxford, UK)\\
 \textbf{Clinical Diagnosis Ranking:} 2\\
 \textbf{Feature selection:} Manual\\
 \textbf{Selected features:} Age, months since baseline, gender, race, marital status, diagnosis, cognitive tests (MMSE, CDRSB, ADAS11, ADAS-Cog13, RAVLT immediate, learning, forgetting and percent forgetting, FAQ) and APOE status.\\
 \textbf{Missing data:} Random forest method deals with missing data automatically, by finding optimal splits with existing data only.\\
 \textbf{Confounder correction:} None\\
 \textbf{Method category:} Machine learning\\
 \textbf{Prediction method:} For the prediction of clinical diagnosis, two random forest models are trained, the first for transitions from a healthy diagnosis, and the second for transitions from an MCI diagnosis.  For AD individuals, authors assume that the diagnosis will not change.  The training data was generated by ordering each participant's data by time, then associating the feature vector x with diagnosis y for each time horizon available for the participant. ADAS-Cog13 and Ventricles were not predicted.  The PLOS paper describes a method similar to the original, but using different predictors and a single random forest. \\
 \textbf{Publication link:} \url{https://journals.plos.org/plosone/article?id=10.1371/journal.pone.0211558}\\
 \\
  \\
 \textbf{Team:} Tohka-Ciszek (Members: Jussi Tohka, Robert Ciszek Institution: A.I. Virtanen Institute for Molecular Sciences, University of Eastern Finland, Finland)\\
 \textbf{Overall Ranking:} 19\\
 \textbf{Feature selection:} Manual\\
 \textbf{Selected features:}
 \begin{itemize}
 \item D2: diagnosis, gender, education, race, marital and APOE status, age, cognitive tests (CDRSB, ADAS11, ADAS-Cog13, MMSE, RAVLT learning, immediate and perc. forgetting, FAQ, MOCA, all Ecog), MRI volumes (ventricles, hippocampus, whole brain, entorhinal, fusiform, middle temporal, ICV)
 \item D3: diagnosis, age, gender, education, ethnicity, race, marital status, ADAS-Cog13, MMSE, MRI volumes as above
\end{itemize}
 \textbf{Missing data:}
 \begin{itemize}
  \item SMNSR \crule[gray]{0.3cm}{0.3cm}\crule[blue]{0.3cm}{0.3cm}: A sub-model is trained for each subset of features which occurs without missing values. A specific catch-all subset is used for patients for which well performing whole measurement set cannot be found. For this data set, values are imputed using the median of k-nearest neighbours or replaced with -1.
 \item RandomForestLin \crule[magenta]{0.3cm}{0.3cm}\crule[magenta]{0.3cm}{0.3cm}: Imputation using mean values from the timepoints of the same subject (D2) or diagnostic category (for D3).
 \end{itemize}
 \textbf{Confounder correction:} None\\
 \textbf{Method category:} Machine learning\\
 \textbf{Prediction method:}
 \begin{itemize}
 \item Tohka-Ciszek-SMNSR \crule[gray]{0.3cm}{0.3cm}\crule[blue]{0.3cm}{0.3cm}: For ADAS-Cog13 prediction, a Sparse Multimodal Neighborhood Search Regression was used. This method first uses a linear regression model to estimate ADAS-Cog13 from the selected features belonging to the current subject and neighbour subjects, estimated based on a K-nearest neighbour algorithm. The forecasts from this model were passed to a gradient-boosted tree model, providing the final prediction. Ventricles and clinical diagnosis were not predicted.
 \item Tohka-Ciszek-RandomForestLin \crule[magenta]{0.3cm}{0.3cm}\crule[magenta]{0.3cm}{0.3cm}: To predict ADAS-Cog13 and Ventricles, a weighted average of two models was used: 1) a unimodal linear model 2) a linear model taking the response variable from the final time point and predictor variables from a time point before that. For diagnosis prediction, a random forest was trained using ADAS-Cog13, Ventricle/ICV, age and APOE status.
 \end{itemize}
 \textbf{Confidence Intervals:}  Predicted score +/- cross-validation MAE.\\
 \textbf{Repository link:} \url{https://github.com/jussitohka/tadpole (RandomForestLin), https://github.com/rciszek/SMNSR} (SMNSR)\\
  \\
 \textbf{Team:} VikingAI \crule[green]{0.3cm}{0.3cm}\crule[green]{0.3cm}{0.3cm} (Members: Bruno Jedynak$^1$, Kruti Pandya$^1$, Murat Bilgel$^2$, William Engels$^1$, Joseph Cole$^1$, Institutions: $^1$Portland State University, USA, $^2$Laboratory of Behavioral Neuroscience, National Institute on Aging, National Institutes of Health, Baltimore, MD, USA)\\
 \textbf{Overall Ranking:} 3-13\\
 \textbf{Feature selection:} Manual\\
 \textbf{Selected features:} Diagnosis, age, ADAS-Cog13, CDRSOB, MMSE, RAVLT immediate, tau,\\
 ventricles/ICV, hippocampus/ICV. Features were normalized prior to model fitting.\\
 \textbf{Missing data:} Method automatically deals with missing data through Bayesian inference\\
 \textbf{Confounder correction:} None\\
 \textbf{Method category:} Data-driven disease progression model\\
 \textbf{Prediction method:} For the prediction of ADAS-Cog13 and Ventricles, a latent-time parametric model was used, estimating a linear subject-specific model over the age of subjects, resulting in a disease progression score (DPS). The trajectories of biomarkers were assumed to be either sigmoidal functions or a sum of logistic basis functions over the DPS space. Some feature-specific parameters are estimated from the features' histograms, while the rest are optimized. Priors over the parameters to be optimized are set a-priori. Two submissions were made:
 \begin{itemize}
  \item VikingAI-Sigmoid: sigmoidal function as biomarker trajectory
 \item VikingAI-Logistic: the sum of 15 logistic basis functions as the biomarker trajectory
 \end{itemize}
 \textbf{Confidence Intervals:} Bayesian predictive intervals\\
   \\
 In addition to the above entries, the organisers also included several benchmark algorithms as well as some extra predictions: (1) two ensemble predictions averaging all the predictions submitted by the participants, and (2) 62 randomised predictions which can tell how likely it is that top submissions obtained their scores due to chance, by building a null distribution of the performance metrics. The\\
 Source code of some benchmarks (\emph{BenchmarkLastVisit}, \emph{BenchmarkMixedEffectsAPOE}, \emph{BenchmarkSVM}) was offered to participants before the conference deadline, as a starting point for making predictions.\\
  \\
 \textbf{Benchmark:}  BenchmarkLastVisit \crule[orange]{0.3cm}{0.3cm}\crule[orange]{0.3cm}{0.3cm} (Authors: Daniel Alexander, Razvan Marinescu, Institutions: University College London, UK, Massachusetts Institute of Technology, USA)\\
 \textbf{Overall Ranking:} 40\\
 \textbf{Feature selection:} None\\
 \textbf{Selected features:} ADAS-Cog13, ventricle volume, diagnosis\\
 \textbf{Missing data:} Not required\\
 \textbf{Confounder correction:} None\\
 \textbf{Method category:} Regression\\
 \textbf{Prediction method:} For ADAS-Cog13 and Ventricles, the last available measure is used, otherwise the average for the current diagnostic group is used. Confidence intervals are set to default widths of 2 for ADAD13 and 0.001 for Ventricles/ICV. For prediction of clinical diagnosis, the last available diagnosis is used with probability 100\%, and 0\% probability for the other diagnoses.\\
 \textbf{Confidence Intervals:} hard-coded\\
 \textbf{Repository link:} \url{https://github.com/noxtoby/TADPOLE/blob/master/evaluation}\\
  \\
 \textbf{Benchmark:}  BenchmarkMixedEffects \crule[orange]{0.3cm}{0.3cm}\crule[orange]{0.3cm}{0.3cm} (Author: Razvan Marinescu, Daniel Alexander, Institution: University College London, UK, Massachusetts Institute of Technology, USA)\\
 \textbf{Overall Ranking:} 10-18\\
 \textbf{Feature selection:} None\\
 \textbf{Selected features:} ADAS-Cog13, ventricle volume, diagnosis (, APOE status)\\
 \textbf{Missing data:} Automatic, since model is univariate.\\
 \textbf{Confounder correction:} APOE status was used as covariate in the linear mixed effects model\\
 \textbf{Method category:} Regression\\
 \textbf{Prediction method:} Linear Mixed Effects Model with age at visit as the predictor variable. Model was fitted independently for ADAS-Cog13 and Ventricles. Predictions for clinical diagnosis were derived from the corresponding ADAS-Cog13 forecasts, using three Gaussian likelihood models for CN, MCI and AD. The likelihoods for diagnostic classes were finally converted to probabilities by normalisation. Default values were used for confidence intervals. Two predictions were made:
 \begin{itemize}
 \item BenchmarkMixedEffectsAPOE: the slope of the population trajectory was stratified by APOE status
 \item BenchmarkMixedEffects: as above but without APOE
 \end{itemize}
 \textbf{Confidence Intervals:} hard-coded\\
 \textbf{Repository link:} \url{https://github.com/noxtoby/TADPOLE/blob/master/evaluation}\\
  \\
 \textbf{Benchmark:}  BenchmarkSVM \crule[orange]{0.3cm}{0.3cm}\crule[orange]{0.3cm}{0.3cm} (Author: Esther Bron, Institution: Erasmus MC)\\
 \textbf{Overall Ranking:} 34-35\\
 \textbf{Feature selection:} Manual\\
 \textbf{Selected features:} Diagnosis, age, ADAS-Cog13, Ventricles, ICV, APOE\\
 \textbf{Missing data:} Fill-in using average value of biomarker from past visits of the same subject, otherwise population average.\\
 \textbf{Confounder correction:} None\\
 \textbf{Method category:} Machine Learning\\
 \textbf{Prediction method:} For the prediction of clinical diagnosis, a probabilistic SVM was used based on the selected features, while for the prediction of ADAS-Cog13 and Ventricles, a Support Vector Regressor (SVR) was used. All SVM/SVRs used linear kernels. Default values were used for confidence intervals.\\
 \textbf{Confidence Intervals:} hard coded\\
 \textbf{Repository link:} \url{https://github.com/noxtoby/TADPOLE/blob/master/evaluation}\\
  \\
 \textbf{Benchmark:}  RandomisedBest \crule[gray]{0.3cm}{0.3cm}\crule[gray]{0.3cm}{0.3cm} (Author: Razvan Marinescu, Institutions: University College London, UK, Massachusetts Institute of Technology, USA)\\
 \textbf{Overall Ranking:} 15\\
 \textbf{Feature selection:} Manual\\
 \textbf{Selected features:} Diagnosis, age, ADAS-Cog13, Ventricles, ICV\\
 \textbf{Missing data:} Fill-in using last available measurement\\
 \textbf{Confounder correction:} None\\
 \textbf{Method category:} Regression\\
 \textbf{Prediction method:} The method aims to construct a null distribution of values, to check how likely a high score could be obtained by chance alone. Starting from the simplest prediction method, i.e. BenchmarkLastVisit which simply takes the last available measure, 62 randomised predictions were created (as many as the total number of predictions in D2) by adding random perturbations to the predictions. For Diagnosis prediction, the probability of controls and MCI subjects to convert within 1 year was computed from ADNI historical data, and then each control and MCI was randomly chosen to convert with those probabilities. For ADAS-Cog13 and Ventricles, random uniform noise was added to the predictions as follows:
 \begin{itemize}
 \item ADAS: new\_adas $\sim$ last\_available\_adas  + U(0, 7)
 \item Ventricles: new\_ventricles $\sim$ last\_available\_ventricles  + U(0, 0.01)
 \end{itemize}
 The new values new\_adas and new\_ventricles, as well as the new diagnosis, were assigned to all 60 months, thus assuming no change across the 60 month predictions. All 62 different predictions were evaluated, and the RandomisedBest entry shows the best scores obtained by all 62 submissions in each category separately.\\
 \textbf{Confidence Intervals:} hard-coded\\
 \textbf{Repository link:} \url{https://github.com/noxtoby/TADPOLE/blob/master/evaluation}\\

\FloatBarrier
\section{Acknowledgements}

TADPOLE Challenge has been organised by the European Progression Of Neurological Disease (EuroPOND) consortium, in collaboration with the ADNI. We thank all the participants and advisors, in particular Clifford R. Jack Jr. from Mayo Clinic, Rochester, United States and Bruno M. Jedynak from Portland State University, Portland, United States for useful input and feedback. 

The organisers are extremely grateful to The Alzheimer's Association, The Alzheimer's Society and Alzheimer's Research UK for sponsoring the challenge by providing the \pounds 30,000 prize fund and providing invaluable advice into its construction and organisation. Similarly, we thank the ADNI leadership and members of our advisory board and other members of the EuroPOND consortium for their valuable advice and support.

RVM was supported by the EPSRC Centre For Doctoral Training in Medical Imaging with grant EP/L016478/1 and by the Neuroimaging Analysis Center with grant NIH NIBIB NAC P41EB015902. NPO, FB, SK, and DCA are supported by EuroPOND, which is an EU Horizon 2020 project. ALY was supported by an EPSRC Doctoral Prize fellowship and by EPSRC grant EP/J020990/01. PG was supported by NIH grant NIBIB NAC P41EB015902 and by grant NINDS R01NS086905. DCA was supported by EPSRC grants J020990, M006093 and M020533. FB was supported by the NIHR UCLH Biomedical Research Centre and the AMYPAD project, which has received support from the EU-EFPIA Innovative Medicines Initiatives 2 Joint Undertaking (AMYPAD project, grant 115952). EEB was supported by the Dutch Heart
Foundation (PPP Allowance, 2018B011) and Medical Delta Diagnostics 3.0: Dementia and Stroke. The UCL-affiliated researchers received support from the NIHR UCLH Biomedical Research Centre. This project has received funding from the European Union Horizon 2020 research and innovation programme under grant agreement No 666992. Data collection and sharing for this project was funded by the Alzheimer's Disease Neuroimaging Initiative (ADNI) (National Institutes of Health Grant U01 AG024904) and DOD ADNI (Department of Defense award number W81XWH-12-2-0012){\color{black}, and the Alzheimer's Disease Cooperative Study (ADCS), funded by the National Institutes of Health Grant U19 AG010483.} {\color{black}NPO is a UKRI Future Leaders Fellow (MRC MR/S03546X/1).}
 
The work of VikingAI was facilitated in part by the Portland Institute for Computational Science and its resources acquired using NSF Grant DMS 1624776. MB was supported by the Intramural Research Program of the National Institute on Aging, NIH. BTTY, NS and MN were supported by the Singapore National Research Foundation (NRF) Fellowship (Class of 2017). Team SBIA was supported by grant R01 AG054409. Team DIKU has received funding from the European Union's Horizon 2020 research and innovation programme under the Marie Skodowska-Curie grant agreement No 721820.

\section{Author Contributions}

DCA, RVM, NPO, ALY, EEB and SK designed the challenge. RVM performed the evaluation of the algorithms and the analysis of the results, and drafted the manuscript. {\color{black}NPO performed the external validation analysis and drafted the corresponding text.} All authors provided feedback on the manuscript. RVM, NPO, ALY, EB and DCA created the TADPOLE website. RVM and NPO constructed the TADPOLE D1-D4 datasets. RVM, DCA and EB wrote benchmark scripts which were offered to participants before the deadline. DCA and NPO organised and ran the TADPOLE webinars. NF and FB provided valuable suggestions on the challenge design. AT and MW offered access to the ADNI database. All other co-authors participated in the challenge. EB lead the TADPOLE-SHARE effort to make the algorithms openly available for further reuse, in a standardised format.

\section{Conflicts of Interest}

SJK received fees for participation in a Roche Diagnostics advisory board and does paid consulting for DIADEM outside of the submitted work.

\section{Ethical Standards}
ADNI obtained all IRB approvals and met all ethical standards in the collection of data. The ADNI protocol was approved by the Institutional Review Boards of all of the participating institutions.

The fifty-one centers in the DHA clinical trial obtained approval from their local institutional review boards. Written informed consent was obtained from study participants, legally authorized representatives, or both, according to local guidelines. 

The AIBL study, including the follow-up protocol and subsequent amendments and revisions to the protocol, was approved by the institutional human research ethics committees of Austin Health, St Vincent’s Health, Hollywood Private Hospital and Edith Cowan University. All volunteers gave written informed consent before participating in study assessments, and the study was conducted in accordance with the Helsinki Declaration of 1975.

\bibliography{bibliography-melba}

\renewcommand{\theHsection}{A\arabic{section}}

\appendix






\section{Creating the D1-D4 datasets}
\label{creatingD14}

The data used from ADNI consists of: (1) CSF markers of amyloid-beta and tau deposition; (2) various imaging modalities such as magnetic resonance imaging (MRI), positron emission tomography (PET) using several tracers: Fluorodeoxyglucose (FDG, hypometabolism), AV45 (amyloid), AV1451 (tau) as well as diffusion tensor imaging (DTI); (3) cognitive assessments acquired in the presence of a clinical expert; (4) genetic information such as alipoprotein E4 (APOE4) status extracted from DNA samples; and (5) general demographic information. Extracted features from this data were merged together into a final spreadsheet and made available on the LONI ADNI website.

The imaging data has been pre-processed with standard ADNI pipelines. For MRI scans, this included correction for gradient non-linearity, B1 non-uniformity correction and peak sharpening. [ADNI MRI pre-processing]. Meaningful regional features such as volume and cortical thickness were extracted using the Freesurfer cross-sectional and longitudinal pipelines (\cite{reuter2012within}). Each PET image (FDG, AV45, AV1451) had their frames co-registered, averaged across the six five-minute frames, standardised with respect to the orientation and voxel size, and smoothed to produce a uniform resolution of 8mm full-width/half-max (FWHM) (see \url{http://adni.loni.usc.edu/methods/pet-analysis/pre-processing/}). Standardised uptake value ratio (SUVR) measures for relevant regions-of-interest were extracted after registering the PET images to corresponding MR images using the SPM5 software (\cite{friston1994statistical}). Further details have been provided in the ADNI procedures manual. DTI scans were corrected for head motion and eddy-current distortion, skull-stripped, EPI-corrected, and finally aligned to the T1 scans using the pipeline from (\cite{prasad2013tractography}). Diffusion tensor summary measures were estimated based on the Eve white-matter atlas (\cite{oishi2009atlas}). 

In addition to the standard datasets, we also created three leaderboard datasets LB1, LB2 and LB2 which mimick the D1, D2 and D4 datasets. These datasets were used by participants to preliminarily evaluate their algorithms before the competition deadline, and to compare their results on the leaderboard system (\url{https://tadpole.grand-challenge.org/Leaderboard/}).

\section{Statistical testing}
\label{statTesting}

\subsection{Differences in MAUC scores}
\label{statTestingMAUC}

For analysing whether the MAUC scores obtained by top algorithms are significantly different, we performed a bootstrapped hypothesis test (\cite{efron1994introduction}), since the significance test for comparing two AUC scores (\cite{delong1988comparing}) does not extend to multiple classes. Let $A$ and $B$ be two TADPOLE algorithms and $M_A$ and $M_B$ be their associated MAUC scores.  If $M_A > M_B$ on the full D4 test set, we want to confirm if algorithm $A$ was significantly better than $B$, or if this was likely due to chance. We define the null hypothesis $H_0: M_A = M_B$ and the alternative hypothesis $H_1: M_A > M_B$. We then proceed as follows:
\begin{itemize}
\item Sample $N=100$ random bootstraps $D_i$ of the D4 test set with replacement.
\item Compute the $M_A^{D_i}$ and $M_B^{D_i}$ based on the bootstrapped dataset. Repeat for all $N$ bootstraps.
\item Compute the p-value as $\sum_i I[M_A^{D_i} < M_B^{D_i}]/N$, which is the proportion of bootstrapped datasets where $A$ performed worse than $B$. 
\item Accept/reject $H_0$ based on a 5\% significance level. 
\end{itemize}

\subsection{Differences in MAE scores}
\label{statTestingMAE}

For comparing differences in MAE scores, we applied the non-parametric Wilcoxon signed-rank test on paired samples of absolute errors across all visits of the D4 subjects. We chose the non-parametric Wilcoxon test because the input samples are not normally distributed, as they represent absolute errors and are always positive. We report results based on a 5\% significance level.

\subsection{Differences between D2 and D3 forecasts}
\label{statTestingD2D3}

For comparing differences between the scores obtained by two algorithms on D2 vs D3 forecasts, we use an approach similar to comparing MAUC scores (section 8.4.2). 

\subsection{Comparisons with random guessing model}
\label{statTestingRandom}

We used predictions from the \emph{RandomisedBest} model to test whether a TADPOLE algorithm was significantly better performance than random guessing. If the MAE error of a TADPOLE algorithm was $X$ and the performance of a random guess model was $R$, we wanted to test whether $H_0: X = R$ (no difference in performance) or $H_1: X < R$ (there is a difference in performance). For this, we performed the following:
\begin{itemize}
\item Generate 100 random predictions $R_i$, $i=[1, ..., 100]$. 
\item Compute the p-value as $\sum_i I[X < R_i] $
\end{itemize}   

\FloatBarrier

\section{Supplementary Results}
\label{supRes}

\begin{figure}[h]
\includegraphics[width=\textwidth]{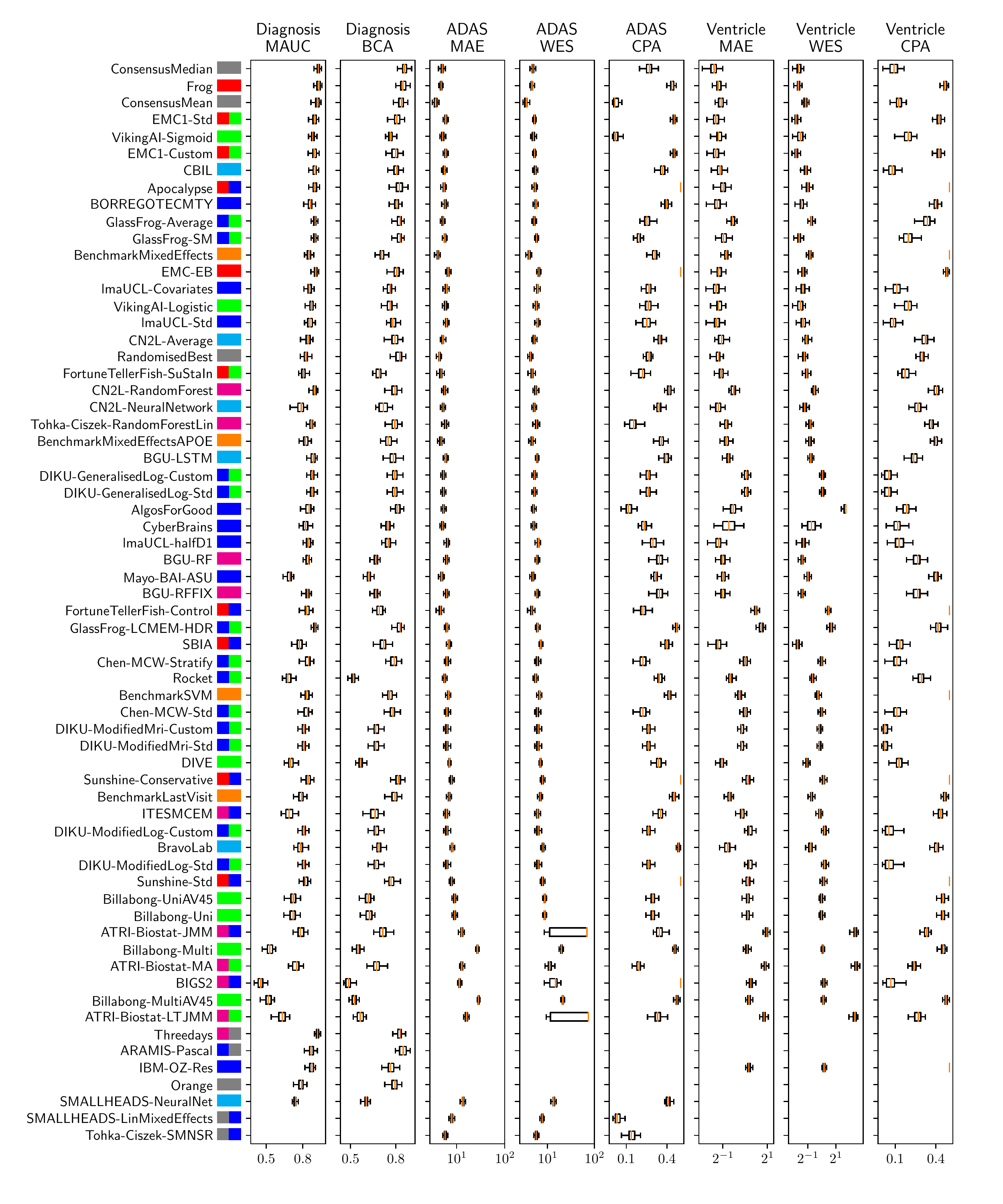}
 \caption{Distribution of performance metrics for clinical diagnosis (MAUC and BCA), ADAS-Cog13 (MAE, WES and CPA) and ventricle volume (MAE, WES and CPA) on the longitudinal D2 prediction set. For each entry, we plot the distribution of performance metrics derived using 50 bootstrap data sets drawn from the D4 test set. The submissions (rows) are in the same order as in Table \ref{tab:resultsD2}. Entries are missing where teams did not make predictions for a particular target variable.}
 \label{confIntD2}
\end{figure}

\begin{figure}[h]
\includegraphics[width=\textwidth]{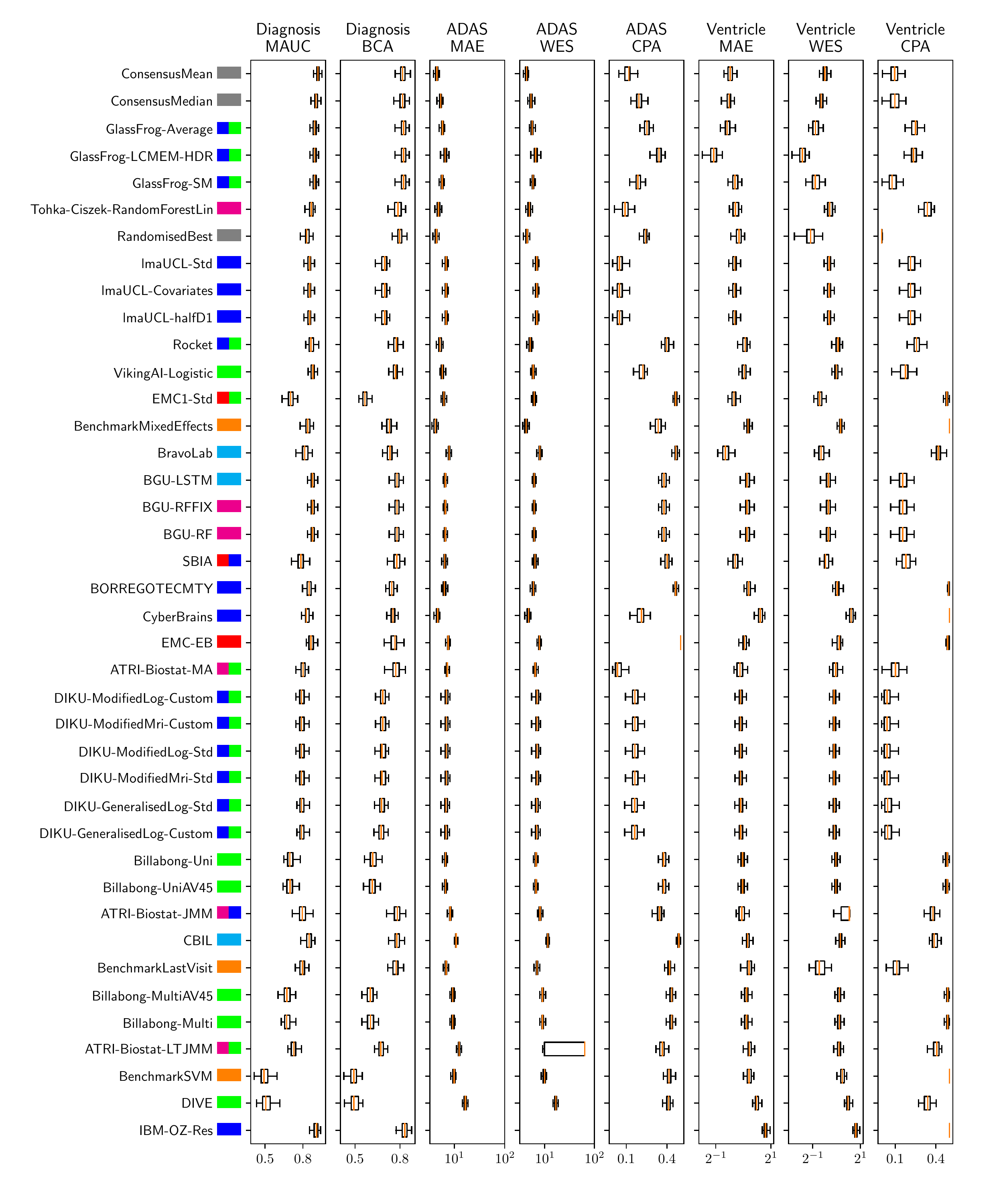}
 \caption{Box plots of performance metrics for clinical diagnosis (MAUC and BCA), ADAS-Cog13 (MAE, WES and CPA) and ventricle volume (MAE, WES and CPA) on the cross-sectional D3 prediction set. The submissions (rows) are in the same order as in Table \ref{tab:resultsD3}. Some entries are missing because teams did not make predictions for those target variables.}
 \label{confIntD3}
\end{figure}

\begin{table}[h]
 \resizebox{1\textwidth}{!}{%
\begin{tabular}{rc|ccc|cccc|cccc}
\toprule
  & Overall & \multicolumn{3}{c|}{Diagnosis} & \multicolumn{4}{c|}{ADAS-Cog13}  & \multicolumn{4}{c}{Ventricles (\% ICV)}\\
                       Submission & Rank &  Rank & MAUC & BCA  & Rank & MAE & WES & CPA & Rank & MAE & WES & CPA\\
\midrule
Billabong-UniAV45 \crule[green]{0.3cm}{0.3cm}\crule[green]{0.3cm}{0.3cm}  & \textbf{1} & \textbf{1} & \textbf{0.719} & \textbf{0.624} & \textbf{1-2} & \textbf{8.71} & \textbf{8.55} & \textbf{0.33} & 3-4 & 3.49 & 3.40 & 0.50\\ 
Billabong-Uni \crule[green]{0.3cm}{0.3cm}\crule[green]{0.3cm}{0.3cm}  & 2 & 2 & 0.717 & 0.621 & \textbf{1-2} & \textbf{8.71} & \textbf{8.55} & \textbf{0.33} & 3-4 & 3.49 & 3.40 & 0.50\\ 
Billabong-MultiAV45 \crule[green]{0.3cm}{0.3cm}\crule[green]{0.3cm}{0.3cm}  & 3 & 3 & 0.661 & 0.562 & 3-4 & 12.95 & 12.71 & 0.42 & \textbf{1-2} & \textbf{3.16} & \textbf{3.08} & \textbf{0.47}\\ 
Billabong-Multi \crule[green]{0.3cm}{0.3cm}\crule[green]{0.3cm}{0.3cm}  & 4 & 4 & 0.658 & 0.552 & 3-4 & 12.95 & 12.71 & 0.42 & \textbf{1-2} & \textbf{3.16} & \textbf{3.08} & \textbf{0.47}\\ 
Simple-SPMC-Plymouth2 \crule[gray]{0.3cm}{0.3cm}\crule[gray]{0.3cm}{0.3cm}  & - & 5 & 0.500 & 0.504 & - & - & - & - & - & - & - & -\\ 
Simple-SPMC-Plymouth1 \crule[gray]{0.3cm}{0.3cm}\crule[gray]{0.3cm}{0.3cm}  & - & 6 & 0.500 & 0.499 & - & - & - & - & - & - & - & -\\ 
  \bottomrule
\end{tabular}
}
\caption{Results on custom prediction sets from two teams: Billabong and SPMC-Plymouth. SPMC-Plymouth predicted fewer subjects due to an incomplete submission, while Billabong used a prediction set similar to D3, but filled in missing data for cognitive tests and MRI with the last available measurement. SPMC-Plymouth only submitted predictions for clinical diagnosis, and obtained an MAUC score of 0.5. Results from Billabong show higher MAUC and BCA in diagnosis prediction compared to D3, but lower performance for ADAS-Cog13 and Ventricle volume prediction. Bold entries show best scores in this category.}
\label{resDcustom}
\end{table}

\begin{figure}[h]
\includegraphics[width=\textwidth]{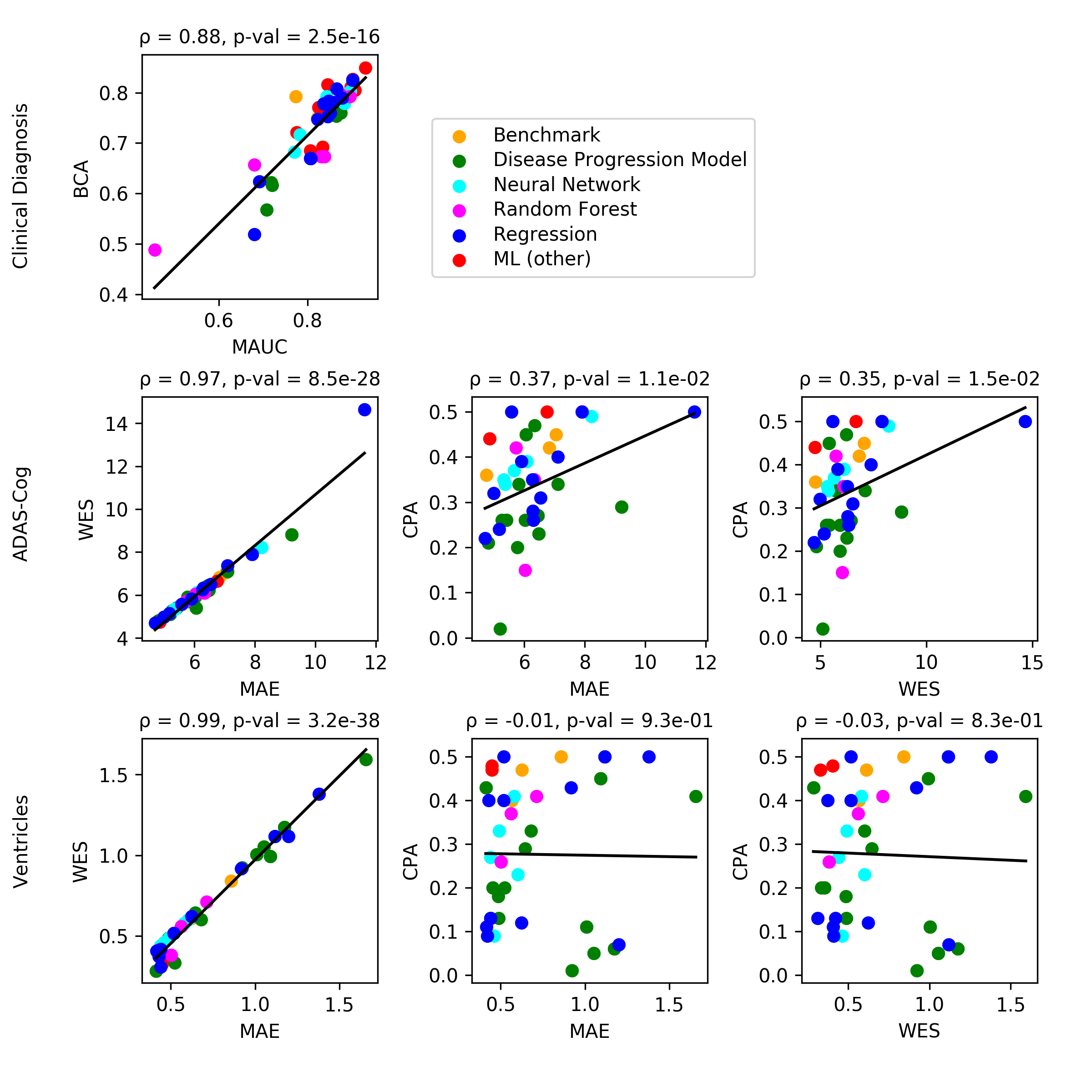}
 \caption{For D2 submissions, we show scatter plots of pairs of performance metrics for (top row) clinical diagnosis, (middle row) ADAS-Cog13 and (bottom row) Ventricles. Each dot is a participant submission, coloured according to the type of prediction algorithm used.   Correlation coefficients and p-values are given above each subplot. A few outlier submissions with ADAS MAE $>$ 20, ADAS WES $>$ 40 or Ventricle WES $>$ 3 were excluded from the analysis.}
 \label{corrMetricsD2}
\end{figure}

\begin{figure}[h]
\includegraphics[width=\textwidth]{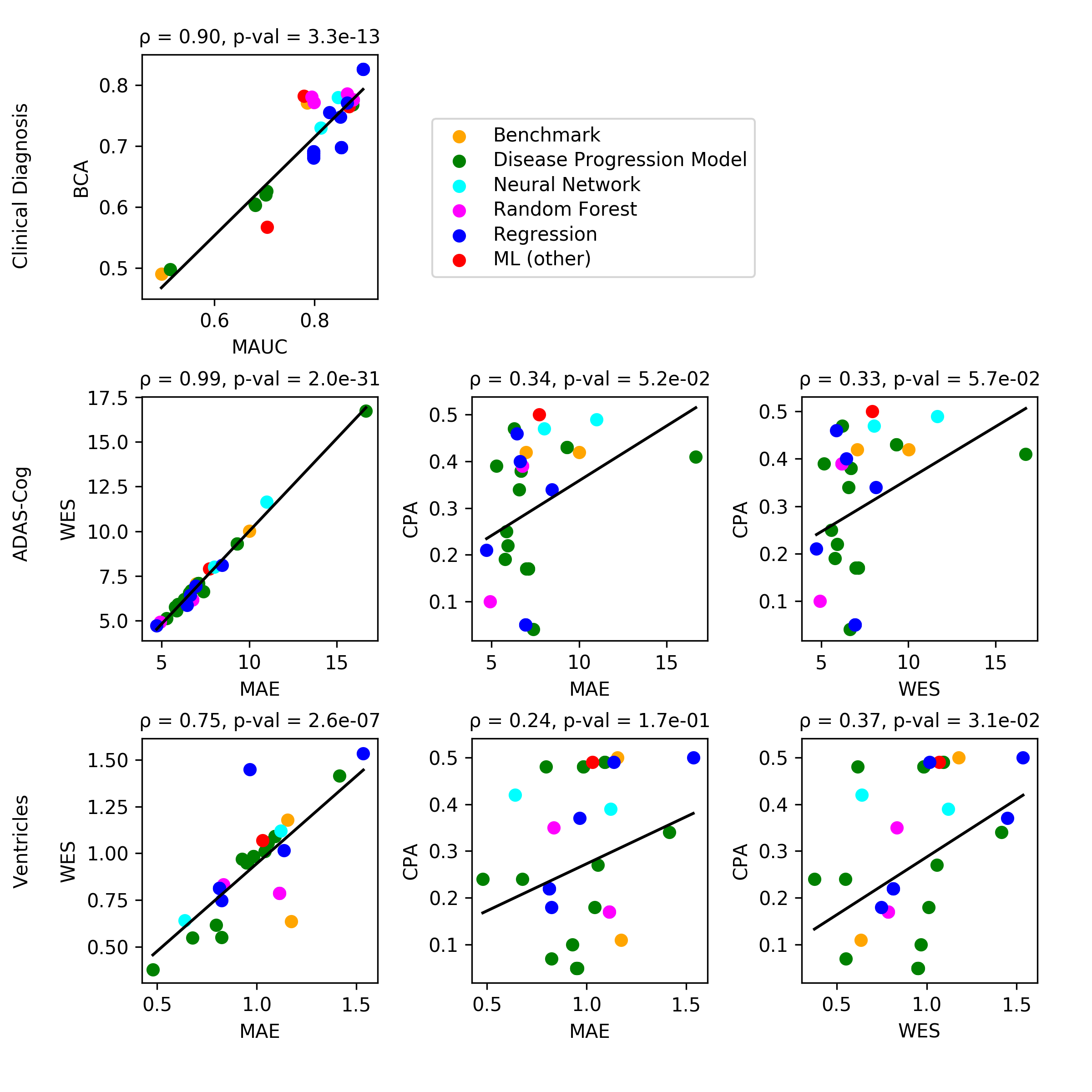}
 \caption{For D3 submissions, we show scatter plots of pairs of performance metrics for (top row) clinical diagnosis, (middle row) ADAS-Cog13 and (bottom row) Ventricles. Each dot is a participant submission, coloured according to the type of prediction algorithm used. Correlation coefficients and p-values are given above each subplot. A few outlier submissions with ADAS MAE $>$ 20, ADAS WES $>$ 40 or Ventricle WES $>$ 3 were excluded from the analysis.}
 \label{corrMetricsD3}
\end{figure}

\begin{figure}[h]
\includegraphics[width=\textwidth]{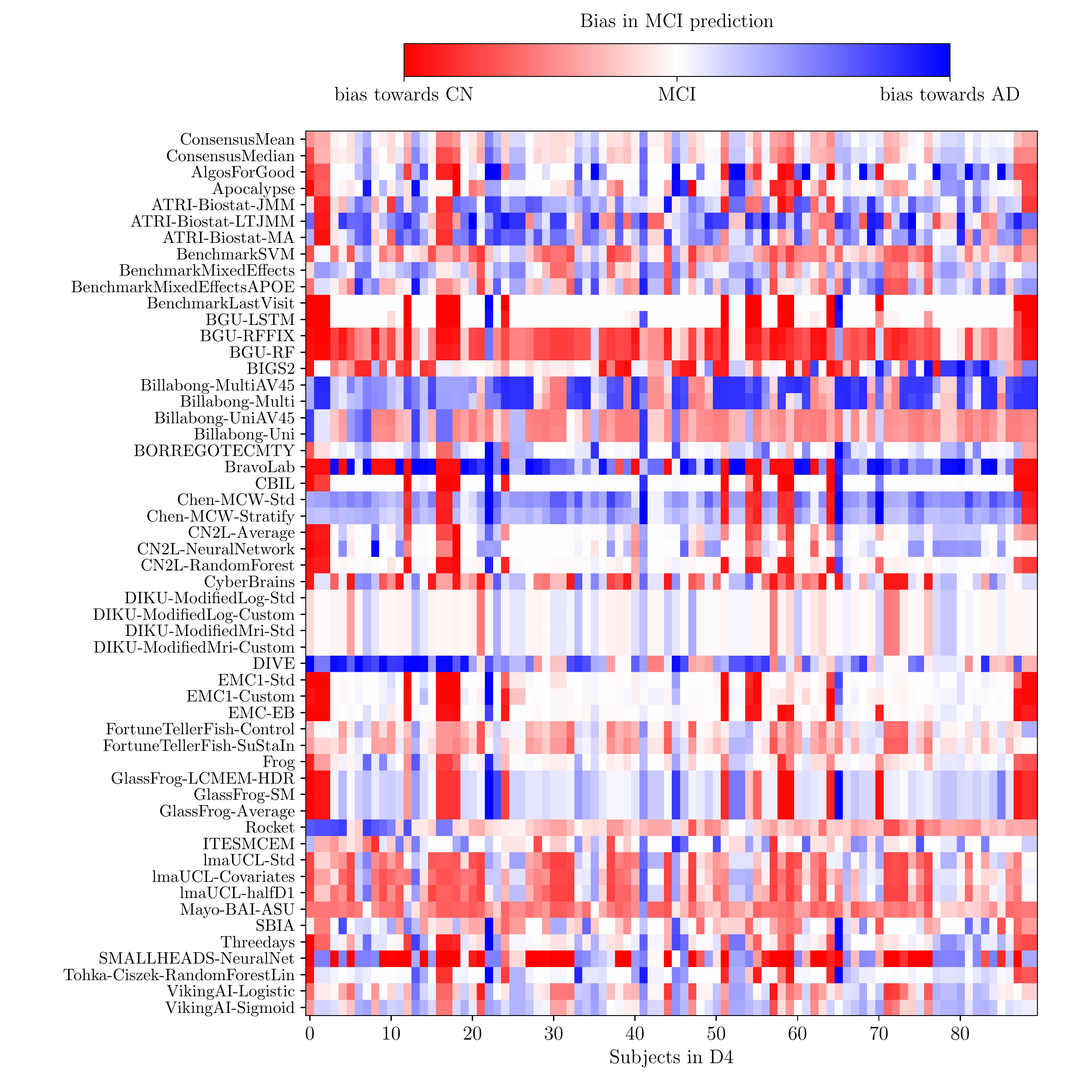}
 \caption{Bias in prediction of clinical diagnosis for MCI subjects only. X-axis shows individual subjects with designated MCI status at the clinical visit in D4, while the Y-axis shows TADPOLE algorithms. Red represents subjects which were predicted as CN with true diagnosis of MCI, while blue represents subjects predicted as AD with true diagnosis of MCI. Some algorithms show systematic biases either towards CN or AD.}
 \label{visGridMAUC}
\end{figure}

\begin{figure}[h]
\includegraphics[width=\textwidth]{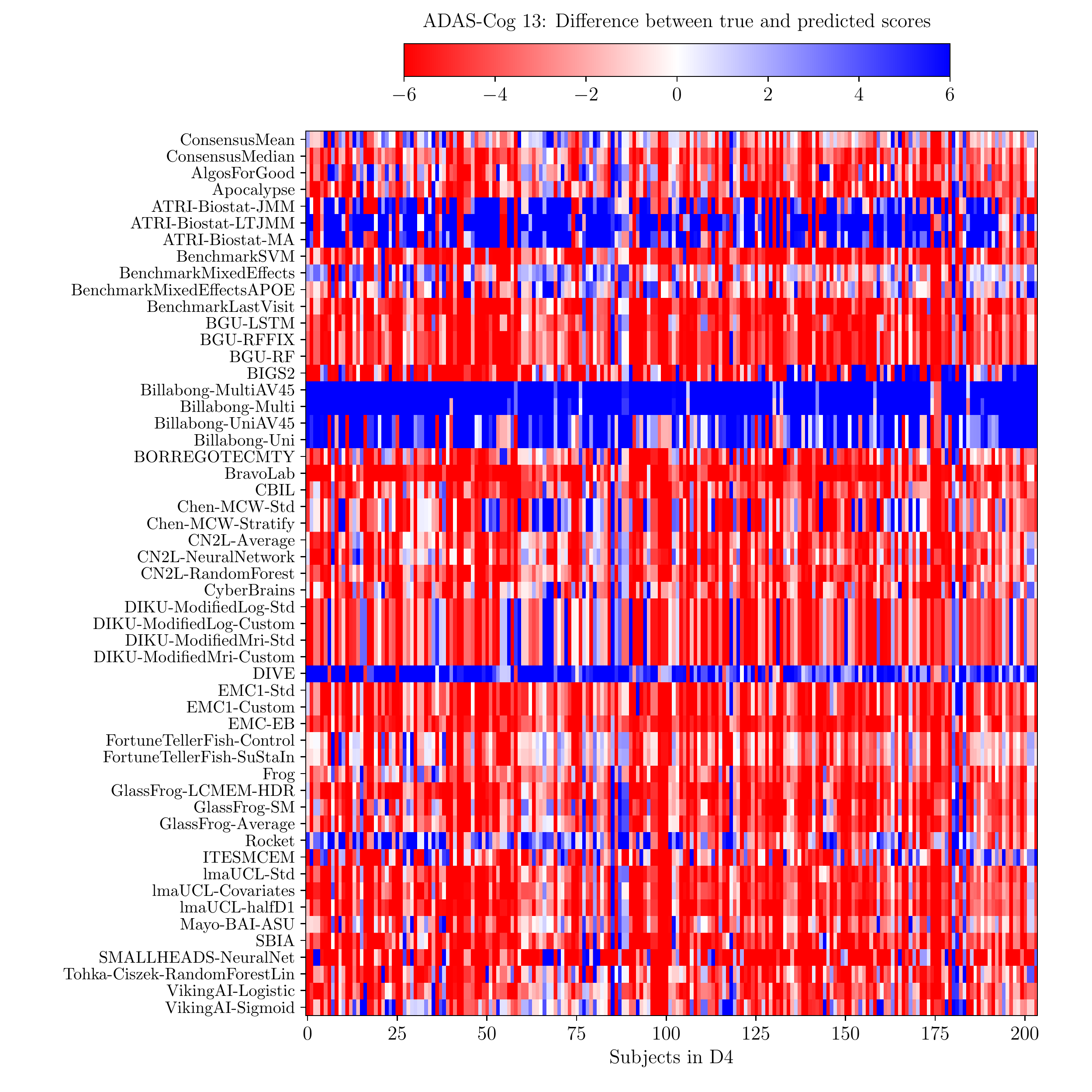}
 \caption{Bias in prediction of ADAS-Cog13. X-axis shows individual subjects with ADAS-Cog measurements in D4, while Y-axis shows TADPOLE algorithms. Red represents under-estimates while blue represents over-estimates. Most algorithms under-estimate ADAS-Cog measurements.}
 \label{visGridADAS}
\end{figure}

\begin{figure}[h]
\includegraphics[width=\textwidth]{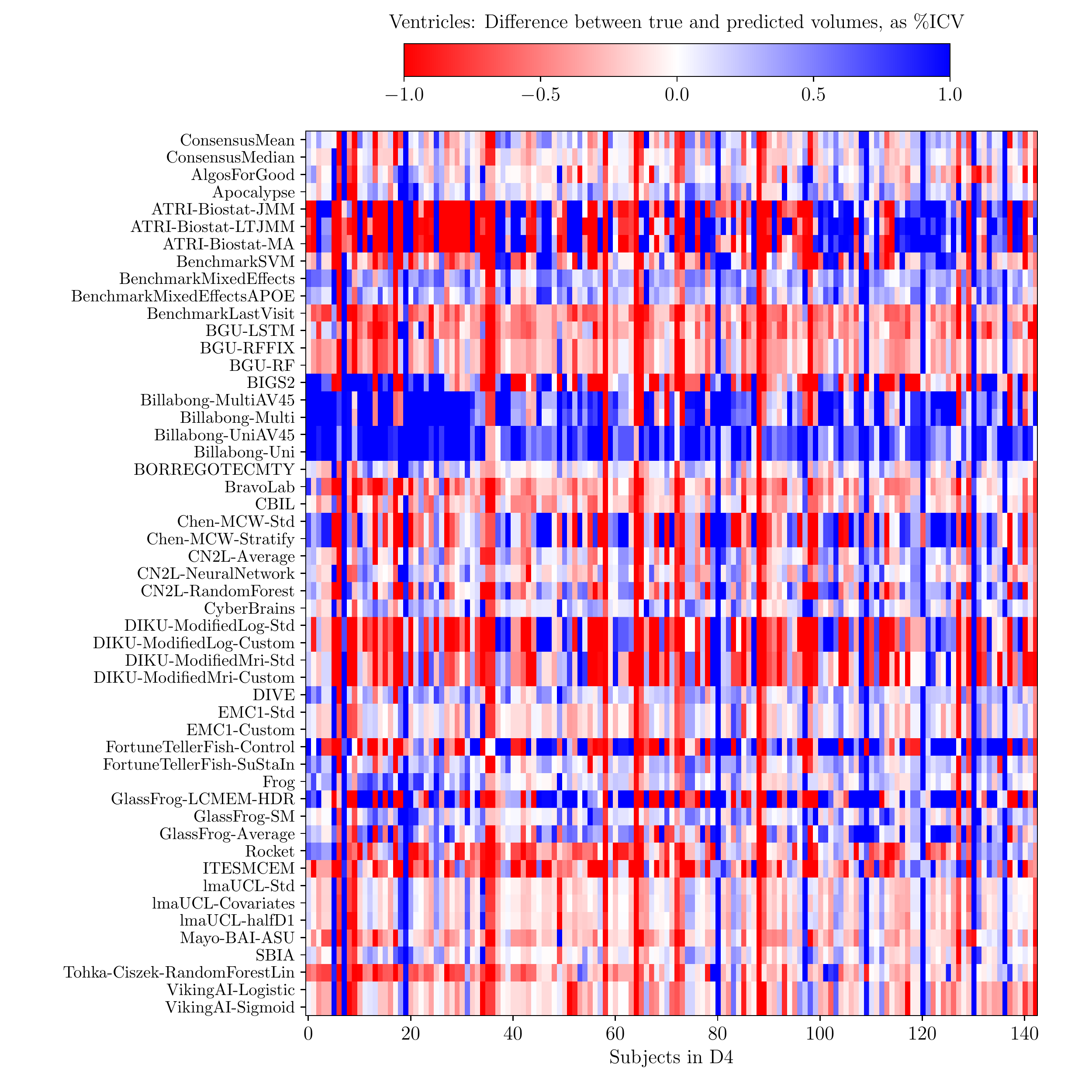}
 \caption{Bias in prediction of ventricle volume. X-axis shows individual subjects with Ventricle volume measurements in D4, while Y-axis shows TADPOLE algorithms. Red represents under-estimates while blue represents over-estimates. Some algorithms systematically under-estimate or over-estimate ventricle volume. }
 \label{visGridVENTS}
\end{figure}

\begin{figure}[h]
\includegraphics[width=\textwidth]{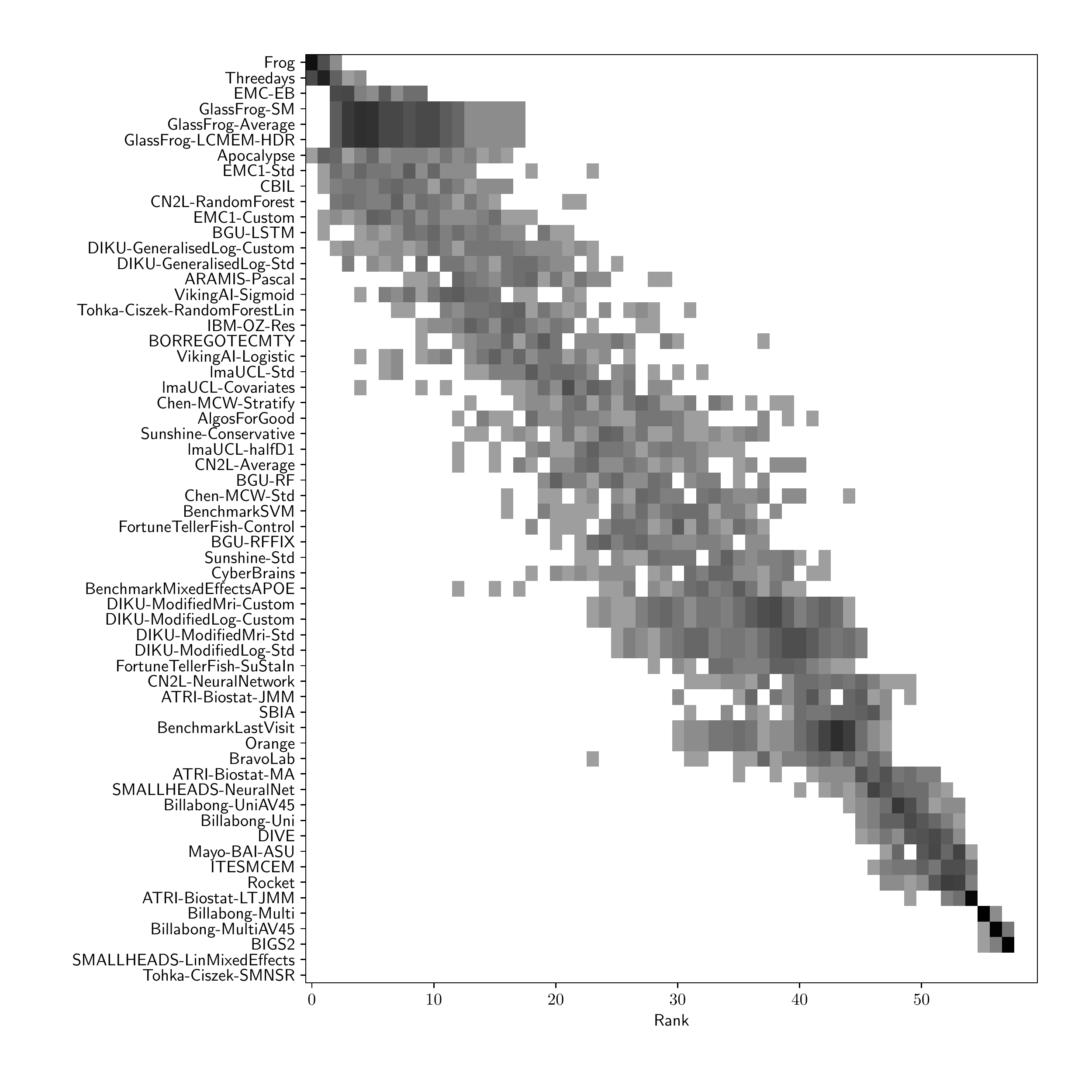}
 \caption{Distribution of ranks in clinical diagnosis MAUC for TADPOLE submissions using the longitudinal prediction set (D2), obtained from $N=50$ bootstraps of the test set (D4). More precisely, we computed the MAUC ranks given a specific bootstrap of the test set, and then for each TADPOLE submission (Y-axis) we plotted the number of times it achieved a specific rank. Figures \ref{visRanksADAS} -- \ref{visRanksVENTSD3} use the same methodology.}
 \label{visRanksMAUC}
\end{figure}

\begin{figure}[h]
\includegraphics[width=\textwidth]{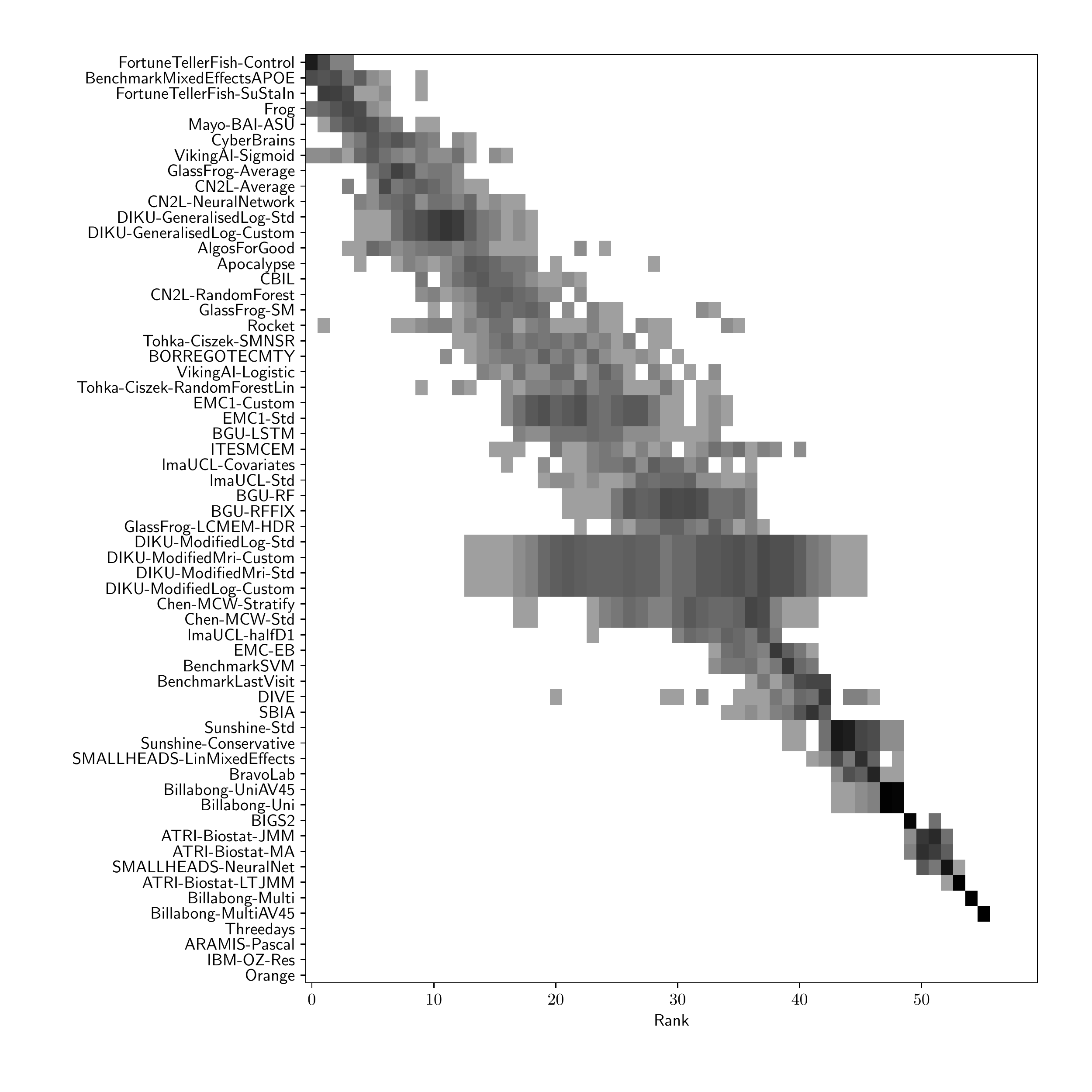}
 \caption{Distribution of ranks in ADAS-Cog13 MAE for TADPOLE submissions using the longitudinal prediction set (D2)}
 \label{visRanksADAS}
\end{figure}

\begin{figure}[h]
\includegraphics[width=\textwidth]{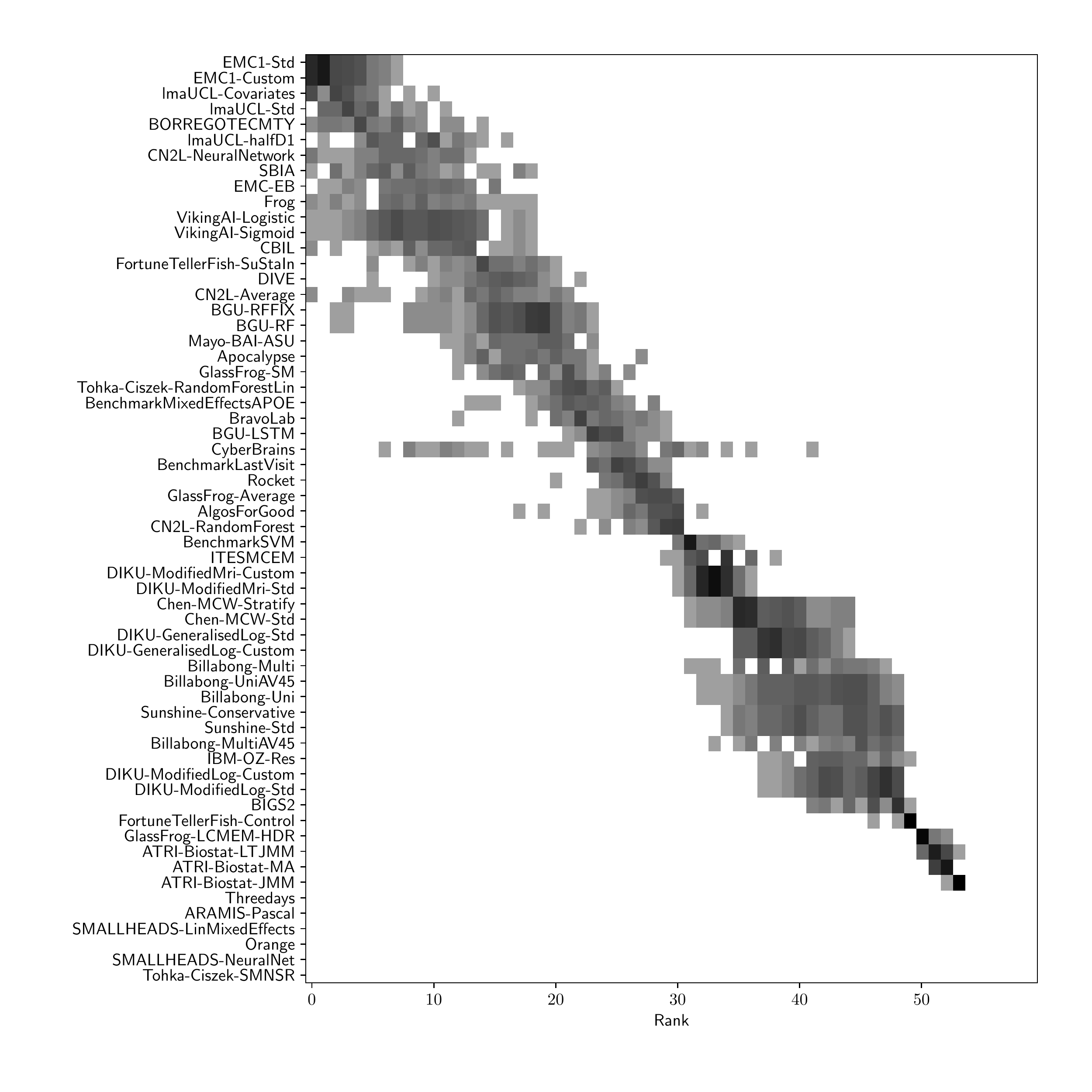}
 \caption{Distribution of ranks in Ventricle Volume MAE for TADPOLE submissions using the longitudinal prediction set (D2).}
 \label{visRanksVENTS}
\end{figure}

\begin{figure}[h]
\includegraphics[width=\textwidth]{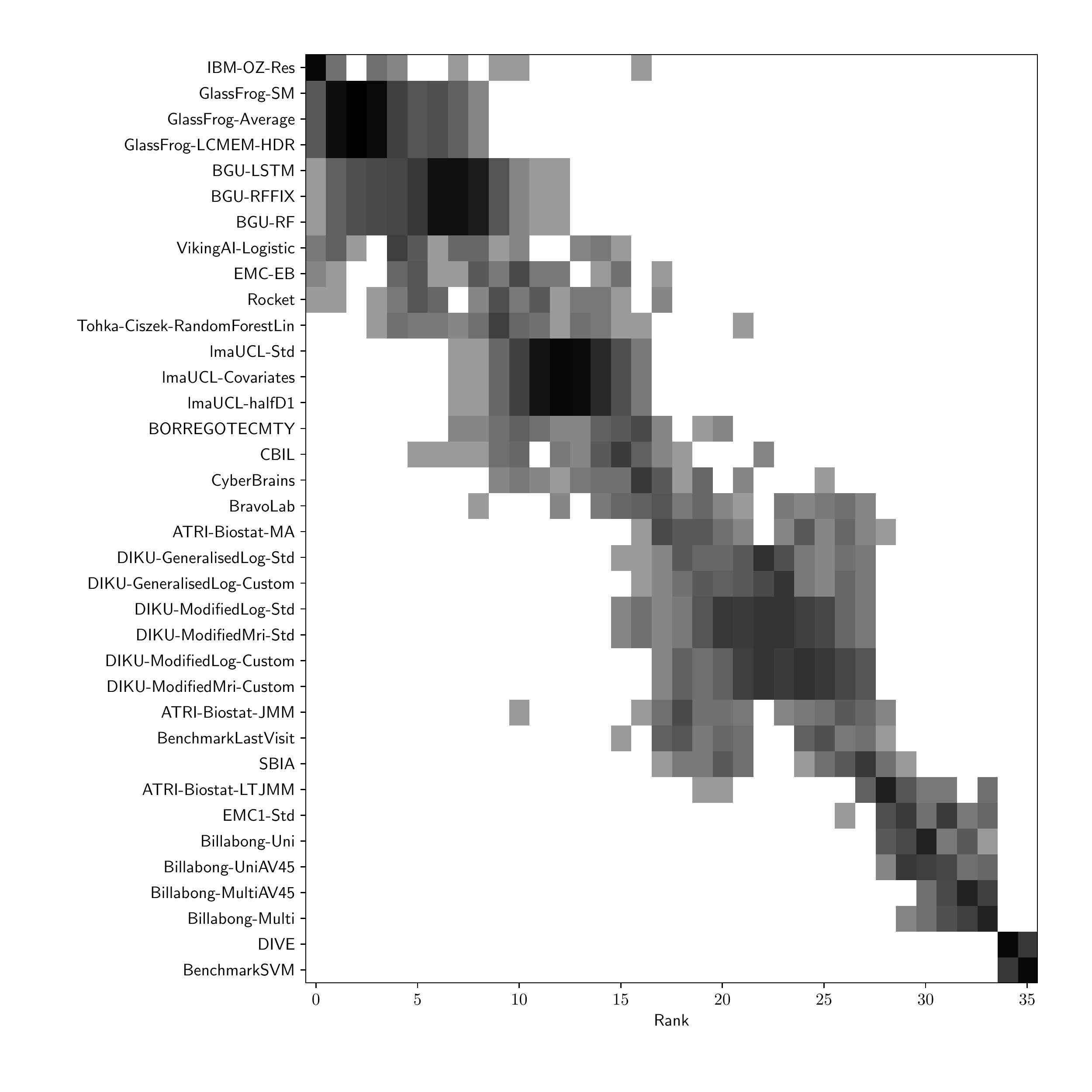}
 \caption{Distribution of ranks in clinical diagnosis MAUC for TADPOLE submissions using the longitudinal prediction set (D3).}
 \label{visRanksMAUCD3}
\end{figure}

\begin{figure}[h]
\includegraphics[width=\textwidth]{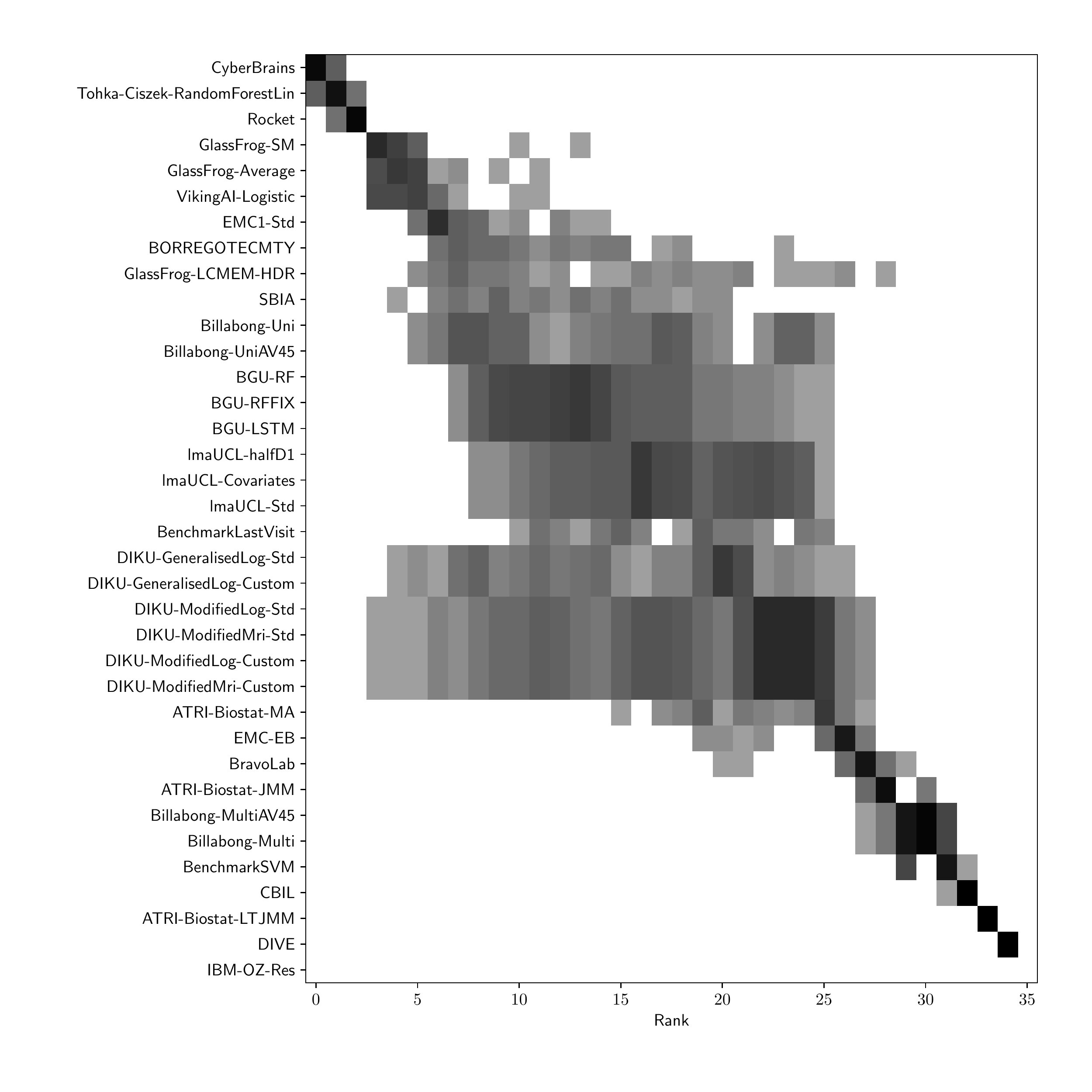}
 \caption{Distribution of ranks in ADAS-Cog13 MAE for TADPOLE submissions using the longitudinal prediction set (D3).}
 \label{visRanksADASD3}
\end{figure}

\begin{figure}[h]
\includegraphics[width=\textwidth]{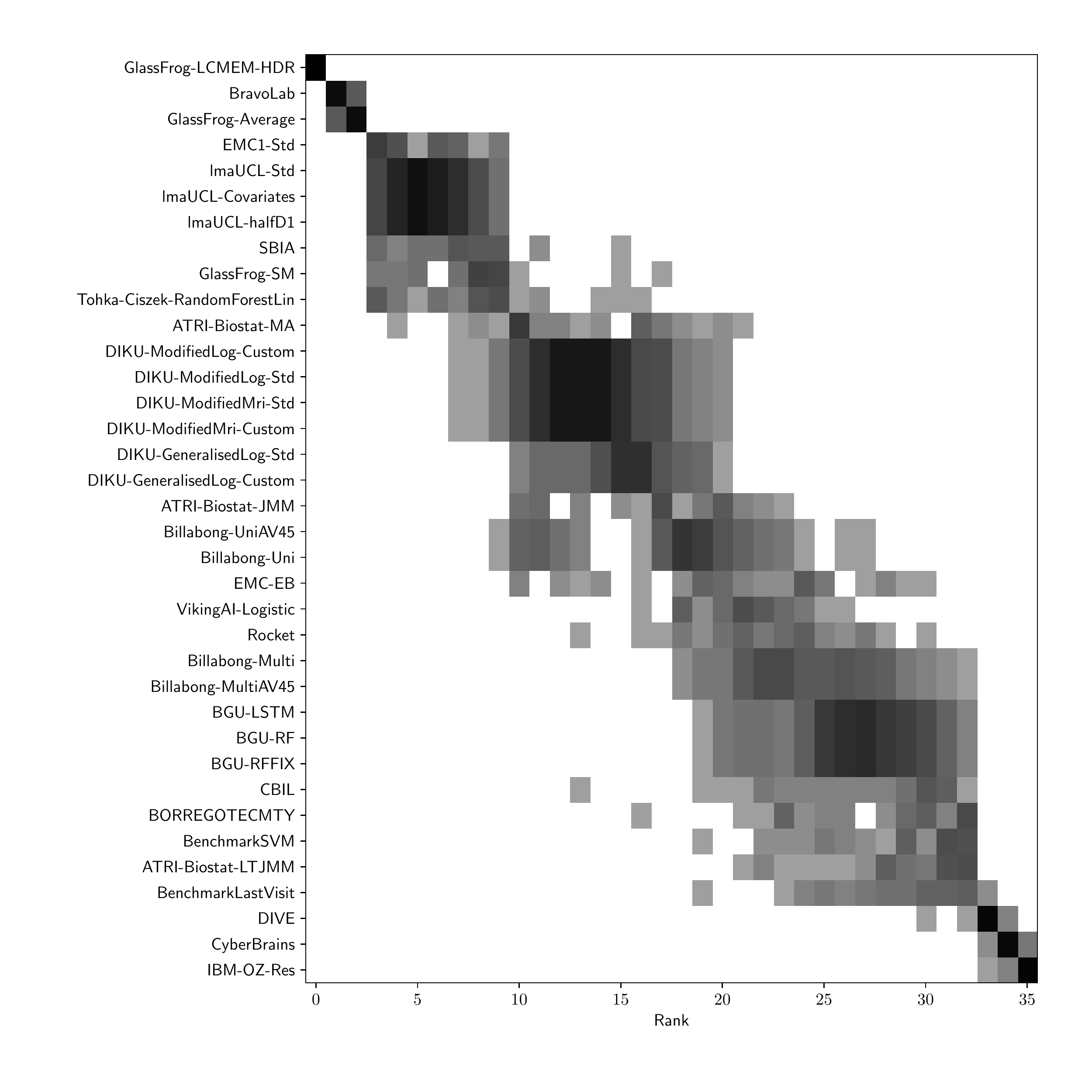}
 \caption{Distribution of ranks in Ventricle Volume MAE for TADPOLE submissions using the longitudinal prediction set (D3).}
 \label{visRanksVENTSD3}
\end{figure}

{\color{black}
\section{External validation}\label{secExternal}

This section reports an external validation experiment to verify, on an independent data set, the general finding that TADPOLE participants' algorithms perform better than simple baselines and that consensus approaches perform as well or better than individual entries. To this end, we evaluate our primary performance metrics for each prediction task on external data for the \emph{ConsensusMean} and \emph{ConsensusMedian} algorithms, all benchmark algorithms, and the challenge-winning submission (Team Frog).

\subsection{Data}
External data comes from a clinical trial and an observational study. Data from the completed DHA clinical trial (\cite{quinn2010DHA}); \url{https://www.adcs.org/dha/}) was obtained from the Alzheimer's Disease Cooperative Study (\url{https://www.adcs.org/}), including MMSE and ADAS-Cog scores from 402 AD participants over 18 months. A subset of approximately 90 also had T1-weighted MRI from which we calculated Ventricle volume (normalised by ICV) using FreeSurfer version 6.0.0. The ADAS-Cog11 scores in DHA were mapped to ADAS-Cog13 using a simple regression model trained on ADNI data. Clinical data was obtained from the Australian Imaging, Biomarkers, and Lifestyle Flagship Study of Ageing (AIBL) (\cite{ellisAIBL2009}), including MMSE and diagnosis from 
857 participants (608 CN, 144 MCI, 105 AD) 
over a 5-year period (average followup interval $1.5 \pm 1.9$ years). The data from both studies was assembled into three external datasets as follows:
\begin{itemize}
	\item \textbf{D5} (analogous to D3): The external cross-sectional prediction set containing \emph{baseline data} from the DHA clinical trial and AIBL observational study. 
	\item \textbf{D6} (analogous to D4): The external test set containing data from the n=361 DHA participants with follow-up visits providing at least one of the following variables: (i) ADAS-Cog13 score, (ii) Ventricle Volume.
	\item \textbf{D7} (analogous to D4): The external test set containing data from the n=399 AIBL participants (318 CN, 47 MCI, 34 AD) with follow-up visits, which include only clinical diagnosis.
\end{itemize}

We could not run all participants' methods directly on the external data sets, since the external evaluation necessarily occurred after the original challenge was complete.  Instead, we obtain predictions for each subject in the external data sets by copying predictions from the closest matching TADPOLE data point. We used a weighted L2-norm to find the closest match for each D5 case among TADPOLE subjects with the same sex and diagnosis. We eliminated 93 subjects from D5 for whom we could not find a sufficiently close match in D3, i.e. within $7$ years in age, $3$ points on the MMSE test, $7$ points on ADAS-Cog13, and Ventricles volume within $\pm29\%$. These cutoff values are $\sigma/\sqrt{2}$ for each variable, where $\sigma$ is the standard deviation of that variable in D5 (DHA and AIBL participants at baseline). The prediction of each quantity (diagnosis, ADAS-Cog13, ventricle volume) for a matched D5 case using a particular algorithm is then the prediction of the same quantity using the same algorithm on the matched D3 case; the matches do not vary among algorithms.

\subsection{External test results}
In D3 we found acceptable matches for 268 of 361 DHA participants and all 399 AIBL participants in D5. Supplementary Table \ref{resExternal2} shows prediction performance metrics for the test sets (D6 and D7), together with corresponding metrics from the internal test set D4 (reproduced from Figure \ref{resultsD2}).

Similar trends in predictive performance among methods arise on the external test sets as on the internal: consensus methods perform better overall than the best individual method, and substantially outperform all benchmarks; the representative strongly performing submission overall outperforms benchmarks. Results for diagnosis directly reflect that trend; consensus methods now outperform even Frog and Frog remains substantially ahead of all benchmarks. For ADAS-Cog13, as with the internal test set, Frog fails to outperform the simple benchmarks and only consensus methods approach viable (better than simple default) performance. Slightly anomalous results arise for ventricle volume prediction (Frog attains the worst MAE), which likely arises from the small sample; consensus methods still perform best. In comparison with internal performance metrics, clinical diagnosis MAUC is lower (indicating lower predictive accuracy) for D7 than D4;  ADAS-Cog13 MAE is higher (lower predictive accuracy) on D6 than D4; ventricle volume MAE is slightly lower (higher predictive accuracy) on D6 than D4. In general, we expect the imperfect matching process to reduce performance on the unseen external data set compared to the internal data set, as we observe in diagnosis and ADAS-Cog13 prediction. However, the elimination of D5 subjects for which no good match is found may push average performance up by avoiding difficult/unusual subjects, as we observe in ventricle-volume prediction. 

\begin{table}[h]
 \resizebox{1\textwidth}{!}{%
{\color{black}\begin{tabular}{l|cc|cc|cc}
\toprule
                          & \multicolumn{2}{c|}{Diagnosis}      & \multicolumn{2}{c|}{ADAS-Cog13}  & \multicolumn{2}{c}{Ventricles, \% ICV} \\
Algorithm                 & MAUC (D7) & MAUC (D4)  & MAE (D6) & MAE (D4)  & MAE (D6) & MAE (D4)  \\ \midrule
ConsensusMedian           & 0.864     & 0.925      & 6.39     & 5.12      & 0.35     & 0.38      \\ 
ConsensusMean             & 0.863     & 0.920      & 7.45     & 3.75      & 0.39     & 0.48      \\ 
Frog                      & 0.835     & 0.931      & 9.18     & 4.85      & 0.92     & 0.45      \\
BenchmarkLastVisit        & 0.787     & 0.774      & 8.88     & 7.05      & 0.70     & 0.63      \\
BenchmarkSVM              & 0.773     & 0.836      & 9.77     & 6.82      & 0.57     & 0.86      \\
BenchmarkMixedEffects     & 0.740     & 0.846      & 10.09    & 4.19      & 0.42     & 0.56      \\
BenchmarkMixedEffectsAPOE & 0.740     & 0.822      & 9.61     & 4.75      & 0.42     & 0.57      \\

\end{tabular}
}}
\caption{\textcolor{black}{External validation results using consensus models and benchmark methods on D6 (ADAS-Cog13 and Ventricles) and D7 (Diagnosis), together with internal test results on D4 (as in Figure \ref{resultsD2}). The increase in performance of consensus methods over individual methods and benchmarks remains similar. Cognitive test scores remain difficult to predict.}}
\label{resExternal2}
\end{table}

\subsection{Conclusion}

External test set results reaffirm the strong performance of consensus methods in comparison to individual TADPOLE entries and baselines, as well as the particular difficulty in predicting cognitive test scores, such as ADAS-Cog13. 
}

\end{document}